\documentclass[12pt]{article} 
\usepackage[bookmarks=false]{hyperref}
\usepackage{amsmath,amssymb,amsfonts,amsxtra, mathrsfs, makeidx,graphics,graphicx,amsthm, youngtab}

\numberwithin{equation}{section}

\newcommand{\be}{\begin{equation}}
\newcommand{\ee}{\end{equation}}
\newcommand{\beq}{\begin{equation}}
\newcommand{\eeq}{\end{equation}}
\newcommand{\ba}{\begin{array}}
\newcommand{\ea}{\end{array}}
\newcommand{\bi}{\begin{itemize}}
\newcommand{\ei}{\end{itemize}}
\def\bea#1\eea{\allowdisplaybreaks \begin{eqnarray}#1\end{eqnarray}}
\newcommand{\ben}{\begin{enumerate}}
\newcommand{\een}{\end{enumerate}}
\newcommand{\bean}{\begin{eqnarray*}}
\newcommand{\eean}{\end{eqnarray*}}
\newcommand{\eref}[1]{(\ref{#1})}

\newcommand{\fref}[1]{Figure~\ref{#1}}
\newcommand{\nn}{\nonumber}

\newcommand{\tr}{\mathop{\rm Tr}}
\newcommand{\PE}{\mathop{\rm PE}}

\newcommand{\BC}{\mathbb{C}}
\newcommand{\BR}{\mathbb{R}}

\newcommand{\BZ}{\mathbb{Z}}

\newcommand{\BH}{\mathbb{H}}

\newcommand{\BU}{\mathbf{1}}

\newcommand{\comment}[1]{}

\newcommand{\CM}{{\cal M}}

\newcommand{\CN}{{\cal N}}

\newcommand{\CP}{{\cal P}}

\newcommand{\ie}{{\it i.e.}}
\newcommand{\eg}{{\it e.g.}}

\newcommand{\ud}{\mathrm{d}}

\begin{document}

\begin{titlepage}

\begin{flushright}
IMPERIAL/TP/11/AH/07 \\ MPP-2011-93
\end{flushright}
\vskip 1.5cm
\begin{center}
{\Large \bfseries
Complete Intersection Moduli Spaces in $\CN=4$ Gauge Theories in Three Dimensions
}

\vskip 1.2cm

Amihay Hanany$^1$ and Noppadol Mekareeya$^2$

\bigskip\bigskip

$^1$ Theoretical Physics Group, Imperial College London, \\
Prince Consort Road, London,  SW7 2AZ,  UK
\bigskip

$^2$ Max-Planck-Institut f\"ur Physik (Werner-Heisenberg-Institut), \\
F\"ohringer Ring 6, 80805 M\"unchen, Deutschland
\vskip 1.5cm

\textbf{Abstract}
\end{center}

We study moduli spaces of a class of three dimensional $\CN=4$ gauge theories which are in one-to-one correspondence with a certain set of ordered pairs of integer partitions.  It was found that these theories can be realised on brane intervals in Type IIB string theory and can therefore be described using linear quiver diagrams.  Mirror symmetry was known to act on such a theory by exchanging the partitions in the corresponding ordered pair, and hence the quiver diagram of the mirror theory can be written down in a straightforward way.  The infrared Coulomb branch of each theory can be studied using moment map equations for a hyperK\"ahler quotient of the Higgs branch of the mirror theory.  We focus on three infinite subclasses of these singular hyperK\"ahler spaces which are complete intersections. The Hilbert series of these spaces are computed in order to count generators and relations, and they turn out to be related to the corresponding partitions of the theories.  For each theory, we explicitly discuss the generators of such a space and relations they satisfy in detail.  These relations are precisely the defining equations of the corresponding complete intersection space.

\bigskip

\end{titlepage}

\setcounter{tocdepth}{2}
\tableofcontents

\section{Introduction}
An infinite class of three dimensional gauge theories with $\CN=4$ supersymmetry was recently proposed by Gaiotto and Witten \cite{Gaiotto:2008ak}.\footnote{Aspects of this class of theories are also discussed in \cite{Nakajima}.}   It was found that these theories can be realised in Type IIB string theory using brane configurations discussed in \cite{Hanany:1996ie}, and hence can naturally be described by linear quiver diagrams.  In this paper, we restrict ourselves to theories whose quiver diagrams contain only unitary groups. Such a class of theories has several interesting features \cite{Gaiotto:2008ak} (see also, \eg, \cite{Benvenuti:2011ga, Nishioka:2011dq, Assel:2011xz, Gulotta:2011si, Dey:2011pt})  Three of the remarkable properties are as follows.
\ben
\item When realised on brane intervals proposed in \cite{Hanany:1996ie}, each theory is naturally in one-to-one correspondence with a certain ordered pair $(\sigma, \rho)$ of partitions of an integer $N$.  This theory is referred to in the literature as $T^{\sigma}_{~\rho} (SU(N))$.  The brane construction and a certain condition that is statisfied by $\sigma$ and $\rho$ are discussed in \cite{Gaiotto:2008ak, Nishioka:2011dq, Assel:2011xz}.  We shall briefly summarise these known results in the next section.
\item Each of these theories has a non-trivial fixed point in the low energy limit \cite{Gaiotto:2008ak}.
\item As conjectured by mirror symmetry \cite{Intriligator:1996ex}, at the fixed point the theory possesses a dual description (which is known as the {\it mirror theory}).  Mirror symmetry exchanges the Higgs branch of the original theory with the infrared Coulomb branch of the mirror theory and vice-versa.  Given a theory in this class, the mirror theory also belongs to the same class. In particular, the mirror symmetry acts on each theory by exchanging the partitions in the corresponding ordered pair.  In other words, the mirror theory of $T^{\sigma}_{~\rho} (SU(N))$ is $T^{\rho}_{~\sigma} (SU(N))$.
\een

In this paper, we focus on certain infinite subclasses of theories whose infrared Coulomb branches\footnote{For brevity, we shall refer to the `infrared Coulomb branch' simply as the `Coulomb branch'.} are {\it complete intersection} singular hyperK\"ahler cones.  By a complete intersection, we mean an algebraic variety which has a finite number of generators subject to a finite number of relations such that the dimension of the variety is equal to the number of generators minus the number of relations.  In particular, these subclasses are as follows.
\ben
\item The $(1)-(2)-\cdots -(n-1)-[n]$ theory,
\item The $(1)-(2)- \cdots -(m-1)-(m)-[n]$ theory (with $n > m$),
\item The $(k)-(2k)- \cdots -(nk-k)-[nk]$ theory.
\een
The shorthand notation above deserves some explanations.  The round brackets $(m)$ denote $U(m)$ gauge symmetry; this corresponds to a circular node in the quiver diagram.  On the other hand, the square brackets $[n]$ denote $U(n)$ global symmetry; this corresponds to a square node in the quiver diagram.  Finally, a dash denotes the bi-fundamental hypermultiplets.  We shall use this shorthand notation throughout the paper.

Let us mention some aspects of geometry which are of our concerns.  Assuming mirror symmetry, we take the Coulomb branch of a theory to be equal to the Higgs branch of the mirror theory.  (We emphasise that this is not a test of mirror symmetry, but rather a use of it as a working assumption.)  Since the metric of the latter does not get any quantum correction, the hyperK\"ahler quotient is directly given by the solutions of moment map equations (\ie~the $F$ and $D$ terms\footnote{The $R$-symmetry of the theory is $SO(4) = SU(2)_C \times SU(2)_H$, where $SU(2)_C$ acts on the Coulomb branch and $SU(2)_H$ acts on the Higgs branch.  The moment map equations, which consist of $F$ and $D$ terms, transform as triplets under $SU(2)_H$.}) of such a Higgs branch quotiented by the gauge symmetries.  Such spaces which are complete intersections constitute an interesting class of hyperK\"ahler geometry for the reason that there are finite number of relations between the generators and in many cases they can be written down explicitly.  These relations are precisely the defining equations of the corresponding algebraic varieties.

In fact, complete intersection moduli spaces are rather commonplace in gauge theory and string theory literature.  We list certain well-known examples below.
\bi
\item The orbifolds $\BC^2/\Gamma$, where $\Gamma$ are discrete ADE subgroups of $SU(2)$ \cite{Benvenuti:2006qr, Feng:2007ur}.
\item The moduli space of 1 $SU(2)$ instanton on $\BR^4$.  This space is in fact $\BC^2 \times \BC^2/\BZ_2$ (see \eg~\cite{Benvenuti:2010pq}).
\item The moduli spaces of tri-vertex theories with one or two external legs and arbitrary genus.  In particular, for a theory with genus $g$ and one external leg, the moduli space is $\BC^2/\hat{D}_{g+1}$ \cite{Hanany:2010qu}. 
\item The conifold \cite{Benvenuti:2006qr}, the cone over del Pezzo surfaces $dP_n$ with $n > 4$ \cite{Benvenuti:2006qr}, the cone over suspended pinch point, SPP or $L^{121}$ (see the defining equation in \cite{Morrison:1998cs}).  We emphasise that the constructions of these spaces involve K\"ahler quotients, but not hyperK\"ahler ones.
\item The moduli spaces of 4d $\CN=1$ supersymmetric QCD with $SU(N_c)$ gauge group and $ N_c$ flavours \cite{Gray:2008yu, Chen:2011wn}, $SU(N_c)$ gauge group with $1$ flavour and 1 adjoint chiral multiplet \cite{Hanany:2008sb}, $SO(N_c)$ gauge group and $N_c$ flavours \cite{Hanany:2008kn}, $Sp(N_c)$ gauge group with $N_c+1$ flavours \cite{Hanany:2008kn}.  Note that the constructions of these spaces involve K\"ahler quotients, but not hyperK\"ahler ones.
\ei

In order to study the moduli space of a given gauge theory, we compute a partition function which counts the number of chiral operators on such a space.  This partition function is known as the {\it Hilbert series}. For a complete intersection, the Hilbert series takes a special form from which the number of generators and relations can be read off immediately.  It therefore also provides an immediate check whether or not the space is a complete intersection.   In this paper, the Hilbert series of the Coulomb branch are calculated from the Higgs branch of the mirror theory.  Hence, the relevant computations can be done in a similar fashion to those discussed in \cite{Benvenuti:2010pq, Hanany:2010qu}.  

The plan of the subsequent parts of the paper is as follows.  In Section \ref{sec:HW}, we summarise the brane constructions of the $T^\sigma_{~\rho} (SU(N))$ theories and stating the consistency condition that needs to be satisfied by the partition $\sigma$ and $\rho$.  In the following Sections \ref{sec:12mn} and \ref{sec:k2knk}, we discuss the aforementioned infinite subclasses of theories in detail.  We analyse a number of examples in each subclass in the following order:  A brane construction, dimensions of the Higgs and Coulomb branches, the mirror theory, the Hilbert series of the Coulomb branch, and the generators and relations.  Derivations of various relations are given in Appendix \ref{app:derivations}.

Let us now discuss the brane configurations of such theories.

\section{Brane constructions of the $T^\sigma_{~\rho}(SU(N))$ theories} \label{sec:HW}
In this section, we give a summary on the brane configurations of the $T^\sigma_{~\rho}(SU(N))$ theories.  We also refer the reader to the original paper \cite{Gaiotto:2008ak} and the papers \cite{Nishioka:2011dq, Assel:2011xz} which give extensive reviews on this topic.

We start our discussion by considering supersymmetric configurations proposed in \cite{Hanany:1996ie}.  There are three types of branes such that their worldvolumes span the following directions.
\bi
\item D3-branes with worldvolume spanned by $x^0,~x^1,~x^2,~x^3$~,
\item D5-branes with worldvolume spanned by $x^0,~x^1,~x^2$ together with $x^4,~x^5,~x^6$~,
\item NS5-branes with worldvolume spanned by $x^0,~x^1,~x^2$ together with $x^7,~x^8,~x^9$~,
\ei
The D3-branes end on 5-branes in such a way that the worldvolume in the $x^3$ direction is finite.  We focus on dynamics of a gauge theory living on the D3-branes, and so macroscopically the theory is $(2+1)$ dimensional.  Since such brane configurations preserve 8 supercharges, the concerned gauge theory possesses $\CN=4$ supersymmetry.

Assume for simplicity that the Fayet-Iliopoulos parameters and the mass terms are set to zero.  A D3-brane can be stretched between two NS5-branes, or two D5-branes, or one NS5-brane and one D5-brane.    In the first case, the moduli in the positions of a D3-brane in the $7-8-9$ directions, together with the dual photon, parametrise the Coulomb branch.  In the second case,  the moduli in the positions of a D3-brane in the $4-5-6$ directions, together with the scalar coming from the component of the gauge field in the finite $x^3$-direction, parametrise the Higgs branch.  In the third case, there is no moduli in the D3-brane transverse directions.  Moreover, for a supersymmetric configuration, the number of D3-branes connecting a D5-brane with an NS5-brane is either zero or one \cite{Hanany:1996ie}.

Let us define a {\it net number} of D3-branes ending on a 5-brane to be the number of D3-branes ending on it from the right minus the number ending on it from the left.  Define also the {\it linking number of an NS5-brane} as the total number of D5-branes to the left plus the net number of D3-branes ending on this NS5-brane, and define the {\it linking number of a D5-brane} as the total number of NS5-branes to the right minus the net number of D3-branes ending on this NS5-brane.\footnote{These definitions are according to \cite{Assel:2011xz}. They differ from that in \cite{Hanany:1996ie} by unimportant signs and shifts.}  As discussed in \cite{Hanany:1996ie}, the linking number of a 5-brane is the D3-brane charge measured at infinity on that 5-brane, and it must be invariant under various brane manipulations.  An important consequence of this statement is that {\it whenever a D5-brane and an NS5-brane pass through each other, a D3-brane is created or annihilated} \cite{Hanany:1996ie}.

We take the brane ordering to be such that for the NS5-branes the linking numbers are non-decreasing from left to right, and for the D5-branes the linking numbers are non-decreasing from right to left.  Examples of these configurations are depicted in \fref{fig:1235quiv} and \fref{fig:246quiv}.  Such a brane ordering implies that for any $U(n_c)$ gauge node in the quiver diagram, the total rank $n_f$ of the nodes directly connected to this $U(n_c)$ node satisfies the condition \cite{Gaiotto:2008ak}:
\bea
n_f \geq 2n_c~;
\eea 
this condition guarantee that the quiver gauge theory has a superconformal fixed point.

Note that when all D5-branes are moved to the right of all NS5-branes, \eg~ diagrams (c) of \fref{fig:1235quiv} and \fref{fig:246quiv}.  The linking numbers for an NS5-brane and a D5-brane are respectively the net number and minus the net number of D3-branes ending on such a 5-brane.  Therefore computations involving linking numbers most easily done once all D5-branes are moved to the right of all NS5-branes.

\paragraph{Partitions of an integer.}  Let us denote by $N$ the sum of the linking numbers of all D5-branes (which is equal to that of all NS5-branes).  For our previous examples, $N=5$ in \fref{fig:1235quiv} and $N=6$ in \fref{fig:246quiv}.  Such a construction naturally corresponds to an ordered pair of partitions $(\sigma, \rho)$ of $N$ according to the following rules.
\bi
\item The partition $\sigma$ corresponds to a collection of the linking numbers of each D5-brane reading from the left to right.
\item The partition $\rho$ corresponds to a collection of the linking numbers of each NS5-brane reading from the right to left.
\ei
When all D5-branes are moved to the right of all NS5-branes, the number $N$ is simply the total D3-brane segments connecting a collection of NS5-branes on the left with a collection of D5-branes on the right.   The partitions $\sigma$ and $\rho$ are the collections of net numbers of D3-branes ending on each D5-brane and NS5-brane respectively.

In our examples, $\sigma = (1, 1, 1, 1, 1)$ and $\rho = (2, 1, 1,1)$ in \fref{fig:1235quiv} and $\sigma = (1,1,1,1,1,1)$ and $\rho = (2,2,2)$ in \fref{fig:246quiv}.  Each partition can also be written in terms of Young diagram such that the number in the $i$-th slot is equal to the number of boxes in the $i$-th row (reading from top to bottom), \eg, for \fref{fig:1235quiv}, 
\bea
\sigma = (1,1,1,1,1) = {\tiny \yng(1,1,1,1,1)}~, \qquad  \rho = (2,1,1,1) = {\tiny \yng(2,1,1,1)}~.
\eea
Observe that
\bi
\item The total number of rows in the partition $\sigma$ is the number of D5-branes.
\item The total number of rows in the partition $\rho$ is the number of NS5-branes. 
\ei

\paragraph{Condition on the partitions.}  In order to guarantee that the brane configuration is supersymmetric and does not break into disconnected configurations, the partitions $\sigma$ and $\rho$ are chosen such that they satisfy the condition \cite{Nishioka:2011dq, Assel:2011xz} that {\it the total number of boxes up to any $i$-th row of $\sigma^T$ is strictly greater than the total number of boxes up to the $i$-th row of $\rho$}. Here $\sigma^T$ denotes the transpose of the Young diagram $\sigma$.   This statement can be written using the shorthand notation
\bea
\sigma^T > \rho~. \label{consistency}
\eea

To illustrate this, let us consider various examples that violate condition \eref{consistency}:
\ben
\item {\bf The partitions $\sigma = (1,1),~\rho = (2)$.} Therefore $\sigma^T = (2)$.  This corresponds to a setup in which one NS5-brane on the left and connected to two D5-branes on the right, each by one D3-brane.  By moving both D5-branes to the left of the NS5-brane, the D5-branes are completely detached from the NS5-branes, \ie~ the original brane configuration breaks into disconnected ones.
\item {\bf The partitions $\sigma = (2),~\rho = (2)$.} Therefore $\sigma^T = (1,1)$. This corresponds to a setup in which one NS5-brane is connected to one D5-brane by two D3-branes.  This however is not a supersymmetric configuration and is not of our interest.
\item {\bf The partitions $\sigma = (1,1,1),~\rho = (3)$.} Therefore $\sigma^T = (3)$.  This corresponds to a setup in which one NS5-brane on the left and connected to three D5-branes on the right, each by one D3-brane.  A similar situation to Example 1 occurs, \ie~ the original brane configuration breaks into disconnected ones.
\item {\bf The partitions $\sigma = (2,1),~\rho = (3)$.} Therefore $\sigma^T = (2,1)$. This corresponds to a setup in which three D3-branes connecting one NS5-brane on the left with two D5-brane on the right.  This configuration is not supersymmetric.
\item {\bf The partitions $\sigma = (3),~\rho = (3)$.} Therefore $\sigma^T = (1,1,1)$. This corresponds to a setup in which one NS5-brane is connected to one D5-brane by three D3-branes.  This is not a supersymmetric configuration.
\een

\paragraph{Mirror symmetry.}  As pointed out in \cite{Hanany:1996ie}, the mirror symmetry acts on such brane configuration by exchanging the D5-branes with the NS5-branes and exchange the $x^4,x^5,x^6$ directions with the  $x^7 , x^8 , x^9$.  It follows from the above construction that the mirror symmetry acts on the $T^\sigma_{~\rho}(SU(N))$ theory by exchanging $\sigma$ and $\rho$.  Note that the consistency condition \eref{consistency} is equivalent to $\rho^T > \sigma$; hence the mirror theory of $T^\sigma_{~\rho}(SU(N))$ also belongs to the same class as $T^\sigma_{~\rho}(SU(N))$.  In other words, {\it the mirror theory of $T^\sigma_{~\rho}(SU(N))$ is $T^\rho_{~\sigma}(SU(N))$.}  The corresponding brane and quiver diagrams for the mirror theory can be obtained from the rules discussed above, with the words `left' and `right' exchanged.  For example, the mirror configurations of  \fref{fig:1235quiv} and \fref{fig:246quiv} are depicted in \fref{fig:mirror1235} and \fref{fig:mirror246} respectively.

Having been summarising the general setup of the $T^\sigma_{~\rho}(SU(N))$, we are now focusing on the theories which are of the main interest of this paper.

\section{The $(1)-(2)- \cdots -(m)-[n]$ theory and its mirror} \label{sec:12mn}
In this section, we consider the $(1)-(2)- \cdots -(m)-[n]$ theory (with $n >m$) and its mirror.  
Subsequently, we compute the Hilbert series of the Coulomb branch of the former and find that the space is a complete intersection.
As a warm-up exercise, we begin the section by considering a special case of $(1)-(2)- \cdots -(n-1)-[n]$ theory, whose Higgs and Coulomb branches are identical.

\subsection{Special case: The $(1)-(2)- \cdots -(n-1)-[n]$ theory} \label{sec:123n}
\begin{figure}[htbp]
\begin{center}
\includegraphics[height=2.4 in]{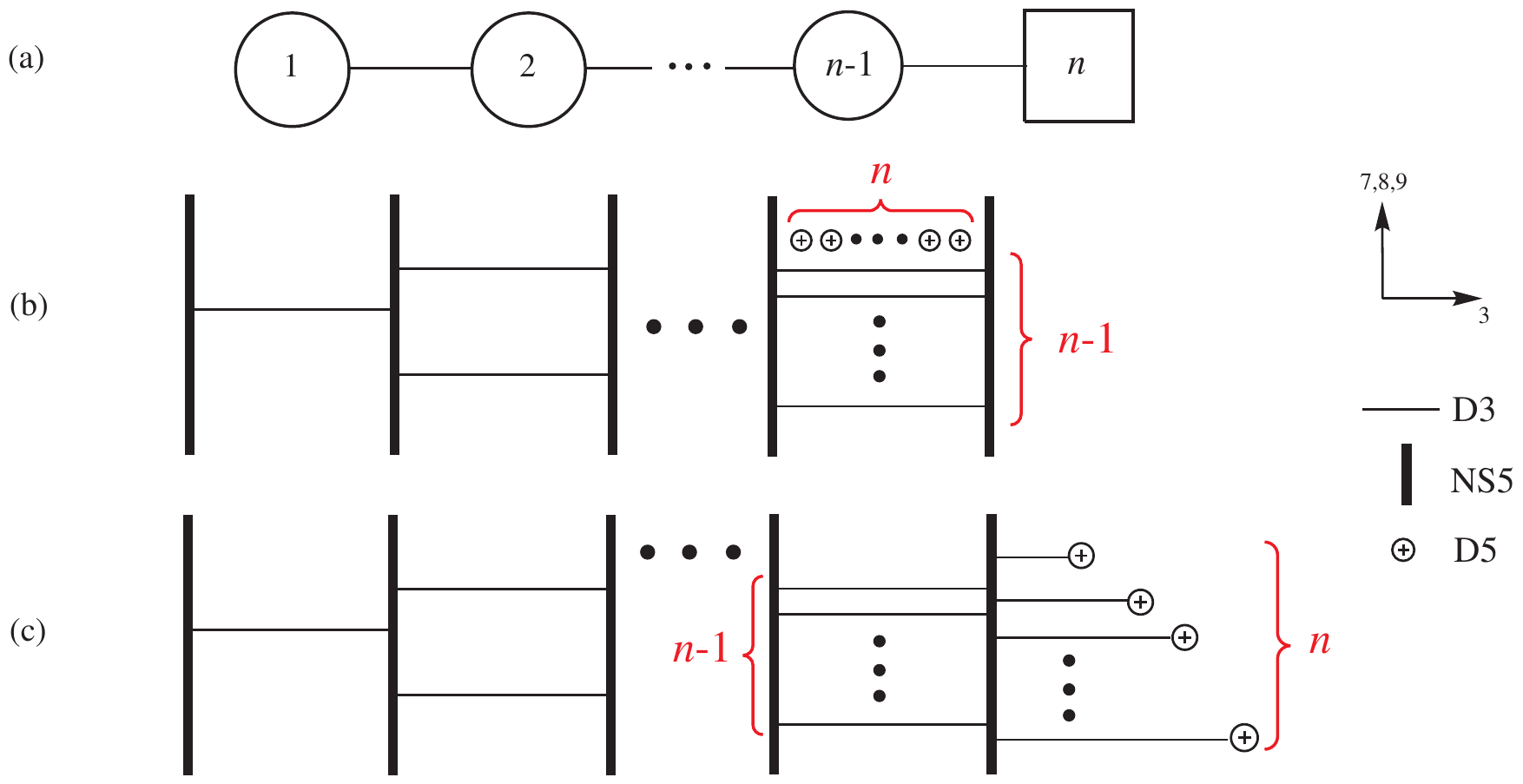}
\caption{(a) The quiver diagram of the $(1)-(2)- \cdots -(n-1)-[n]$ theory. (b) The corresponding brane configuration. (c) The D5-branes are moved to the right of all NS5-branes.  The D3-branes are created according to \cite{Hanany:1996ie}.  The partitions $\sigma = (1, \ldots,1)$ and $\rho = (1, \ldots,1)$ (with $n$ one's) are in one-to-one correspondence with this diagram.}
\label{fig:123nquiv}
\end{center}
\end{figure}

Let us consider the $(1)-(2)- \cdots -(n-1)-[n]$ theory.  
The quiver diagram and the corresponding brane configuration are depicted in \fref{fig:123nquiv}.

From diagram (c), this theory can be identified with $T^\sigma_{\rho}(SU(n))$, where $\sigma$ and $\rho$ are the following partitions of $n$:
\bea
\sigma = \rho = {\scriptsize \young(~,~,\vdots,~)}~{\scriptsize \text{($n$ boxes)}} =(\underbrace{1,1, \ldots,1}_{n~\text{one's}})~. \label{par123nquiv}
\eea

The dimension of the moduli space can be computed from the quiver diagram.
The quaternionic dimension of the Higgs branch of this theory is
\bea
\dim_{\BH} \text{Higgs}_{(1)-(2)- \cdots -(n-1)-[n]} = \sum_{i=1}^{n-1} i(i+1) - \sum_{i=1}^{n-1} i^2 = \frac{1}{2}n(n-1) ~. \label{dimhiggs123n}
\eea
On the other hand, the quaternionic dimension of the Coulomb branch of this theory is
\bea
\dim_{\BH} \text{Coulomb}_{(1)-(2)- \cdots -(n-1)-[n]} = \sum_{i=1}^{n-1} i = \frac{1}{2}n(n-1) ~. \label{dimcou123n}
\eea
Observe that the dimensions of the Higgs and the Coulomb branches are equal.  
This is in agreement with the known property by mirror symmetry that the Higgs and Coulomb branches are identical (see, \eg~ \cite{Gaiotto:2008ak, Benvenuti:2011ga, Nishioka:2011dq}).

\subsection*{The mirror theory}
\begin{figure}[htbp]
\begin{center}
\includegraphics[height=3 in]{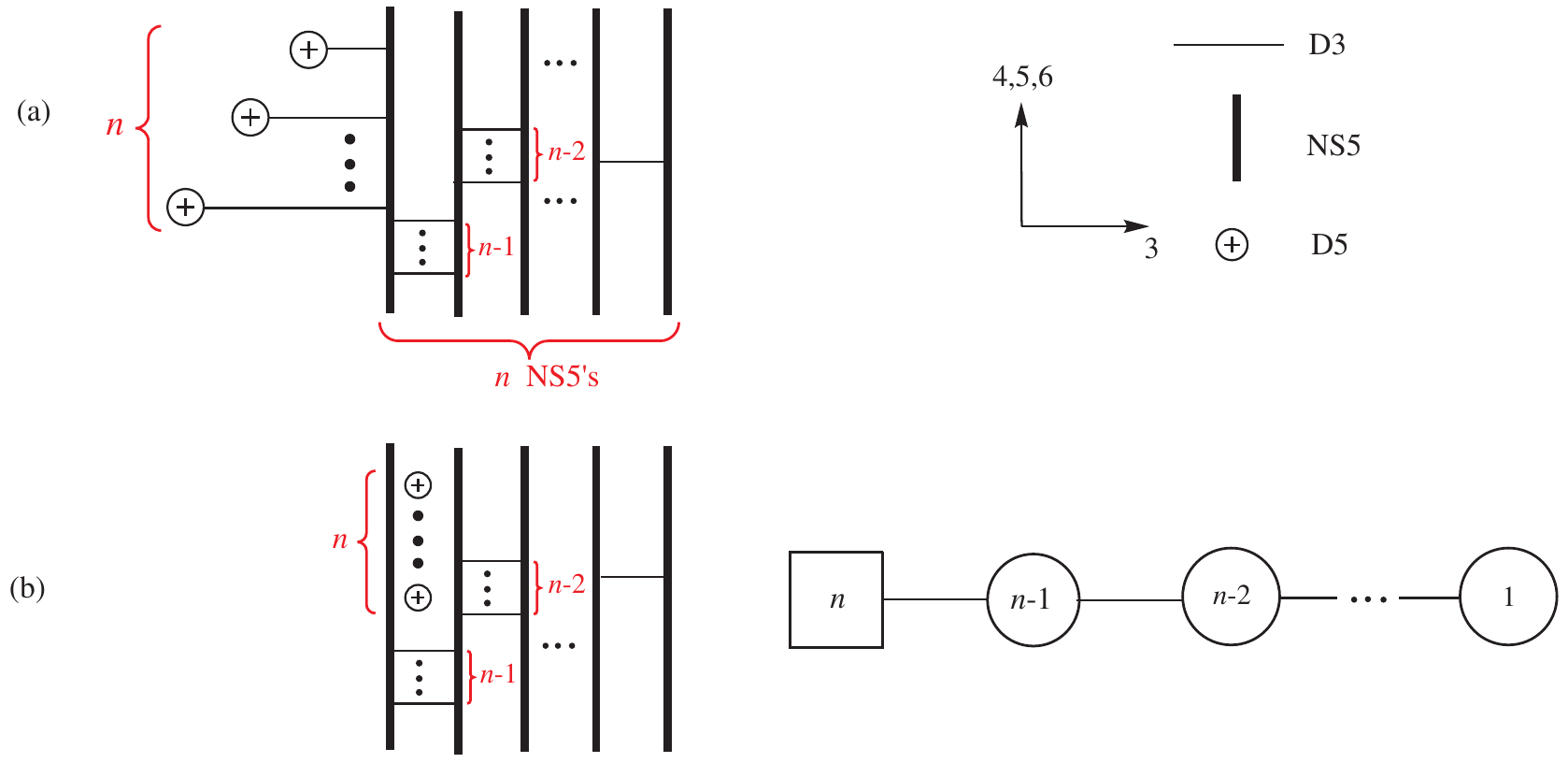}
\caption{{\it The mirror of the $(1)-(2)- \cdots -(n-1)-[n]$ theory.} (a) From diagram (c) in \fref{fig:123nquiv}, the NS5-branes and the D5-branes are exchanged and the directions $x^4, x^5, x^6$ are rotated into $x^7, x^8, x^9$ and {\it vice-versa}. (b) The D5-branes are moved across the NS5-branes. The D3-brane creation and annihilation are according to \cite{Hanany:1996ie}. The corresponding quiver diagram is also given next to the brane configuration.   Observe that this is actually the $(1)-(2)- \cdots -(n-1)-[n]$ theory.  Thus, the theory is self-mirror.}
\label{fig:mirror123n}
\end{center}
\end{figure}

Now let us consider the mirror of the $(1)-(2)- \cdots -(n-1)-[n]$ theory.  The brane configuration of the mirror theory can be obtained as described in \cite{Hanany:1996ie} and is depicted in \fref{fig:mirror123n}.  It is easy to see that this theory is {\it self-mirror}; the quiver diagram depicted in \fref{fig:123nquiv} (a) and \fref{fig:mirror123n} (b) are identical -- one is simply written in a reverse fashion from the other. This leads to the result that the Higgs and Coulomb branches of the $(1)-(2)- \cdots -(n-1)-[n]$ theory (and, of course, its mirror) are identical.

From the point of view of the NS5-brane theory, the end of a D3-brane looks like a magnetic monopole.  Using diagram (c) of \fref{fig:124quiv}, we can interpret the Coulomb branch (and hence the Higgs branch) of the $(1)-(2)- \cdots -(n-1)-[n]$ theory as the moduli space of $SU(n)$ monopoles: one with magnetic charge $(1,-1,0 \ldots,0)$, two with magnetic charge $(0,1,-1,0, \ldots,0)$, ..., $(n-1)$ with magnetic charge $(0, \ldots,0,1,-1)$, in the presence of $n$ fixed Dirac monopoles with magnetic charge $(0,\ldots,0,1)$ represented by the four rightmost semi-infinite D3-branes.

\subsubsection{The Hilbert series of the Higgs (or Coulomb) branch of the $(1)-(2)- \cdots -(n-1)-[n]$ theory}
We claim that the Hilbert series of the Higgs branch (or Coulomb) branch of the $(1)-(2)- \cdots -(n-1)-[n]$ theory is
\bea
H_{(1)-(2)- \cdots -[n]} (t, x_1, \ldots, x_{n-1})= \PE \left[ [1,0, \ldots, 0,1]_{SU(n)} t^2 \right] \prod_{q=2}^n (1-t^{2q})~, \label{hs123n}
\eea
where $x_1, \ldots, x_{n-1}$ are the $SU(n)$ global fugacities.   We prove this formula inductively in Appendix \ref{app:HS123n}.

The Hilbert series indicates that the Coulomb branch of the $(1)-(2)- \cdots -(n-1)-[n]$ theory is indeed a {\it complete intersection}. There are $n^2-1$ generators at order $t^2$ in the adjoint representation of $SU(n)$, and one relation at each order $t^{2q}$ with $q=2, \ldots,n$.  These altogether give $(n^2-1)-(n-1)=n(n-1)$ complex dimensional space or, equivalently, $\frac{1}{2}n(n-1)$ quaternionic dimensional space -- in agreement with \eref{dimhiggs123n} and \eref{dimcou123n}.
\begin{figure}[htbp]
\begin{center}
\includegraphics[height=0.6 in]{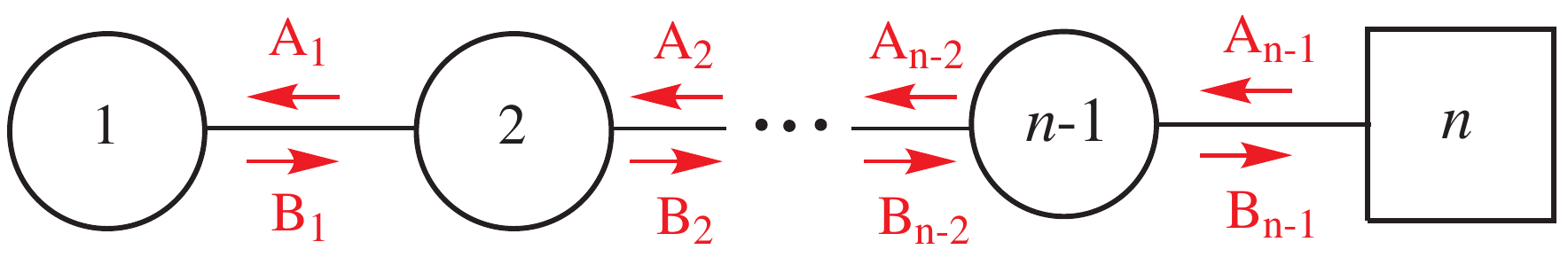}
\caption{The quiver diagram of the mirror of the $(1)-(2)- \cdots -(n-1)-[n]$ theory, with the labels of bi-fundamental chiral multiplets.}
\label{fig:mir123nfield}
\end{center}
\end{figure}

\paragraph{The $F$-terms.} From \fref{fig:mir123nfield}, the F-term constraints for the bi-fundamental chiral fields are (see also (3.4) of \cite{Gaiotto:2008ak}):
\bea
0 &=& (B_1)_{a_2} (A_1)^{a_2}~, \label{Fterms123na} \\
0 &=& -(A_i)^{a_{i+1}}_{~b_{i}} (B_i)^{b_{i}}_{~b_{i+1}} + (B_{i+1})^{a_{i+1}}_{~b_{i+2}} (A_{i+1})^{b_{i+2}}_{~b_{i+1}} \quad (\text{with $1\leq i \leq n-2$})~, \label{Fterms123nb}
\eea
where $a_r, b_r = 1, \ldots, r$ are the indices corresponding to the group $U(r)$.

\paragraph{The generators.} The generators of this theory are 
\bea
M^{a_n}_{~b_n} = ( A_{n-1})^{a_n}_{~a_{n-1}} (B_{n-1})^{a_{n-1}}_{~b_n}~.
\eea
Note that it follows from the $F$-terms \eref{Fterms123na} and \eref{Fterms123nb} that
\bea
\tr M = M^{a_n}_{~a_n} = 0~. 
\eea
Thus, $M$ transforms in the adjoint representation of $SU(n)$.

\paragraph{The relations.} Note that the $F$-term constraints \eref{Fterms123na} and \eref{Fterms123nb} implies that the $(i+1) \times (i+1)$ matrices $A_i \cdot B_i$ and $B_{i+1} \cdot A_{i+1}$~$(\text{with $1\leq i \leq n-2$})$ are nilpotent, and therefore the $n\times n$ matrix $M$ is also nilpotent.\footnote{Here we use the following lemma:  Let $A$ be an $m \times n$ matrix and let $B$ be an $n \times m$ matrix.  If $AB$ is nilpotent (\ie~ all eigenvalues of $AB$ are zero), then $BA$ is also nilpotent.  The proof of this lemma is rather amusing. Suppose that $(AB)^k = 0$ for some positive integer $k$.  Then we can rewrite this relation as $ 0= A (BA)^{k-1} B$.  After multiplying $B$ on the left and multiplying $A$ on the right, we have $0=(BA)^{k+1}$.  In other words, $BA$ is nilpotent. \label{fn:lemma}} Thus, all eigenvalues of $M$ are zero.  The relations are therefore
\bea
\tr (M^p) = 0 \quad (\text{with $1 \leq p \leq n$})~. \label{relation123n}
\eea
These are in agreement with the Hilbert series \eref{hs123n}.  These relations can also be checked using {\tt STRINGVACUA} package \cite{StringvacuaBlock}; the method is described in \cite{Gray:2009fy}.

\subsection{Example: The $(1)-(2)-[4]$ theory and its mirror}
\begin{figure}[htbp]
\begin{center}
\includegraphics[height=2.4 in]{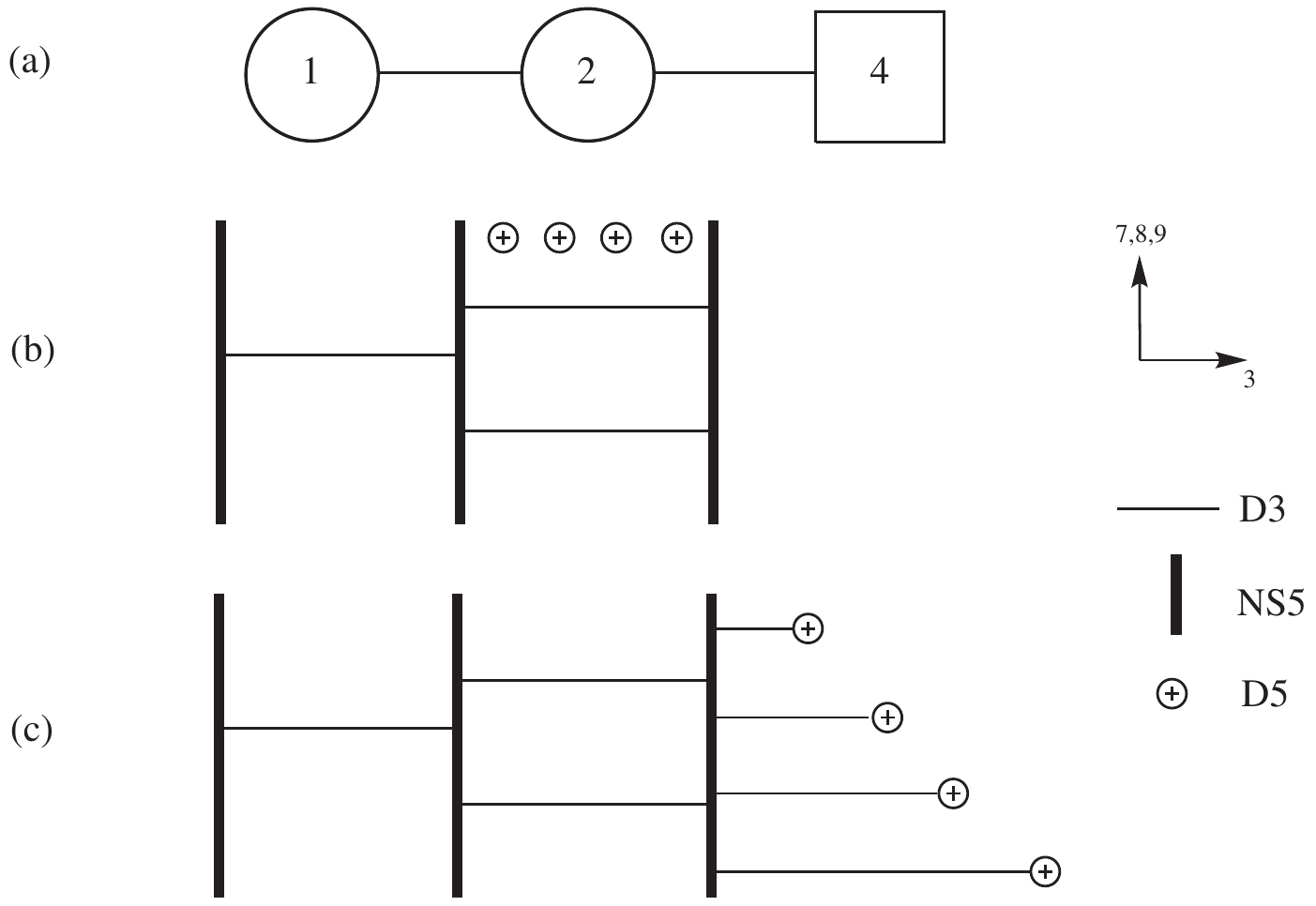}
\caption{(a) The quiver diagram of the $(1)-(2)-[4]$ theory. (b) The corresponding brane configuration. (c) The D5-branes are moved to the right of all NS5-branes.  The D3-branes are created according to \cite{Hanany:1996ie}.  The partitions $\sigma = (1,1,1,1)$ and $\rho = (2,1,1)$ are in one-to-one correspondence with this diagram.}
\label{fig:124quiv}
\end{center}
\end{figure}

Let us consider the $(1)-(2)-[4]$ theory.  The quiver diagram and the corresponding brane configuration are depicted in \fref{fig:124quiv}.

From diagram (c), this theory can be identified with $T^\sigma_{\rho}(SU(4))$, where
\bea
\sigma = {\tiny \yng(1,1,1,1)} = (1,1,1,1)~, \qquad \qquad \rho = {\tiny \yng(2,1,1)} = (2,1,1)~, \label{par124quiv}
\eea
and the number 4 in $SU(4)$ indicates the total number of boxes in each partition and this is also the number of D5-branes present in \fref{fig:124quiv}.  

One can compute the dimension of the moduli space from the quiver diagram.  
The quaternionic dimension of the Higgs branch of this theory is
\bea
\dim_{\BH} \text{Higgs}_{(1)-(2)-[4]} = (1\times 2) + (2 \times 4) - 1^2 - 2^2 = 5~. \label{dimhiggs124}
\eea
On the other hand, the quaternionic dimension of the Coulomb branch of this theory is
\bea
\dim_{\BH} \text{Coulomb}_{(1)-(2)-[4]} = 1+2 = 3~. \label{dimcou124}
\eea

\paragraph{The moduli space of monopoles.}  From the point of view of the NS5-brane theory, the end of a D3-brane looks like a magnetic monopole.  Using diagram (c) of \fref{fig:124quiv}, we can interpret the Coulomb branch of the $(1)-(2)-[4]$ theory as the moduli space of $SU(3)$ monopoles, one with magnetic charge $(1,-1,0)$ and two with magnetic charges $(0,1,-1)$, in the presence of four fixed monopoles with magnetic charge $(0,0,1)$ represented by the four rightmost semi-infinite D3-branes.

\subsection*{The mirror theory}

\begin{figure}[htbp]
\begin{center}
\includegraphics[height=2.8 in]{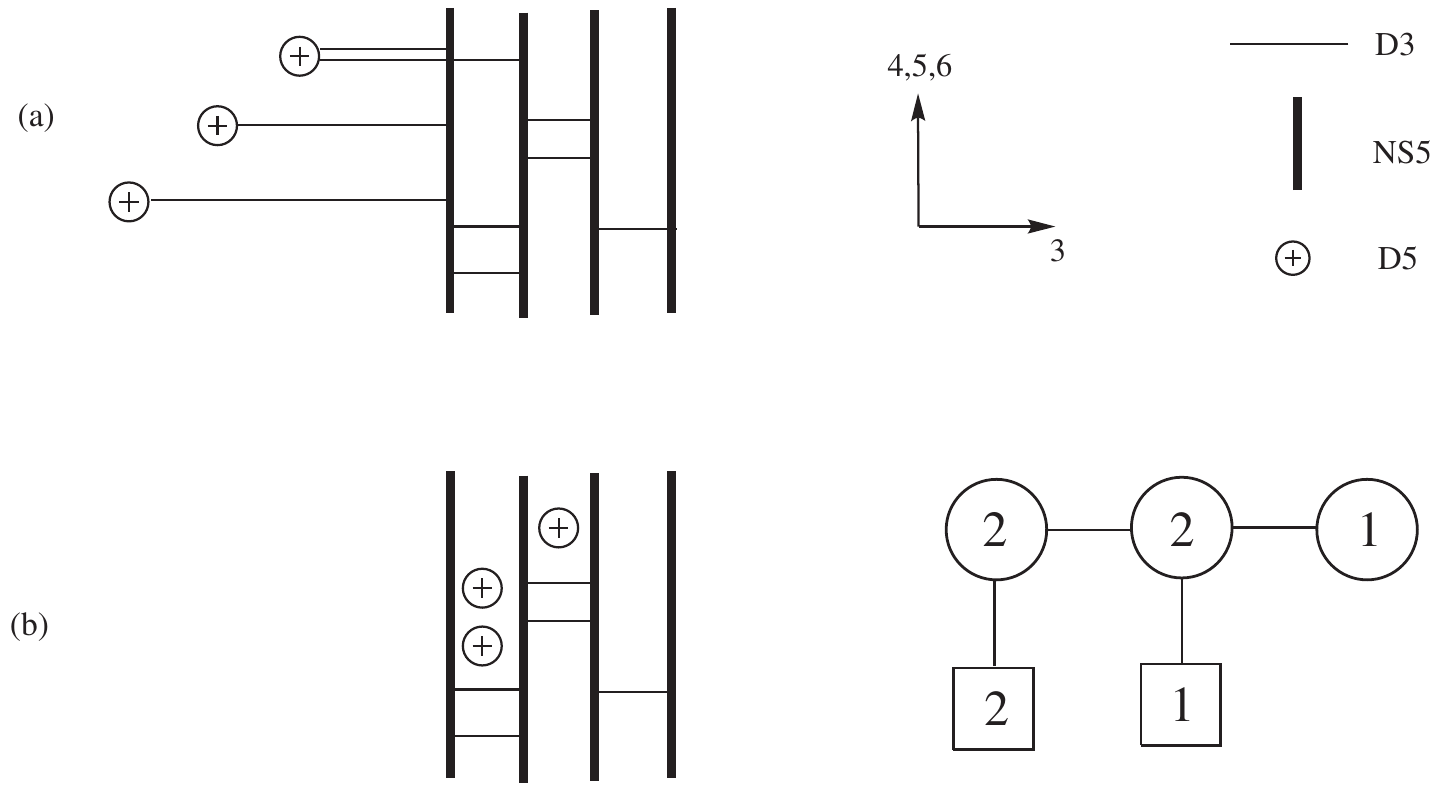}
\caption{{\it The mirror of the $(1)-(2)-[4]$ theory.} (a) From diagram (c) in \fref{fig:124quiv}, the NS5-branes and the D5-branes are exchanged and the directions $x^4, x^5, x^6$ are rotated into $x^7, x^8, x^9$ and {\it vice-versa}. The partitions $\sigma = (2,1,1)$ and $\rho = (1,1,1,1)$ are in one-to-one correspondence with this diagram. (b) The D5-branes are moved across the NS5-branes. The D3-brane creation and annihilation are according to \cite{Hanany:1996ie}. The corresponding quiver diagram is also given next to the brane configuration.}
\label{fig:mirror124}
\end{center}
\end{figure}

Now let us consider the mirror of the $(1)-(2)-[4]$ theory.  The brane configuration of the mirror theory can be obtained as described in \cite{Hanany:1996ie}.  Starting From diagram (c) in \fref{fig:124quiv}, we exchange the NS5-branes and the D5-branes and rotate the directions $x^4, x^5, x^6$ into $x^7, x^8, x^9$ and {\it vice-versa}.  The resulting brane configuration is depicted in \fref{fig:mirror124} (a).  In order to obtain diagram (b), we use the fact that the D5-branes may cross NS5-branes, but whenever a D5-brane moves across an NS5-brane, a D3-brane segment stretched between them is created or annihilated in such a way that the linking numbers remain constant.  From diagram (c) of \fref{fig:mirror124}, it can be seen that 
\bea
\sigma = (2,1,1)~, \qquad \qquad \rho  = (1,1,1,1)~,
\eea
\ie~the partitions $\sigma$ and $\rho$ from \eref{par124quiv} get exchanged.

One can compute the dimension of the moduli space from the quiver diagram.  
The quaternionic dimension of the Higgs branch of the mirror theory is
\bea
\dim_{\BH} \text{Higgs}_{\text{mirror}} = (2 \times 2)+(2 \times 2)+(2 \times 1) + (2 \times 1) - 1^2 - 2^2-2^2 = 3~. \label{dimhiggsmir124}
\eea
The quaternionic dimension of the Coulomb branch of the mirror theory is
\bea
\dim_{\BH} \text{Coulomb}_{\text{mirror}} = 2+2+1 = 5~. \label{dimcoumir124}
\eea
The results are in agreement with the exchange of the Coulomb and Higgs branches of the $(1)-(2)-[4]$ theory predicted by mirror symmetry.

Subsequently, we discuss the Coulomb branch of the $(1)-(2)-[4]$ theory in detail.  We find that this branch is a complete intersection whose generators and relations can be explicitly written down.    On the other hand,  the Higgs branch of this theory is not a complete intersection -- the Hilbert series of this is presented in Appendix \ref{App:Higgs124}.

\subsubsection{The Coulomb branch of the $(1)-(2)- [4]$ theory}
\begin{figure}[htbp]
\begin{center}
\includegraphics[height=1 in]{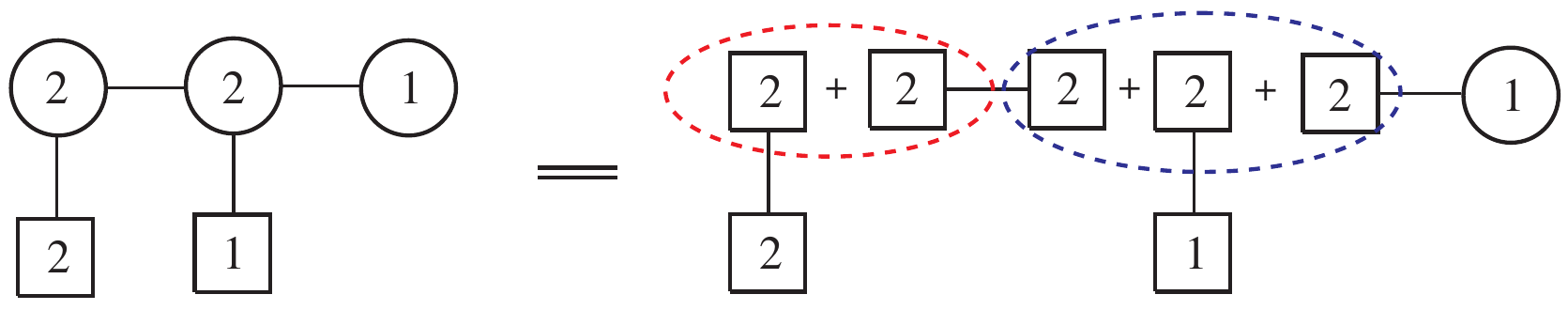}
\caption{The gluing process to obtain the mirror of the $(1)-(2)- [4]$ theory.}
\label{fig:gluemir124}
\end{center}
\end{figure}

By mirror symmetry, the Coulomb branch of the $(1)-(2)- [4]$ theory is identical to the Higgs branch of its mirror.  
The Hilbert series of the Higgs branch of the mirror theory can be obtained by gluing process \cite{Benvenuti:2010pq,Hanany:2010qu} schematically depicted in \fref{fig:gluemir124}.

%The Hilbert series of the Higgs branches of each theory on the right hand side are given below:
%\bea
%g^H_{[2]-[2]} = \PE[ ]
%\eea

\paragraph{Computation of Hilbert series.}
Let $q_1$ be the global $U(1)$ fugacity in the mirror of the $(1)-(2)- [4]$ theory and let $(q_2,x)$ be the $U(1) \times SU(2) = U(2)$ global fugacity.
The Hilbert series of the Higgs branch of the mirror of the $(1)-(2)- [4]$ theory (or, equivalently, the Coulomb branch of the $(1)-(2)- [4]$ theory) is given by
\bea
&& H^{C}_{(1)-(2)- [4]} (t, q_1, x, q_2)=   \int \ud \mu_{U(2)} (z_1,z_2) \ud \mu_{U(2)} (w_1,w_2) \times \nn \\ 
&& \qquad \qquad  H_{[2]-[2]} (t, q_2,x,z_1,z_2) H_{\text{glue}} (t,z_1,z_2) H_{[2]-[2]} (t, z_1,z_2,w_1,w_2)  \times \nn \\
&& \qquad \qquad H_{\text{glue}} (t,w_1,w_2) H_{[2]-[1]} (t, w_1, w_2 ,q_1) H_{(1)-[2]} (t, w_1,w_2) ~, 
\eea
where $(z_1, z_2)$ and $(w_1, w_2)$ are two sets of $U(2)$ fugacities corresponding of red and blue ellipses in \fref{fig:gluemir124} respectively.  
The Haar measure is given by
\bea \label{HaarU2}
\int \ud \mu_{U(2)} (z_1,z_2) = \frac{1}{2} \oint_{|z_1| =1} \frac{\ud z_1}{z_1} \oint_{|z_2| =1} \frac{\ud z_1}{z_2} \left(\frac{1}{z_1}-\frac{1}{z_2} \right)(z_1-z_2)~.
\eea
The gluing factor is given by\footnote{The {\bf plethystic exponential} $\PE$ of a multi-variable function $f(t_1, . . . , t_n)$ that vanishes at the origin, $f(0,...,0) = 0$, is defined as $\PE \left[ f(t_1, t_2, \ldots, t_n) \right] = \exp \left( \sum_{k=1}^\infty \frac{1}{k} f(t_1^k, \ldots, t_n^k) \right)$.}

\bea
H_{\text{glue}} (t,w_1,w_2) = \frac{1}{\PE \left[ (w_1+w_2)\left( \frac{1}{w_1}+\frac{1}{w_2} \right) t^2 \right]}~.
\eea
The Hilbert series $H_{[2]-[2]}$ is given by
\bea
H_{[2]-[2]} (t, z_1,z_2,w_1,w_2) &=& \PE \left[ (z_1+z_2)\left(\frac{1}{w_1}+\frac{1}{w_2} \right) t + (w_1+w_2)\left(\frac{1}{z_1}+\frac{1}{z_2} \right) t\right]~, \nn \\
H_{[2]-[2]} (t, q_2,x,z_1,z_2) &=& \PE \left[ q_2 \left(x+\frac{1}{x} \right)\left(\frac{1}{z_1}+\frac{1}{z_2} \right) t + \frac{1}{q_2}\left(x+\frac{1}{x} \right) (z_1+z_2) t\right]~. \nn \\
\eea
The Hilbert series $H_{[2]-[1]}$ is given by
\bea
H_{[2]-[2]} (t, w_1, w_2 ,q_1) &=& \PE \left[  q_1 \left(\frac{1}{w_1}+\frac{1}{w_2} \right) t + \frac{1}{q_1} (w_1+w_2)  t\right]~.
\eea
The Hilbert series $H_{(1)-[2]} (t, w_1,w_2)$ is given by \eref{hs123n}:
\bea
H_{(1)-[2]} (t, w_1,w_2) = (1-t^4) \PE \left[ \left \{ (w_1+w_2) \left(\frac{1}{w_1}+\frac{1}{w_2} \right) -1 \right \} t^2 \right]~. \label{hs12w1w2}
\eea

\paragraph{The Hilbert series.} The result is
\bea
H^{C}_{(1)-(2)- [4]} (t, q_1, x, q_2)= (1-t^6)(1-t^8) \PE \left[ ([2]_x +[0]_x) t^2 + \left \{ \left( \frac{q_1}{q_2} + \frac{q_2}{q_1} \right) [1]_x \right \} t^3 \right] ~.  \label{HSCou124} \nn \\
\eea
Note that the representation $[2]_x +[0]_x$ of $SU(2)$ is in fact the adjoint representation of the $U(2)$ global symmetry, and the representation  $\left( \frac{q_1}{q_2} + \frac{q_2}{q_1} \right) [1]_x$ is the bi-fundamental representation of $U(1) \times U(2)$ global symmetry.

The Hilbert series indicates that the Coulomb branch of the $(1)-(2)- [4]$ theory is indeed a complete intersection. There are $4$ generators at order $t^2$ in the adjoint representation of $U(2)$, and $4$ generators at order $t^3$ in the bi-fundamental representation of $U(2) \times U(1)$.  These generators are subject to one relation at each order of $t^6$ and $t^8$.  These altogether give $4+4-2=6$ complex dimensional space or, equivalently, 3 quaternionic dimensional space -- in agreement with \eref{dimcou124} and \eref{dimhiggsmir124}.

\begin{figure}[htbp]
\begin{center}
\includegraphics[height=1.3 in]{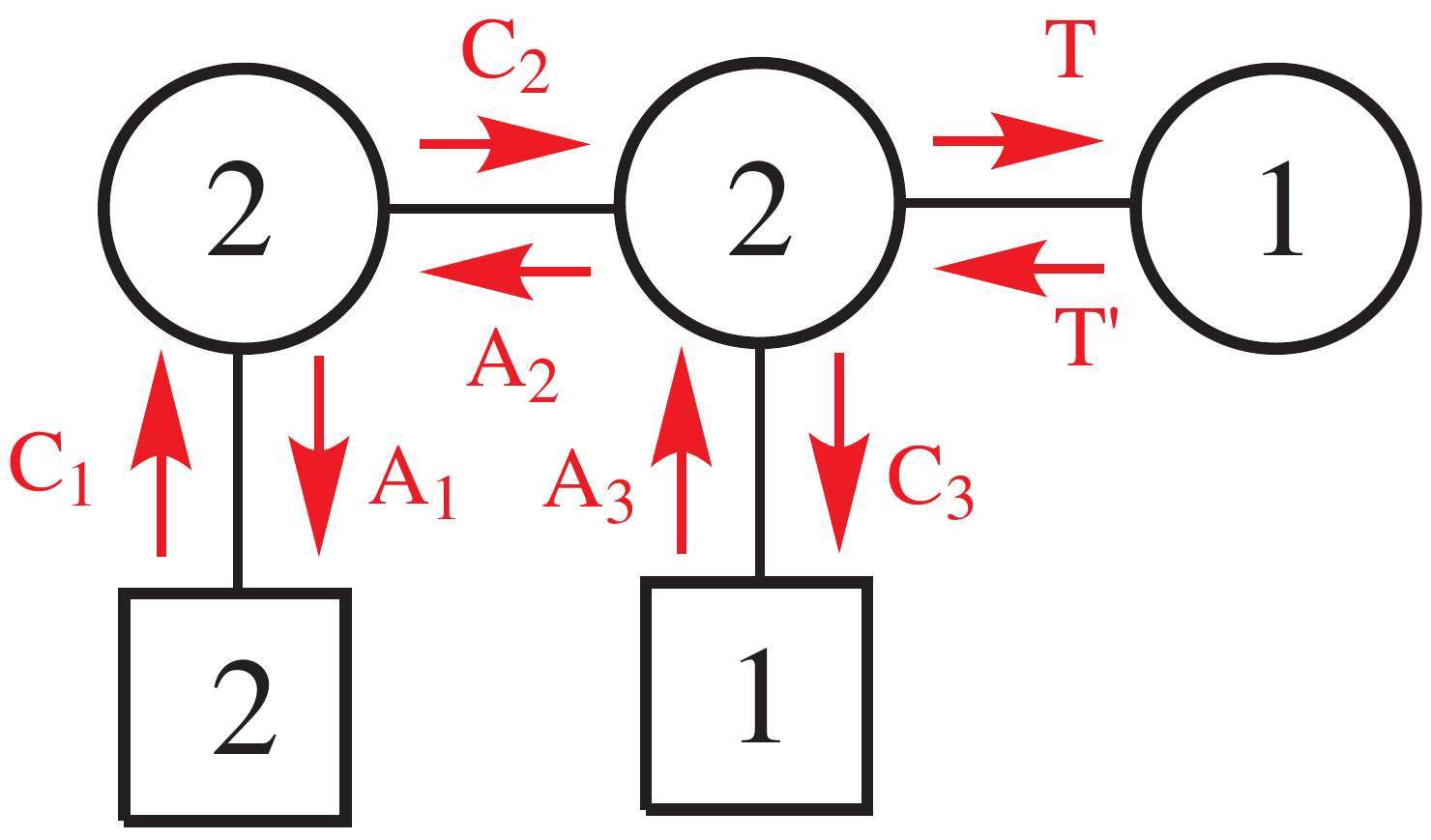}
\caption{The quiver diagram of the mirror of the $(1)-(2)- [4]$ theory, with the bi-fundamental chiral multiplets labelled. Note that $A$'s denote the chiral fields in the anti-clockwise direction, $C$'s denote the chiral fields in the clockwise direction, and $T,~T'$ denote the chiral fields in the tail with reducing ranks.}
\label{fig:mir124field}
\end{center}
\end{figure}

\paragraph{The $F$-terms.}  From \fref{fig:mir124field}, we see that the F-term constraints for the bi-fundamental chiral fields are
\bea
0 &=& (A_1)^a_{~i} (C_1)^i_{~b} + (C_2)^a_{~\alpha} (A_2)^\alpha_{~b} ~, \nn \\
0 &=& -(A_2)^\alpha_{~a} (C_2)^a_{~\beta} + (C_3)^\alpha (A_3)_\beta + T^\alpha T'_\beta ~, \nn\\
0 &=& -T'_\alpha T^\alpha ~, \label{Fterms124}
\eea
where the indices $a, b =1,2$ are the gauge indices corresponding to the leftmost $U(2)$ gauge group, the indices $\alpha, \beta =1,2$ are the gauge indices corresponding to the middle $U(2)$ gauge group, and the indices $i, j =1,2$ are the global indices corresponding to the $U(2)$ global symmetry.

\paragraph{The generators.}  The generators at order $t^2$ can be written as
\bea
M^i_{~j} =  (C_1)^i_{~a} (A_1)^a_{~j}~.
\eea
The generators at order $t^3$ are
\bea
L_i = (A_3)_\alpha (A_2)^\alpha_{~a} (A_1)^a_{~i}  ~, \qquad \quad
R^i = (C_1)^i_{~a} (C_2)^a_{~\alpha} (C_3)^\alpha~.
\eea

\paragraph{The relations.} The relation at order $t^6$ is
\bea
L_i R^i &=& (\tr M)(\tr M^2)~ . \label{rel6124}
\eea
The relation at order $t^8$ is
\bea
 L_i M^i_{~j} R^j =  (\tr M)(\tr M^3)+(\det M)^2~. \label{rel8124}
\eea
We derive these relations in Appendix \ref{app:rel124}.  Note that these relations can be rewritten in other forms which are equivalent to the above, \eg~
\bea
L_i R^i &=& \tr (M^3) + (\tr M)(\det M) ~, \label{rel6124a} \\
%=  (\tr M)^3 -  2(\tr M)(\det M)~, \label{rel6124a} \\
L_i M^i_{~j} R^j &=& \tr (M^4) + \tr(M^2) (\det M)+(\det M)^2~.  \label{rel8124a}
\eea

\subsection{Example: The $(1)-(2)-[5]$ theory and its mirror}
\begin{figure}[htbp]
\begin{center}
\includegraphics[height=2.4 in]{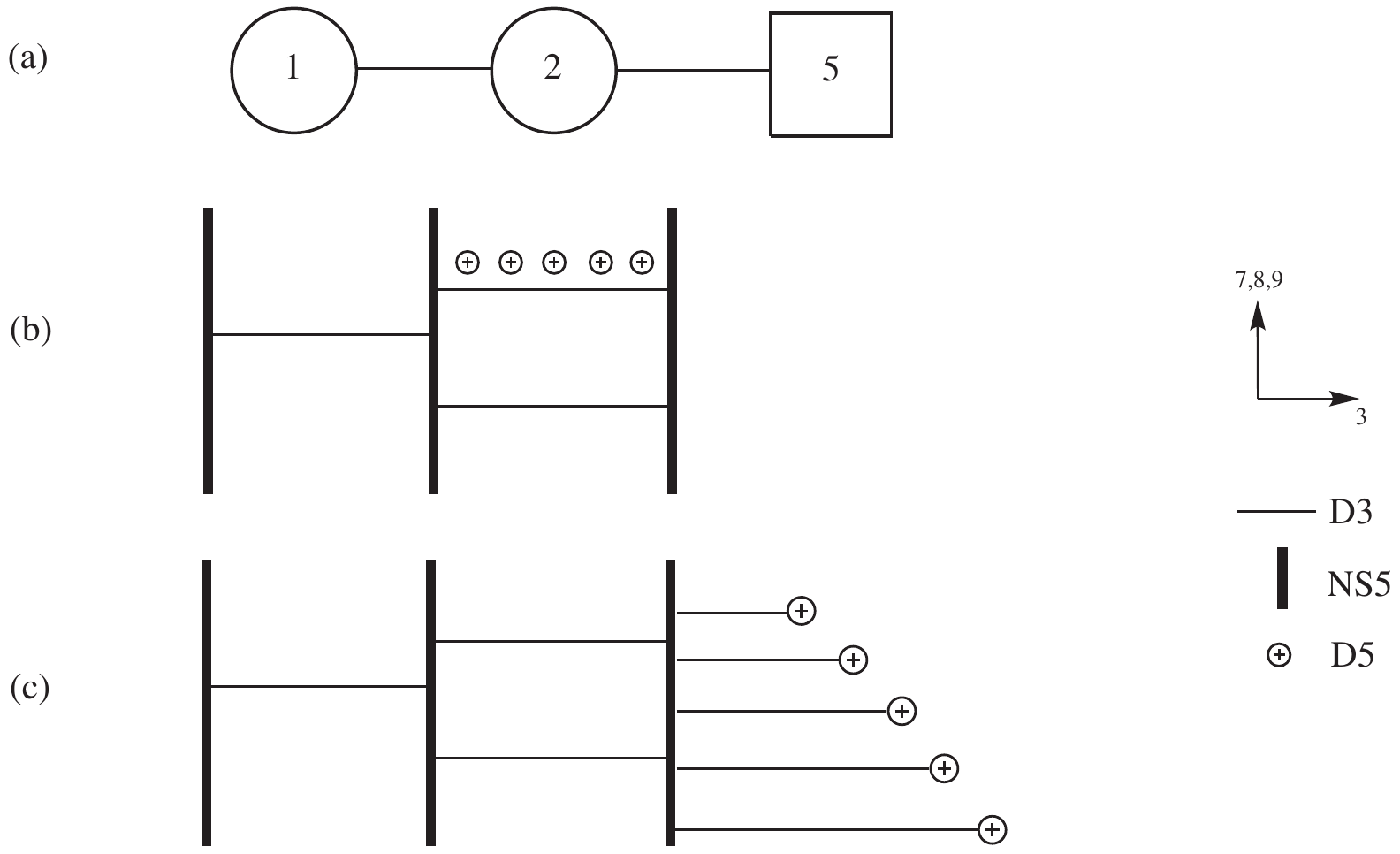}
\caption{(a) The quiver diagram of the $(1)-(2)-[5]$ theory. (b) The corresponding brane configuration. (c) The D5-branes are moved to the right of all NS5-branes.  The D3-branes are created according to \cite{Hanany:1996ie}.  The partitions $\sigma = (1,1,1,1,1)$ and $\rho = (3,1,1)$ are in one-to-one correspondence with this diagram.}
\label{fig:125quiv}
\end{center}
\end{figure}

Let us consider the $(1)-(2)-[5]$ theory.  The quiver diagram and the corresponding brane configuration are depicted in \fref{fig:125quiv}.  From diagram (c), this theory can be identified with $T^\sigma_\rho(SU(5))$, where
\bea
\sigma = (1,1,1,1,1)~, \qquad \rho = (3,1,1)~, \label{par125quiv}
\eea
and the number 5 in $SU(5)$ indicates the total number of boxes in each partition.

One can compute the dimension of the moduli space from the quiver diagram.  
The quaternionic dimension of the Higgs branch of this theory is
\bea
\dim_{\BH} \text{Higgs}_{(1)-(2)-[5]} = (1\times 2) + (2 \times 5) - 1^2 - 2^2  = 7~. \label{dimhiggs125}
\eea
On the other hand, the quaternionic dimension of the Coulomb branch of this theory is
\bea
\dim_{\BH} \text{Coulomb}_{(1)-(2)-[5]} = 1+2 = 3~. \label{dimcou125}
\eea

\paragraph{The moduli space of monopoles.}   Using diagram (c) of \fref{fig:125quiv}, we can interpret the Coulomb branch of the $(1)-(2)-[5]$ theory as the moduli space of $SU(3)$ monopoles, one with magnetic charge $(1,-1,0)$ and two with magnetic charges $(0,1,-1)$, in the presence of five fixed monopoles with magnetic charge $(0,0,1)$ represented by the five rightmost semi-infinite D3-branes.

\subsection*{The mirror theory}

\begin{figure}[htbp]
\begin{center}
\includegraphics[height=2.8 in]{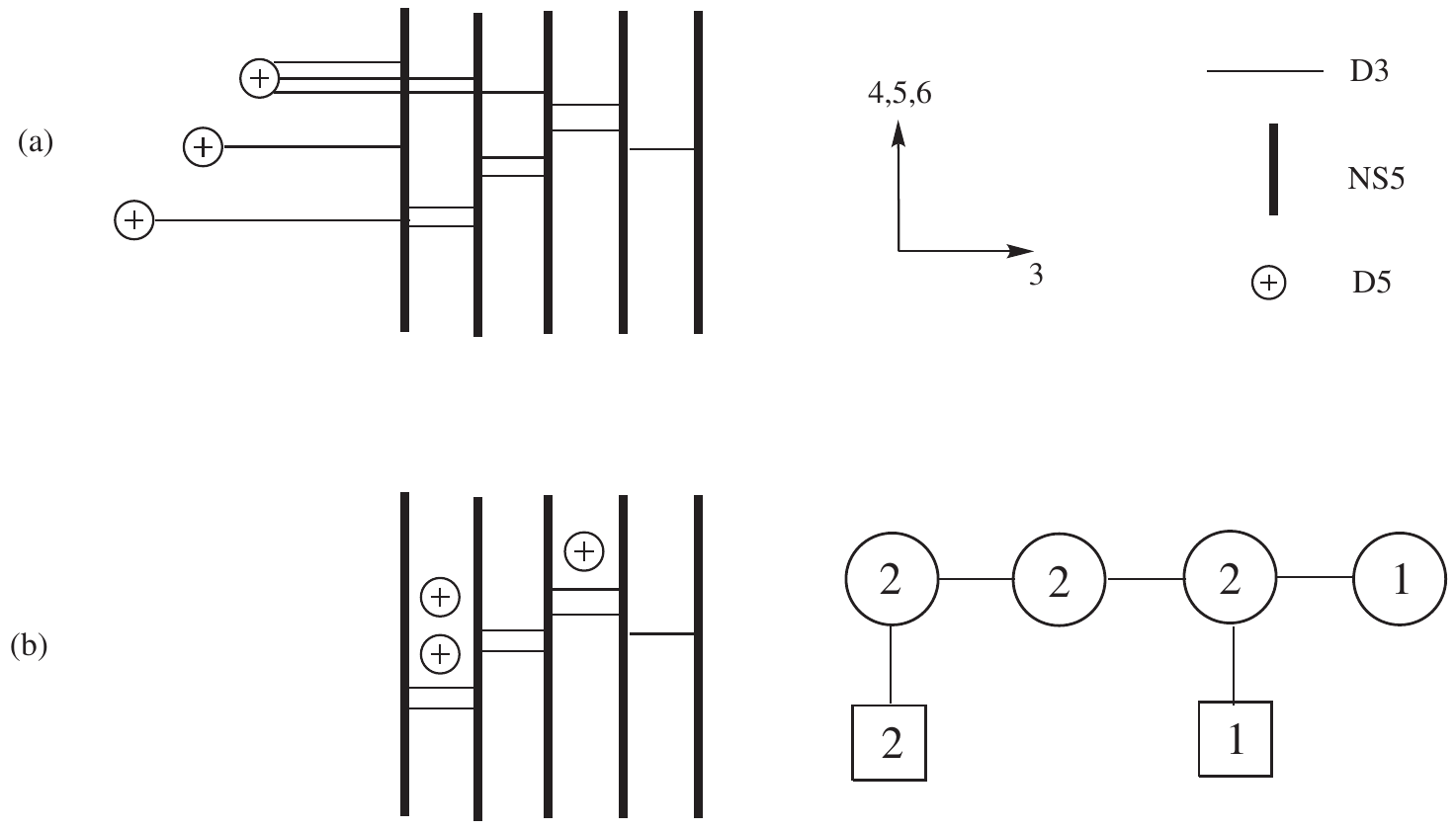}
\caption{{\it The mirror of the $(1)-(2)-[5]$ theory.} (a) From diagram (c) in \fref{fig:124quiv}, the NS5-branes and the D5-branes are exchanged and the directions $x^4, x^5, x^6$ are rotated into $x^7, x^8, x^9$ and {\it vice-versa}. The partitions $\sigma = (3,1,1)$ and $\rho = (1,1,1,1,1)$ are in one-to-one correspondence with this diagram. (b) The D5-branes are moved across the NS5-branes. The D3-brane creation and annihilation are according to \cite{Hanany:1996ie}. The corresponding quiver diagram is also given next to the brane configuration.}
\label{fig:mirror125}
\end{center}
\end{figure}

Now let us consider the mirror of the $(1)-(2)-[5]$ theory.  The brane configuration of the mirror theory can be obtained as described in \cite{Hanany:1996ie}.  Starting From diagram (c) in \fref{fig:125quiv}, we exchange the NS5-branes and the D5-branes and rotate the directions $x^4, x^5, x^6$ into $x^7, x^8, x^9$ and {\it vice-versa}.  The resulting brane configuration is depicted in \fref{fig:mirror125} (a).  In order to obtain diagram (b), we use the fact that the D5-branes may cross NS5-branes, but whenever a D5-brane moves across an NS5-brane, a D3-brane segment stretched between them is created or annihilated in such a way that the linking numbers remain constant.  From diagram (c) of \fref{fig:mirror125}, it can be seen that 
\bea
\sigma = (3,1,1)~, \qquad \qquad \rho  = (1,1,1,1,1)~,
\eea
\ie~the partitions $\sigma$ and $\rho$ from \eref{par125quiv} get exchanged.

One can compute the dimension of the moduli space from the quiver diagram.  
The quaternionic dimension of the Higgs branch of the mirror theory is
\bea
\dim_{\BH} \text{Higgs}_{\text{mirror}} = 3(2 \times 2) +2 (2 \times 1) - (3 \times 2^2)- 1^2= 3~. \label{dimHiggsmir125}
\eea
The quaternionic dimension of the Coulomb branch of the mirror theory is
\bea
\dim_{\BH} \text{Coulomb}_{\text{mirror}} =2+2+2+1 = 7~. \label{dimcoumir125}
\eea
The results are in agreement with the exchange of the Coulomb and Higgs branches of the $(1)-(2)-[5]$ theory predicted by mirror symmetry.

Subsequently, we discuss the Coulomb branch of the $(1)-(2)-[5]$ theory in detail.  We find that this branch is a complete intersection whose generators and relations can be explicitly written down.    On the other hand,  the Higgs branch of this theory is not a complete intersection -- the Hilbert series of this is presented in Appendix \ref{App:Higgs125}.

\subsubsection{The Coulomb branch of the $(1)-(2)-[5]$ theory}
\begin{figure}[htbp]
\begin{center}
\includegraphics[height=1 in]{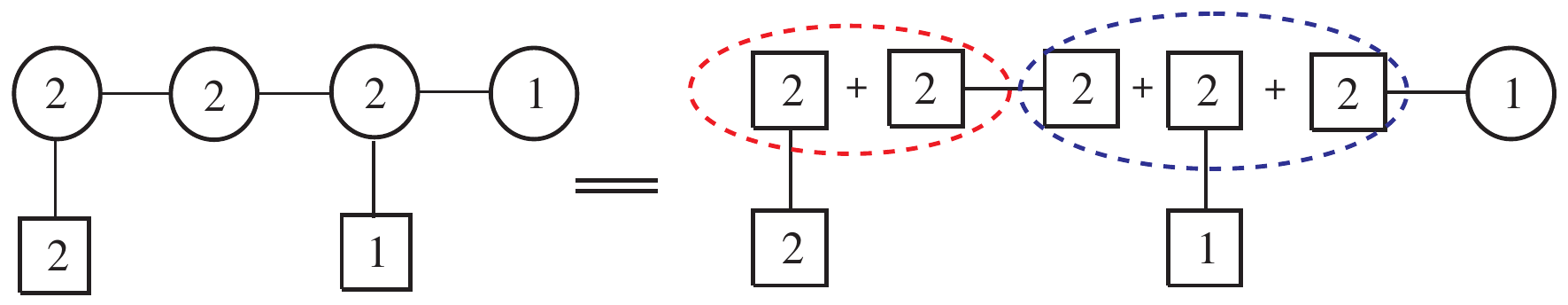}
\caption{The gluing process to obtain the mirror of the $(1)-(2)-[5]$ theory.}
\label{fig:gluemir125}
\end{center}
\end{figure}

By mirror symmetry, the Coulomb branch of the $(1)-(2)-[5]$ theory is identical to the Higgs branch of its mirror.  
The Hilbert series of the Higgs branch of the mirror theory can be obtained by gluing process \cite{Benvenuti:2010pq,Hanany:2010qu} schematically depicted in \fref{fig:gluemir125}.

Let $q_1$ be the global $U(1)$ fugacity in the mirror of the $(1)-(2)-[5]$ theory and let $(q_2,x)$ be the $U(1) \times SU(2) = U(2)$ global fugacity.

The Hilbert series of the Higgs branch of the mirror of the $(1)-(2)-[5]$ theory (or, equivalently, the Coulomb branch of the $(1)-(2)-[5]$ theory) is given by
\bea
H^{C}_{(1)-(2)-[5]} (t, q_1, q_2, x) &=& (1-t^8)(1-t^{10}) \PE \left[ ([2]_x +[0]_x) t^2 + \left \{ \left( \frac{q_1}{q_2} + \frac{q_2}{q_1} \right) [1]_x \right \} t^4 \right] ~. \label{HSCou125} \nn \\
\eea
Note that the representation $[2]_x +[0]_x$ of $SU(2)$ is in fact the adjoint representation of the $U(2)$ global symmetry, and the representation  $\left( \frac{q_1}{q_2} + \frac{q_2}{q_1} \right) [1]_x$ is the bi-fundamental representation of $U(1) \times U(2)$ global symmetry.

The Hilbert series indicates that the Coulomb branch of the $(1)-(2)-[5]$ theory is indeed a complete intersection. There are $4$ generators at order $t^2$ in the adjoint representation of $U(2)$, and $4$ generators at order $t^3$ in the bi-fundamental representation of $U(2) \times U(1)$.  These generators are subject to one relation at each order of $t^8$ and $t^{10}$.  These altogether give $4+4-2=6$ complex dimensional space or, equivalently, 3 quaternionic dimensional space -- in agreement with \eref{dimcou125} and \eref{dimHiggsmir125}.

\begin{figure}[htbp]
\begin{center}
\includegraphics[height=1.3 in]{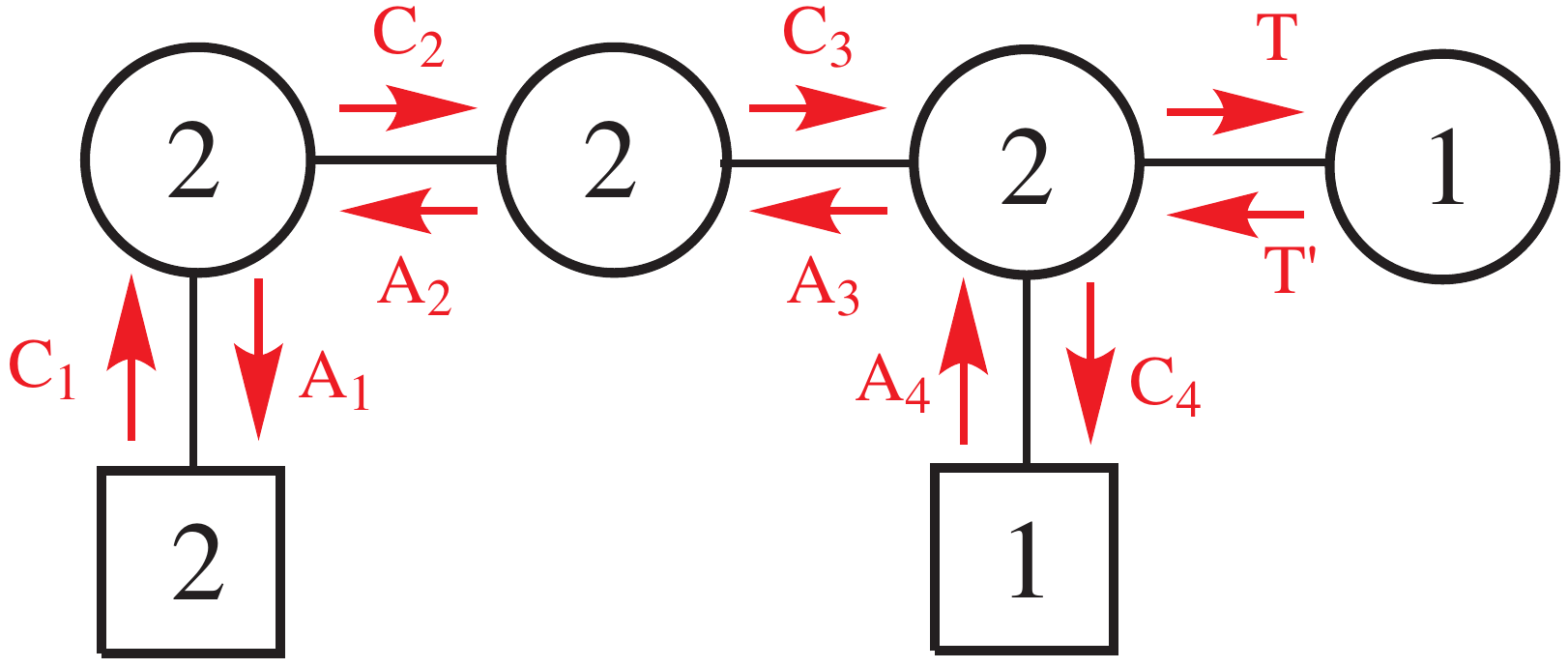}
\caption{The quiver diagram of the mirror of the $(1)-(2)-[5]$ theory, with the bi-fundamental chiral multiplets labelled. Note that $A$'s denote the chiral fields in the anti-clockwise direction, $C$'s denote the chiral fields in the clockwise direction, and $T,~T'$ denote the chiral fields in the tail with reducing ranks.}
\label{fig:mir125field}
\end{center}
\end{figure}

\paragraph{The $F$-terms.}  From \fref{fig:mir125field}, we see that the F-term constraints for the bi-fundamental chiral fields are
\bea \label{Fterms125}
0 &=& (A_1)^a_{~i} (C_1)^i_{~b} + (C_2)^a_{~\alpha} (A_2)^\alpha_{~b} ~, \nn \\
0 &=& -(A_2)^\alpha_{~a} (C_2)^a_{~\beta} + (C_3)^\alpha_{~\mu} (A_3)^\mu_{~\beta}  ~, \nn\\
0 &=&  - (A_3)^\mu_{~\alpha} (C_3)^\alpha_{~\nu} + (C_4)^\mu (A_4)_\nu + T^\mu T'_\nu~, \nn \\
0 &=& -T'_\mu T^\mu~.
\eea
where the indices $a, b =1,2$ are the gauge indices corresponding to the leftmost $U(2)$ gauge group, the indices $\alpha, \beta =1,2$ are the gauge indices corresponding to the next $U(2)$ gauge group on the right, the indices $\mu, \nu =1,2,$ are the gauge indices corresponding to the rightmost $U(2)$ gauge group and the indices $i, j =1,2$ are the global indices corresponding to the $U(2)$ global symmetry.

\paragraph{The generators.}  The generators at order $t^2$ can be written as
\bea
M^i_{~j} =  (C_1)^i_{~a} (A_1)^a_{~j}~.
\eea
The generators at order $t^4$ are
\bea
L_i =  (A_4)_{\mu} (A_3)^\mu_{~\alpha}  (A_2)^\alpha_{~a}  (A_1)^a_{~i}~, \qquad \quad
R^i = (C_1)^i_{~a} (C_2)^a_{~\alpha} (C_3)^\alpha_{~\mu} (C_4)^\mu ~.
\eea

\paragraph{The relations.} The relation at order $t^8$ is
\bea
L_i R^i = - \left[ (\tr M)(\tr M^3) + (\det M)^2 \right] ~. \label{rel8125}
\eea
The relation at order $t^{10}$ is
\bea
L_i M^i_{~j} R^j = -(\tr M) \left[ (\tr M^4)+ (\det M)^2 \right]~. \label{rel10125}
%= - \left[ \tr (M^5) + \tr (M^3) (\det M) + (\tr M) (\det M)^2 \right]~.
\eea
These relations are derived in Appendix \ref{app:rel125}.
Note that these relations can be rewritten in other forms which are equivalent to the above, \eg~
\bea
L_i R^i &=& - \left[ \tr (M^4) + \tr (M^2) (\det M) +(\det M)^2 \right]~, \label{rel8125a} \\
L_i M^i_{~j} R^j &=& - \left[ \tr (M^5) + \tr (M^3) (\det M) + (\tr M) (\det M)^2 \right]~. \label{rel10125a}
\eea

\subsection{Example: The $(1)-(2)-(3)-[5]$ theory and its mirror}
\begin{figure}[htbp]
\begin{center}
\includegraphics[height=2.4 in]{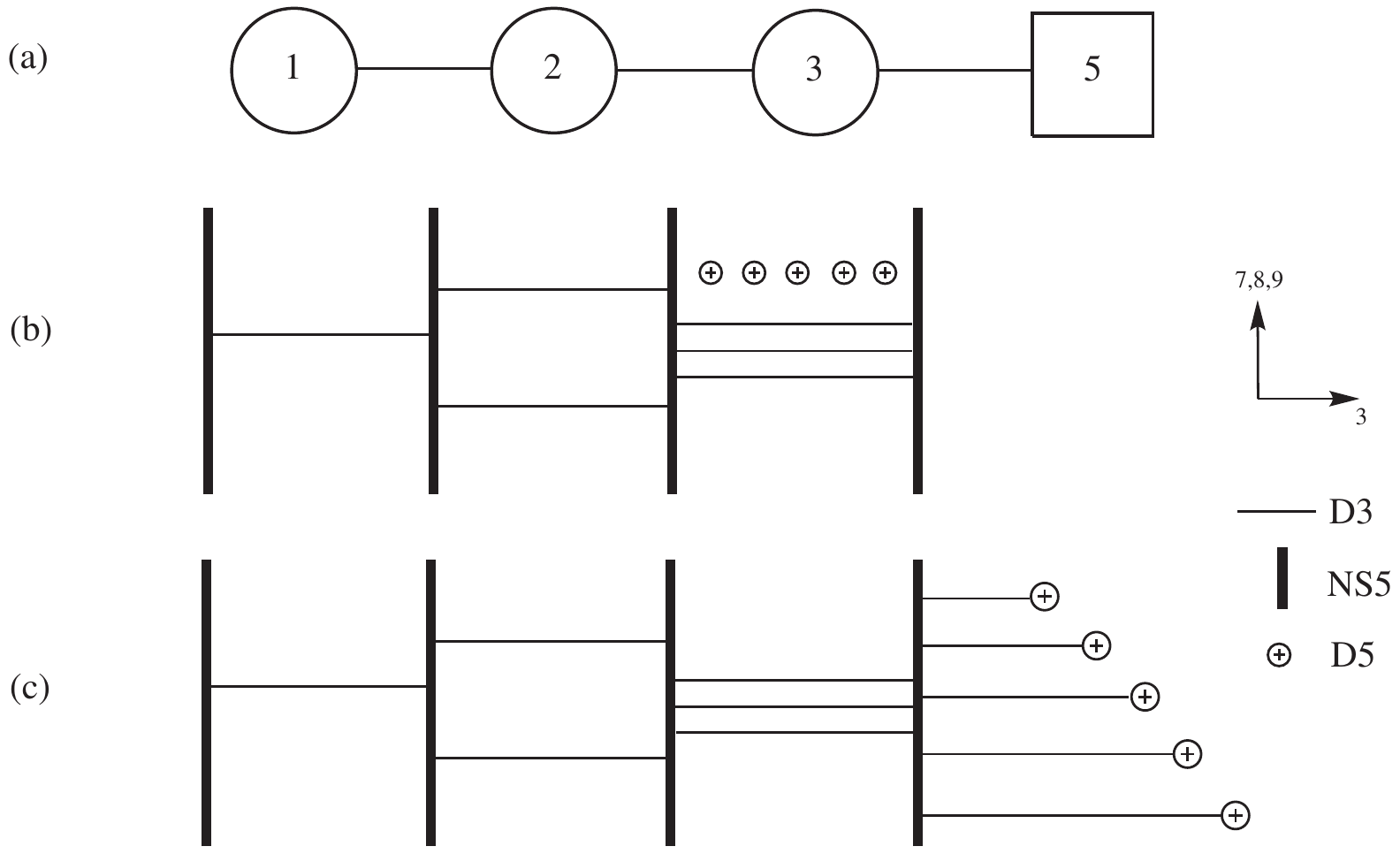}
\caption{(a) The quiver diagram of the $(1)-(2)-(3)-[5]$ theory. (b) The corresponding brane configuration. (c) The D5-branes are moved to the right of all NS5-branes.  The D3-branes are created according to \cite{Hanany:1996ie}.  The partitions $\sigma = (1,1,1,1,1)$ and $\rho = (2,1,1,1)$ are in one-to-one correspondence with this diagram.}
\label{fig:1235quiv}
\end{center}
\end{figure}

Let us consider the $(1)-(2)-(3)-[5]$ theory.  The quiver diagram and the corresponding brane configuration are depicted in \fref{fig:1235quiv}.

This theory can be identified with $T^\sigma_{~\rho}(SU(5))$, where
\bea
\sigma = (1,1,1,1,1)~, \qquad \qquad \rho = (2,1,1,1)~, \label{par1235quiv}
\eea
and the number 5 in $SU(5)$ indicates the total number of boxes in each partition.

One can compute the dimension of the moduli space from the quiver diagram.  
The quaternionic dimension of the Higgs branch of this theory is
\bea
\dim_{\BH} \text{Higgs}_{(1)-(2)-(3)-[5]} = (1\times 2) + (2 \times 3)+ (3 \times 5) - 1^2 - 2^2 -3^2 = 9~. \label{dimhiggs1235}
\eea
On the other hand, the quaternionic dimension of the Coulomb branch of this theory is
\bea
\dim_{\BH} \text{Coulomb}_{(1)-(2)-(3)-[5]} = 1+2+3 = 6~. \label{dimcou1235}
\eea

\paragraph{The moduli space of monopoles.}   Using diagram (c) of \fref{fig:1235quiv}, we can interpret the Coulomb branch of the $(1)-(2)-(3)-[5]$ theory as the moduli space of $SU(4)$ monopoles, one with magnetic charge $(1,-1,0,0)$, two with magnetic charges $(0,1,-1,0)$ and three with magnetic charges $(0,0,1,-1)$, in the presence of five fixed monopoles with magnetic charge $(0,0,0,1)$ represented by the five rightmost semi-infinite D3-branes.

\subsection*{The mirror theory}

\begin{figure}[htbp]
\begin{center}
\includegraphics[height=2.8 in]{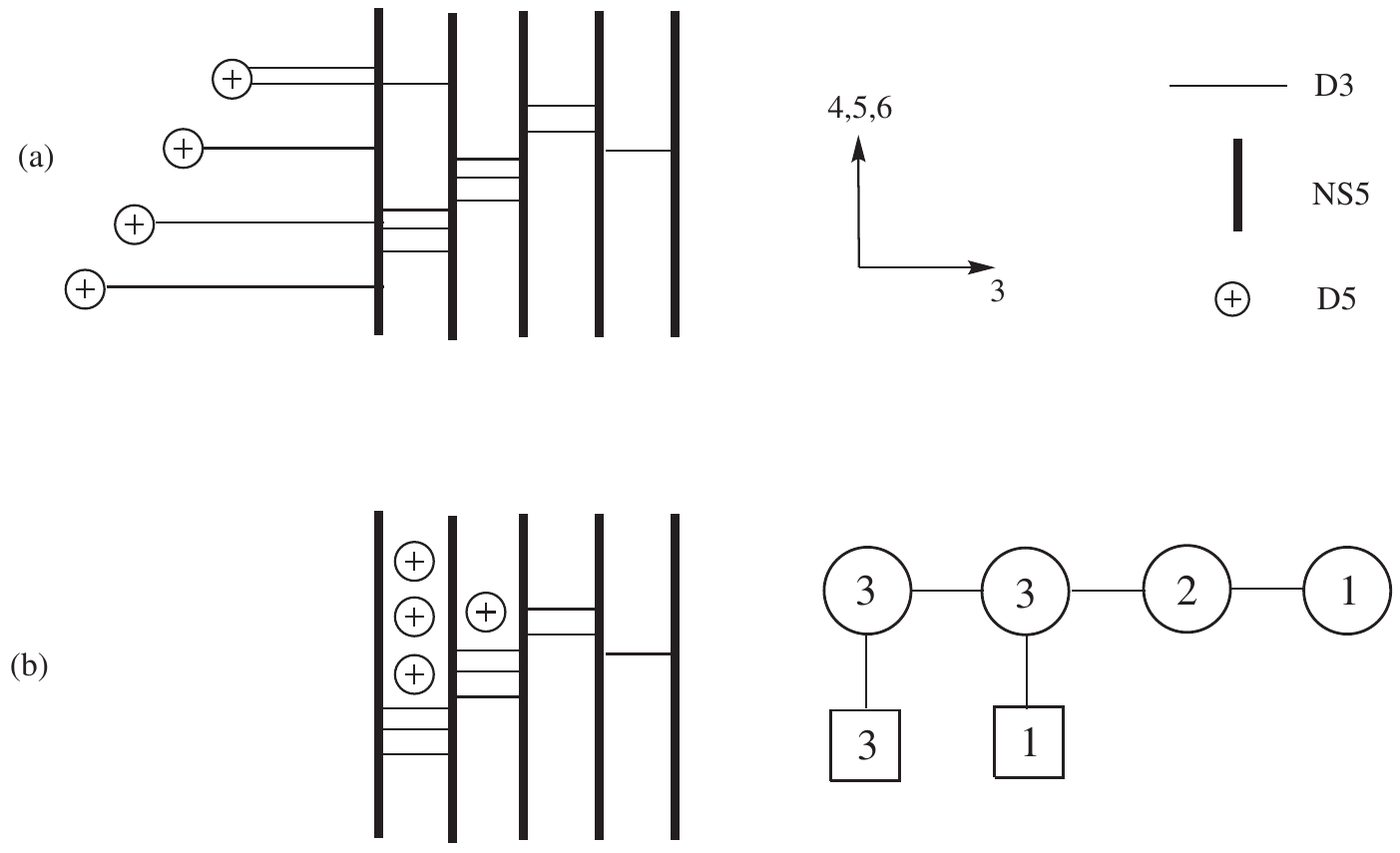}
\caption{{\it The mirror of the $(1)-(2)-(3)-[5]$ theory.} (a) From diagram (c) in \fref{fig:124quiv}, the NS5-branes and the D5-branes are exchanged and the directions $x^4, x^5, x^6$ are rotated into $x^7, x^8, x^9$ and {\it vice-versa}. The partitions $\sigma = (2,1,1,1)$ and $\rho = (1,1,1,1,1)$ are in one-to-one correspondence with this diagram. (b) The D5-branes are moved across the NS5-branes. The D3-brane creation and annihilation are according to \cite{Hanany:1996ie}. The corresponding quiver diagram is also given next to the brane configuration.}
\label{fig:mirror1235}
\end{center}
\end{figure}

Now let us consider the mirror of the $(1)-(2)-(3)-[5]$ theory.  The brane configuration of the mirror theory can be obtained as described in \cite{Hanany:1996ie}.  Starting From diagram (c) in \fref{fig:1235quiv}, we exchange the NS5-branes and the D5-branes and rotate the directions $x^4, x^5, x^6$ into $x^7, x^8, x^9$ and {\it vice-versa}.  The resulting brane configuration is depicted in \fref{fig:mirror1235} (a).  In order to obtain diagram (b), we use the fact that the D5-branes may cross NS5-branes, but whenever a D5-brane moves across an NS5-brane, a D3-brane segment stretched between them is created or annihilated in such a way that the linking numbers remain constant.  From diagram (c) of \fref{fig:mirror1235}, it can be seen that 
\bea
\sigma = (2,1,1,1)~, \qquad \qquad \rho  = (1,1,1,1,1)~,
\eea
\ie~the partitions $\sigma$ and $\rho$ from \eref{par1235quiv} get exchanged.

One can compute the dimension of the moduli space from the quiver diagram.  
The quaternionic dimension of the Higgs branch of the mirror theory is
\bea
\dim_{\BH} \text{Higgs}_{\text{mirror}} = 2(3 \times 3) +(3 \times 1) + (3 \times 2) +(2 \times 1) -3^2-3^2-2^2-1^2= 6~. \nn \\ \label{dimHiggsmir1235}
\eea
The quaternionic dimension of the Coulomb branch of the mirror theory is
\bea
\dim_{\BH} \text{Coulomb}_{\text{mirror}} = 3+3+2+1 = 9~. \label{dimcoumir1235}
\eea
The results are in agreement with the exchange of the Coulomb and Higgs branches of the $(1)-(2)-(3)-[5]$ theory predicted by mirror symmetry.

\subsubsection{The Coulomb branch of the $(1)-(2)-(3)-[5]$ theory}
\begin{figure}[htbp]
\begin{center}
\includegraphics[height=1 in]{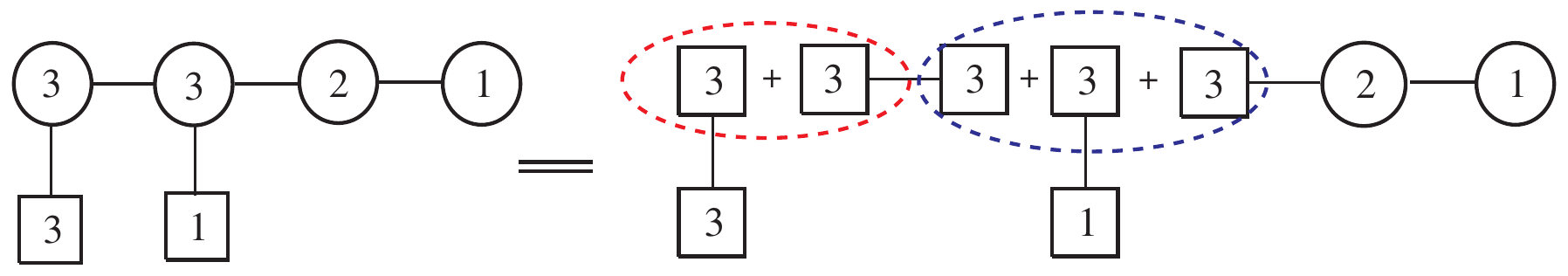}
\caption{The gluing process to obtain the mirror of the $(1)-(2)-(3)-[5]$ theory.}
\label{fig:gluemir1235}
\end{center}
\end{figure}

By mirror symmetry, the Coulomb branch of the $(1)-(2)-(3)-[5]$ theory is identical to the Higgs branch of its mirror.  
The Hilbert series of the Higgs branch of the mirror theory can be obtained by gluing process \cite{Benvenuti:2010pq,Hanany:2010qu} schematically depicted in \fref{fig:gluemir1235}.

%The Hilbert series of the Higgs branches of each theory on the right hand side are given below:
%\bea
%g^H_{[2]-[2]} = \PE[ ]
%\eea

Let $q_1$ be the global $U(1)$ fugacity in the mirror of the $(1)-(2)-(3)-[5]$ theory and let $(q_2,x_1,x_2)$ be the $U(1) \times SU(3) = U(3)$ global fugacity.

The Hilbert series of the Higgs branch of the mirror of the $(1)-(2)-(3)-[5]$ theory (or, equivalently, the Coulomb branch of the $(1)-(2)-(3)-[5]$ theory) is given by
\bea
H^{C}_{(1)-(2)-(3)-[5]} (t, q_1, x_1, x_2, q_2) &=& (1-t^6)(1-t^8)(1-t^{10}) \times \nn \\
&&  \PE \left[ ([1,1] +[0,0]) t^2 + \left( \frac{q_2}{q_1} [1,0] + \frac{q_1}{q_2} [0,1] \right) t^3 \right] ~.  \label{HSCou1235} \nn \\
\eea
Note that the representation $[1,1] +[0,0]$ of $SU(3)$ is in fact the adjoint representation of the $U(3)$ global symmetry, and the representation  $ \frac{q_2}{q_1} [1,0] + \frac{q_1}{q_2} [0,1]$ is the bi-fundamental representation of $U(3) \times U(1)$ global symmetry.

The Hilbert series indicates that the Coulomb branch of the $(1)-(2)-(3)-[5]$ theory is indeed a complete intersection. There are $9$ generators at order $t^2$ in the adjoint representation of $U(2)$, and $6$ generators at order $t^3$ in the bi-fundamental representation of $U(3) \times U(1)$.  These generators are subject to one relation at each order of $t^6$, $t^8$ and $t^{10}$.  These altogether give $9+6-3=12$ complex dimensional space or, equivalently, 6 quaternionic dimensional space -- in agreement with \eref{dimcou1235} and \eref{dimHiggsmir1235}.

\begin{figure}[htbp]
\begin{center}
\includegraphics[height=1.3 in]{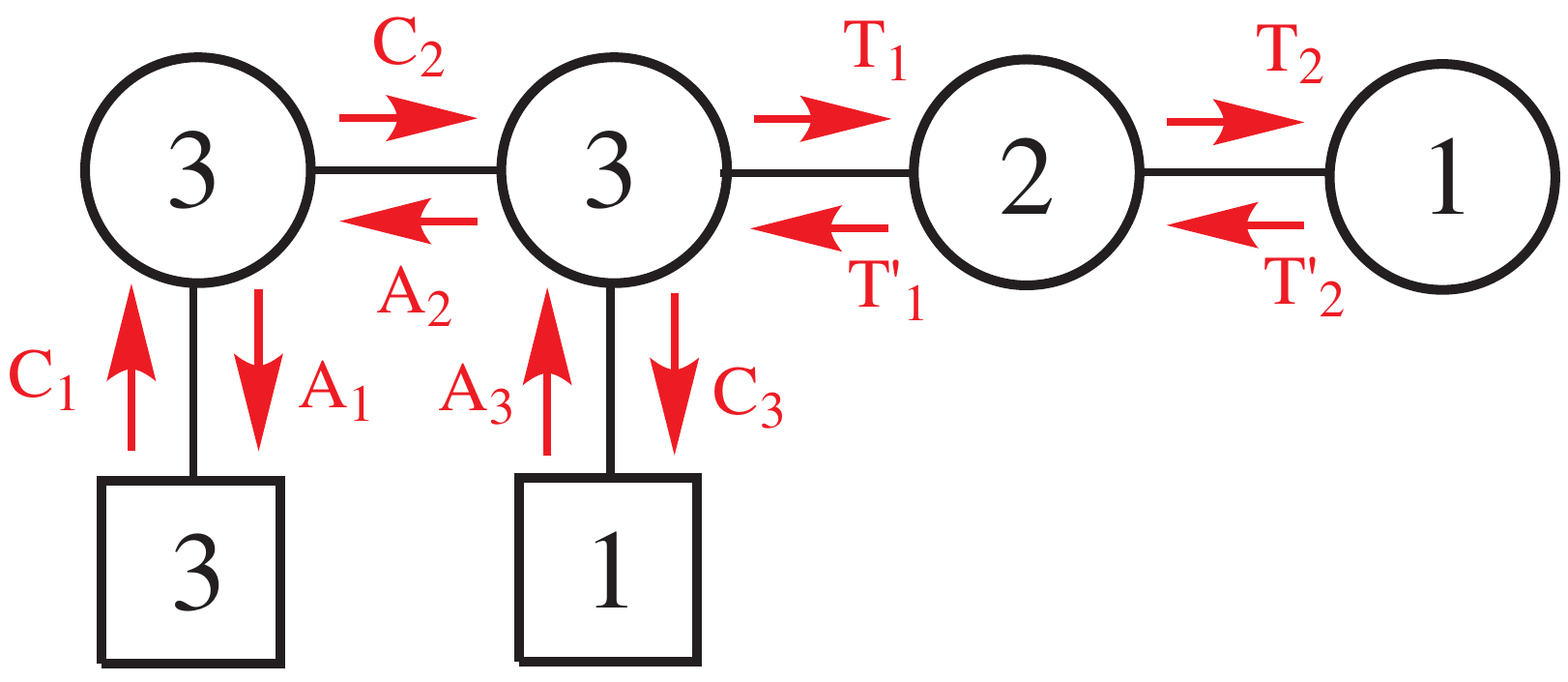}
\caption{The quiver diagram of the mirror of the $(1)-(2)-(3)-[5]$ theory, with the bi-fundamental chiral multiplets labelled. Note that $A$'s denote the chiral fields in the anti-clockwise direction, $C$'s denote the chiral fields in the clockwise direction, and $T, T'$s denote the chiral fields in the tail part.}
\label{fig:mir1235field}
\end{center}
\end{figure}

\paragraph{The $F$-terms.}  From \fref{fig:mir124field}, we see that the F-term constraints for the bi-fundamental chiral fields are
\bea
0 &=& (A_1)^a_{~i} (C_1)^i_{~b} + (C_2)^a_{~\alpha} (A_2)^\alpha_{~b} ~, \nn \\
0 &=& -(A_2)^\alpha_{~a} (C_2)^a_{~\beta} + (C_3)^\alpha (A_3)_\beta + (T_1)^\alpha_{~\mu} (T'_1)^\mu_{~\beta} ~, \nn\\
0 &=& -(T'_1)^\mu_{~\alpha} (T_1)^\alpha_{~\nu} + (T_2)^\mu (T'_2)_\nu ~, \nn \\
0 &=& -(T'_2)_\mu (T_2)^\mu~. \label{fterms1235}
\eea
where the indices $a, b =1,2,3$ are the gauge indices corresponding to the leftmost $U(3)$ gauge group, the indices $\alpha, \beta =1,2,3$ are the gauge indices corresponding to the next $U(3)$ gauge group on the right, the indices $\mu, \nu =1,2,$ are the gauge indices corresponding to the $U(2)$ gauge group, and the indices $i, j =1,2,3$ are the global indices corresponding to the $U(3)$ global symmetry.

\paragraph{The generators.}  The generators at order $t^2$ can be written as
\bea
M^i_{~j} =  (C_1)^i_{~a} (A_1)^a_{~j}~.
\eea
The generators at order $t^3$ are
\bea
L_i = (A_3)_\alpha (A_2)^\alpha_{~a} (A_1)^a_{~i}  ~, \qquad \quad
R^i = (C_1)^i_{~a} (C_2)^a_{~\alpha} (C_3)^\alpha~.
\eea

\paragraph{The relations.} The relation at order $t^6$ is
\bea
L_i R^i &=&  (\tr M)(\tr M^2) + \det M~ . \label{rel61235}
\eea
The relation at order $t^8$ is
\bea
 L_i M^i_{~j} R^j = \frac{1}{2} \left[ \tr (M^4) + (\tr M)^2 (\tr M^2) \right]~. \label{rel81235}
\eea
The relation at order $t^{10}$ is
\bea
L_i M^i_{~j} M^{j}_{~k} R^k &=& \tr (M^5) - (\tr M)(\tr M^4) + \frac{1}{2} \left[5 (\tr M)^2 - (\tr M^2) \right] (\tr M^3) \nn \\
&& - (\tr M)^3 (\tr M^2) - (\tr M)^2 (\det M) ~.
\eea
These relations are derived in Appendix \ref{app:rel1235}.

\subsection{General case: $(1)-(2)- \cdots -(m)-[n]$}
In this subsection, we consider the the $(1)-(2)- \cdots -(m)-[n]$ theory (with $n> m+1$) and its mirror.  As can be seen from the previous examples, this theory can be identified with $T^\sigma_{~\rho}(SU(n))$, where
\bea
\sigma =( \underbrace{1, \ldots,1}_{n~\text{one's}})~, \qquad \qquad \rho = (n-m, \underbrace{1,\ldots,1}_{m~\text{one's}})~, \label{par12mnquiv}
\eea
and $n$ is the total number of boxes in each partition.  It is worth observing that the partition $\rho$, written in terms of Young's diagram, has a hook shape.

Let us compute the dimension of the moduli space.  The quaternionic dimension of the Higgs branch of this theory is
\bea
\dim_{\BH} \text{Higgs}_{(1)-(2)- \cdots -(m)-[n]} &=& \sum_{i=1}^{m-1} i(i+1) + m n - \sum_{j=1}^{m} j^2 \nn \\
&=&  mn -\frac{1}{2} m (m+1)  ~. \label{dimhiggs12mn}
\eea
 The quaternionic dimension of the Coulomb branch this theory is
 \bea
\dim_{\BH} \text{Coulomb}_{(1)-(2)- \cdots -(m)-[n]}  = \sum_{i=1}^m i = \frac{1}{2} m (m+1)~. \label{dimcou12mn}
\eea

The Coulomb branch of the $(1)-(2)- \cdots -(m)-[n]$ theory can be identified with the moduli space of $SU(m+1)$ monopoles in the presence of $n$ fixed monopoles..  Among these $SU(m+1)$ monopoles, $k$ of them has magnetic charge $\alpha_k$ (with $1\leq k \leq m$), where $\alpha_k$ are $(m+1)$-tuples such that
\bea
\alpha_1 = (1,-1,0, \ldots,0), \quad \alpha_2 = (0,1,-1,0,\ldots,0), \quad \ldots, \quad \alpha_{m} = (0,\ldots,0,1,-1)~. \qquad \label{magcharges} 
\eea
The $n$ fixed Dirac monopoles have magnetic charge $(0,0, \ldots,1)$.

\subsection*{The mirror theory}
  
\begin{figure}[htbp]
\begin{center}
\includegraphics[height=1.3 in]{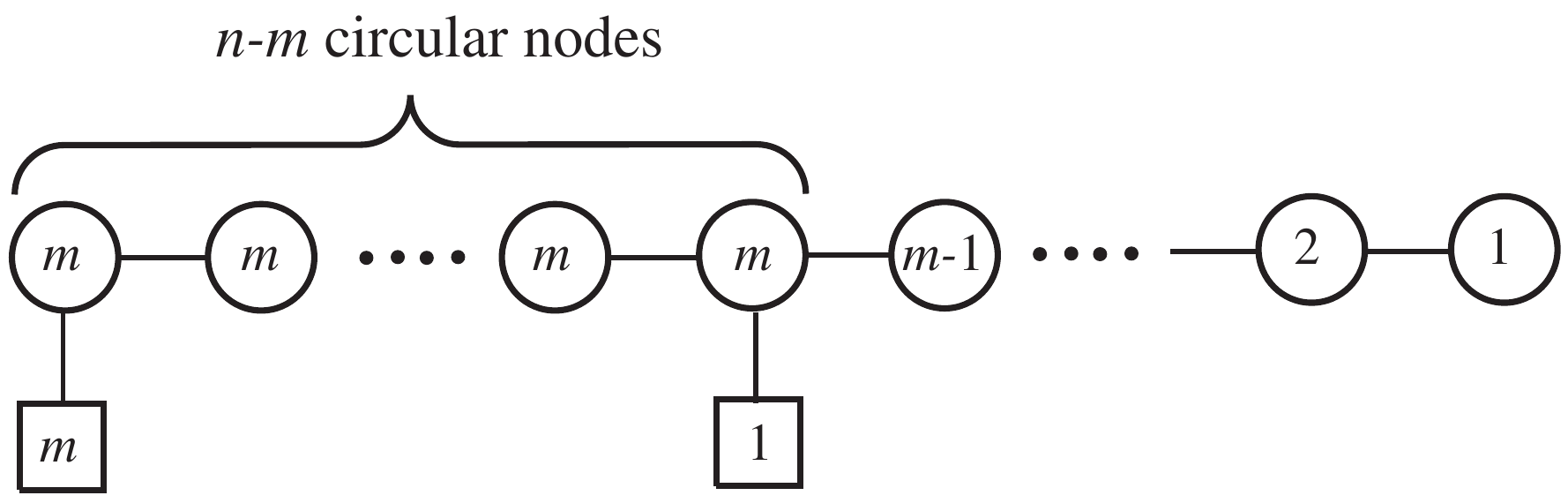}
\caption{The mirror of the $(1)-(2)- \cdots -(m)-[n]$ theory with $n> m+1$.}
\label{fig:gen12mn}
\end{center}
\end{figure}

The mirror of the $(1)-(2)- \cdots -(m)-[n]$ theory (with $n> m+1$) is depicted in \fref{fig:gen12mn}.  This theory can be identified with $T^\rho_{~\sigma}(SU(n))$, where the partitions $\sigma$ and $\rho$ are defined as in \eref{par12mnquiv}.

One can compute the dimension of the moduli space from the quiver diagram.  The quaternionic dimension of the Higgs branch of this theory is
\bea
\dim_{\BH} \text{Higgs}_{\text{mirror}} &=& m^2+m^2(n-m-1)+m+\sum_{i=1}^{m-1} i(i+1)-m^2(n-m) -\sum_{i=1}^{m-1} i^2 \nn \\
&=& \frac{1}{2}m(m+1)~. \label{dimhiggsmir12mn}
\eea
 The quaternionic dimension of the Coulomb branch this theory is
 \bea
\dim_{\BH} \text{Coulomb}_{\text{mirror}}  &=& \sum_{i=1}^{m-1} i +m(n-m)  \nn \\
&=& mn -\frac{1}{2} m (m+1) ~. \label{dimcoumir12mn}
\eea
The results are in agreement with the exchange of the Coulomb and Higgs branches of the theory and its mirror predicted by mirror symmetry.

\subsection*{The Coulomb branch of the $(1)-(2)-\cdots-(m)-[n]$ theory}
In this section, we compute the Hilbert series of the Coulomb branch of the $(1)-(2)-\cdots-(m)-[n]$ theory.  We make use of mirror symmetry and compute this from the Higgs branch of the mirror theory.

Let $q_1$ be the global $U(1)$ fugacity in the mirror of the $(1)-(2)-\cdots-(m)-[n]$ theory and let $(q_2,x_1,\ldots, x_{m-1})$ be the $U(1) \times SU(m) = U(m)$ global fugacity.

The Hilbert series of the Higgs branch of the mirror of the $(1)-(2)-\cdots-(m)-[n]$ theory (or, equivalently, the Coulomb branch of the $(1)-(2)-\cdots-(m)-[n]$ theory) is given by
\bea
&& H^{C}_{(1)-(2)-\cdots-(m)-[n]} (t, q_1, q_2,x_1, \ldots,x_{m-1}) = \prod_{k=0}^{m-1} (1-t^{2n-2k}) \times \nn \\
&&  \qquad \qquad \PE \left[ ([1,0, \ldots,0,1] +[0,\ldots,0]) t^2 + \left( \frac{q_2}{q_1} [1,0, \ldots, 0] + \frac{q_1}{q_2} [0,\ldots,0,1] \right) t^{n-m+1} \right] ~.  \label{HSCou12mn} \nn \\
\eea 
Note that the representation $[1,0, \ldots,0,1] +[0,\ldots,0]$ of $SU(m)$ is in fact the adjoint representation of the $U(m)$ global symmetry, and the representation  $\frac{q_2}{q_1} [1,0, \ldots, 0] + \frac{q_1}{q_2} [0,\ldots,0,1] $ is the bi-fundamental representation of $U(m) \times U(1)$ global symmetry.

The Hilbert series indicates that the Coulomb branch of the $(1)-(2)-\cdots-(m)-[n]$ theory is indeed a {\it complete intersection}. There are $m^2$ generators at order $t^2$ in the adjoint representation of $U(m)$, and $2m$ generators at order $t^{n-m+1}$ in the bi-fundamental representation of $U(m) \times U(1)$.  These generators are subject to one relation at each order of $t^k$ (for $0 \leq k \leq m-1$).  These altogether give $m^2+2m-m = m^2+m$ complex dimensional space or, equivalently, $m(m+1)/2$ quaternionic dimensional space -- in agreement with \eref{dimcou12mn} and \eref{dimhiggsmir12mn}.

\begin{figure}[htbp]
\begin{center}
\includegraphics[height=1.5 in]{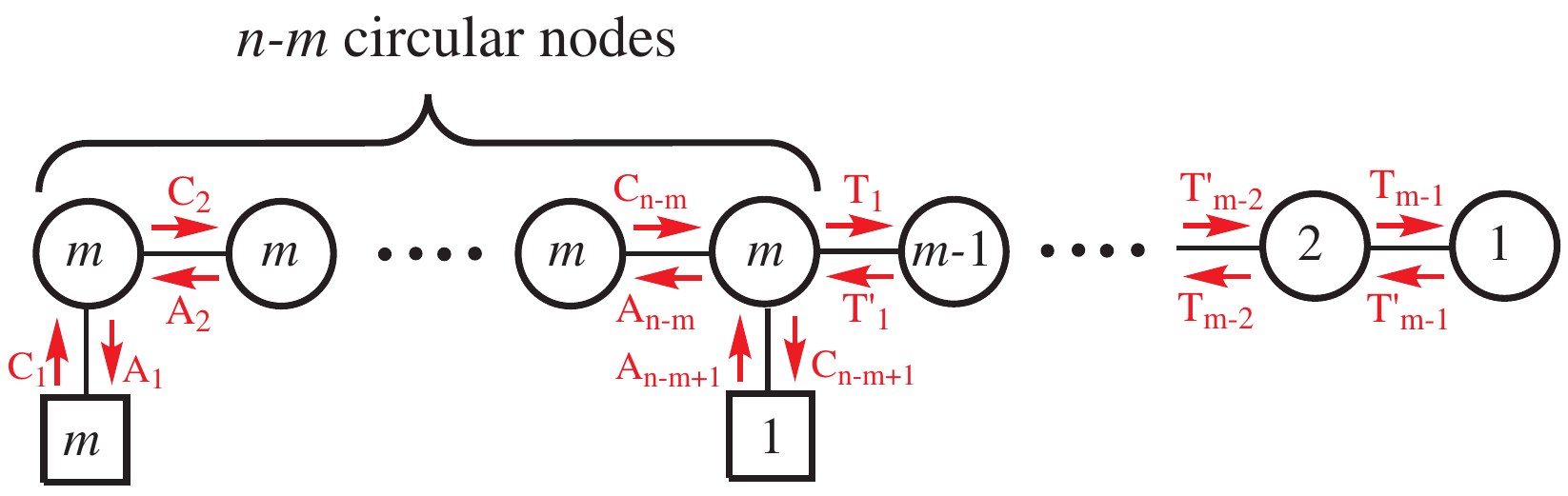}
\caption{The quiver diagram of the mirror of the $(1)-(2)-\cdots-(m)-[n]$ theory, with the bi-fundamental chiral multiplets labelled. Note that $A$'s denote the chiral fields in the anti-clockwise direction, $C$'s denote the chiral fields in the clockwise direction, and $T, T'$s denote the chiral fields in the tail part.}
\label{fig:gen12mnfield}
\end{center}
\end{figure}

\paragraph{The generators.}  Let us consider the generators of the Higgs branch of the mirror theory.  Below we suppress all gauge indices but the global indices $i,j=1, \ldots,m$ are shown explicitly.

The generators at order $t^2$ can be written as
\bea
M^i_{~j} =  (C_1)^i (A_1)_j~.
\eea
The generators at order $t^{n-m+1}$ are
\bea
L_i =  A_{n-m+1}\cdots A_2 (A_1)_{i} ~, \qquad \quad
R^i = (C_1)^{i} C_2 \cdots C_{n-m+1}~.
\eea

\paragraph{The relations.} The relation at order $t^{2n-2k}$ (with $0\leq k \leq m-1$) is
\bea
L_i (M^{m-k-1})^i_{~j} R^j = f_{m,n; k} (M)~,
\eea
where $f_{m,n; k} (M)$ is a function of gauge invariant operators of order $t^{2n-2k}$ constructed only from $M$.
Below we give examples for certain special cases. 
\bi
\item From \eref{relation123n}, we have for $m= n-1$,
\bea
f_{n-1,n; k} (M) = \tr (M^{n-k}) =0 \quad \text{for $0\leq k \leq n-1$}~. \label{f123n}
\eea 
\item From \eref{rel6124a}, \eref{rel8124a} and \eref{rel8125a}, \eref{rel10125a}, we conjecture that for $m=2$ and $k=0,1$,
\bea
f_{2,n; k} (M) &=& (-1)^n  \left[ (\tr M^{n-k})+  \sum_{r=1}^{\lfloor n/2 \rfloor} (\tr M^{n-k-2r})  (\det M)^{r} \right].
\eea
Observe that when $n=3$, we recover \eref{f123n}, since $M$ is nilpotent and $\det M =0$.
\ei

%%%%%%%%%%%%%%%%%%%%%%%%%%%%%%%%%%%%%%%%%%%%
\section{The $(k)-(2k)- \cdots -(nk-k)-[nk]$ theory and its mirror} \label{sec:k2knk}
For $k=1$, we have the $(1)-(2)-\cdots-(n)$ theory, which has been considered in Section 1.  In this section, we examine the two examples, namely the $(2)-(4)-\cdots-(2n-2)-[2n]$ and $(3)-(6)-\cdots-(3n-3)-[3n]$ theories (and their mirrors).  The results for general $k$ can be deduced from these examples.

\subsection{Example: The $(2)-[4]$ theory and its mirror}
\begin{figure}[htbp]
\begin{center}
\includegraphics[height=2.8 in]{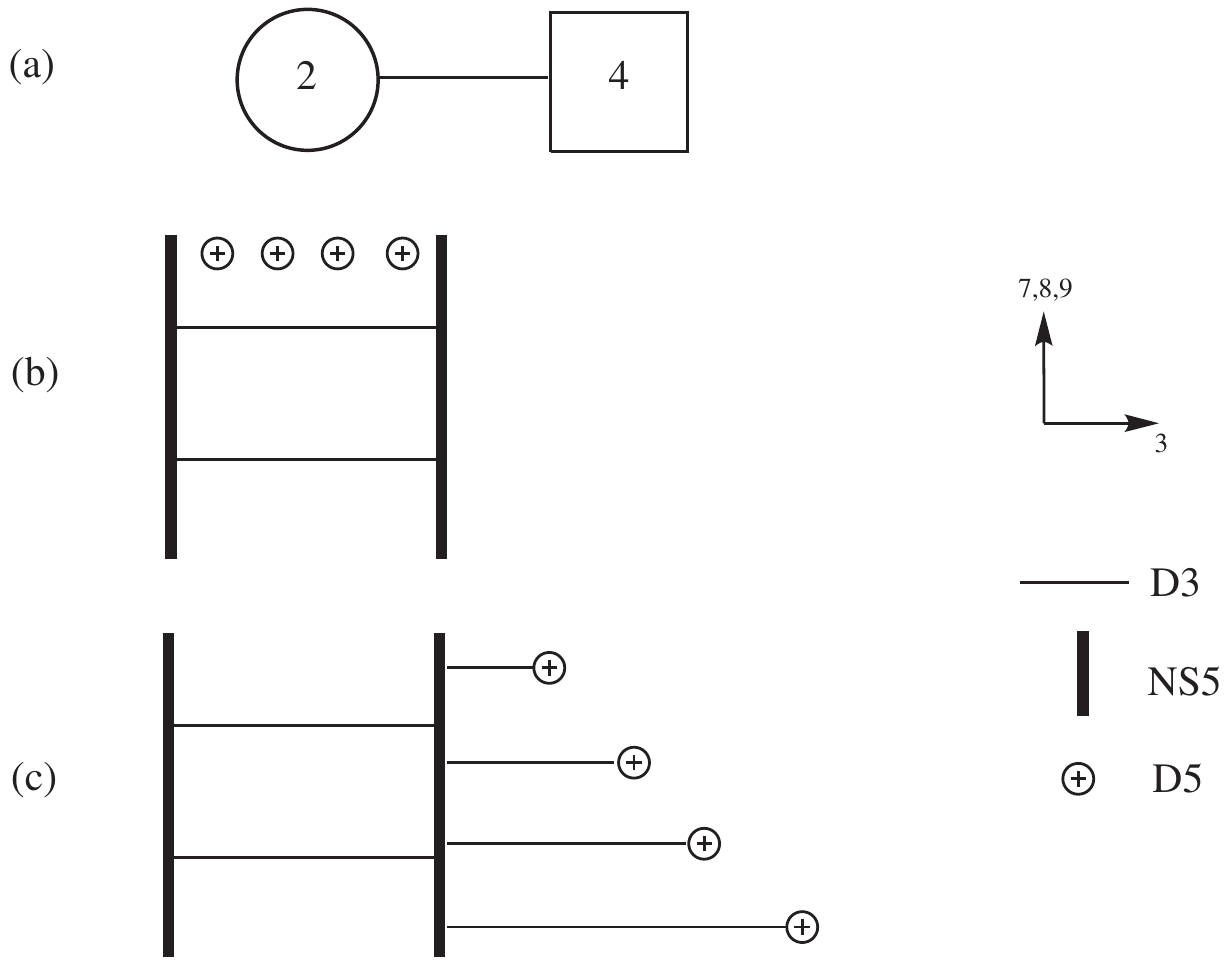}
\caption{(a) The quiver diagram of the $(2)-[4]$ theory. (b) The corresponding brane configuration. (c) The D5-branes are moved to the right of all NS5-branes.  The D3-branes are created according to \cite{Hanany:1996ie}.  The partitions $\sigma = (1,1,1,1)$ and $\rho = (2,2)$ are in one-to-one correspondence with this diagram.}
\label{fig:24quiv}
\end{center}
\end{figure}

The quiver diagram and the corresponding brane configuration are depicted in \fref{fig:24quiv}.  From diagram (c), this theory can be identified with $T^\sigma_{~\rho}(SU(4))$, where
\bea
\sigma = (1,1,1,1)~, \qquad \rho = (2,2)~, \label{par24quiv}
\eea
and the number 4 in $SU(4)$ indicates the total number of boxes in each partition.

One can compute the dimension of the moduli space from the quiver diagram.  
The quaternionic dimension of the Higgs branch of this theory is
\bea
\dim_{\BH} \text{Higgs}_{(2)-[4]} = (2 \times 4) - 2^2 = 4~. \label{dimhiggs24}
\eea
On the other hand, the quaternionic dimension of the Coulomb branch of this theory is
\bea
\dim_{\BH} \text{Coulomb}_{(2)-[4]} = 2 ~. \label{dimcou24}
\eea

\subsection*{The mirror theory}
\begin{figure}[htbp]
\begin{center}
\includegraphics[height=2.8 in]{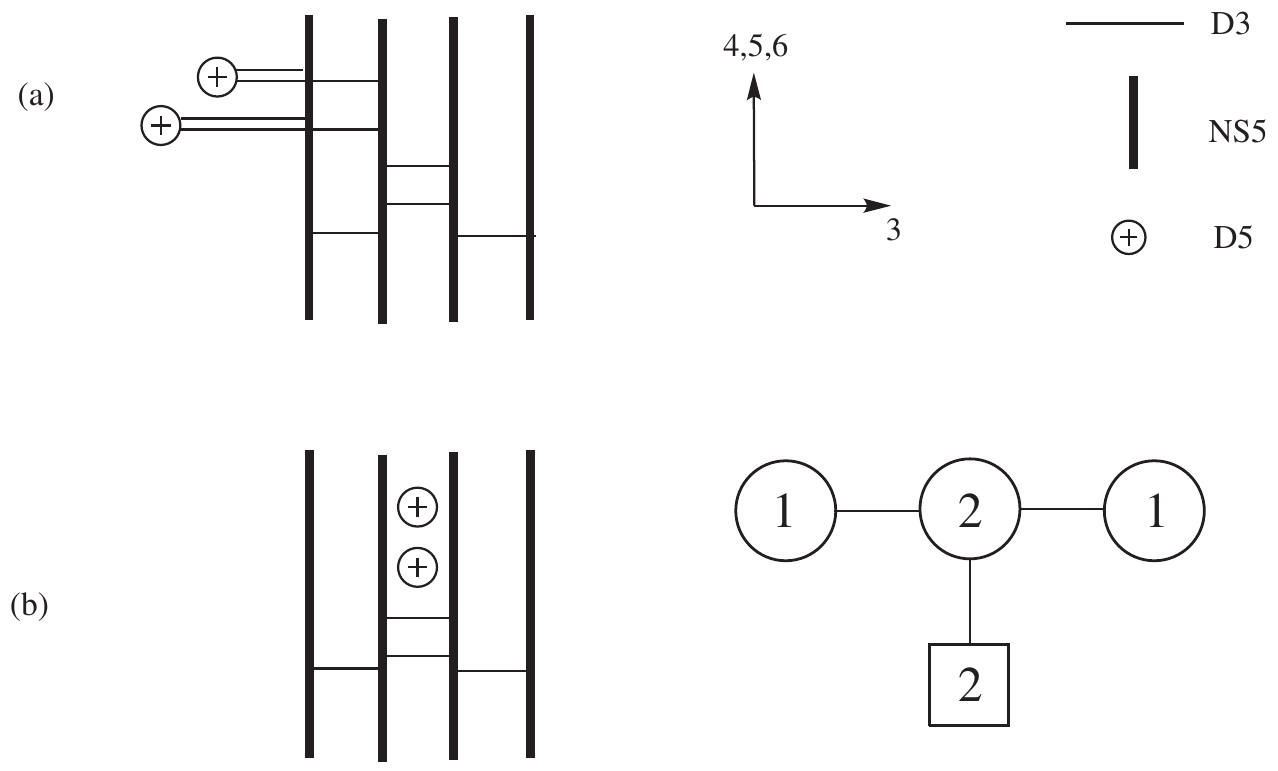}
\caption{{\it The mirror of the $(2)-[4]$ theory.} (a) From diagram (c) in \fref{fig:24quiv}, the NS5-branes and the D5-branes are exchanged and the directions $x^4, x^5, x^6$ are rotated into $x^7, x^8, x^9$ and {\it vice-versa}. The partitions $\sigma = (2,2)$ and $\rho = (1,1,1,1)$ are in one-to-one correspondence with this diagram. (b) The D5-branes are moved across the NS5-branes. The D3-brane creation and annihilation are according to \cite{Hanany:1996ie}. The corresponding quiver diagram is also given next to the brane configuration.}
\label{fig:mirror24}
\end{center}
\end{figure}

Now let us consider the mirror of the $(2)-[4]$ theory.  The brane configuration of the mirror theory can be obtained as described in \cite{Hanany:1996ie} and is depicted in \fref{fig:mirror24}.  From diagram (c), it can be seen that 
\bea
\sigma = (2,2)~, \qquad \qquad \rho  = (1,1,1,1)~,
\eea
\ie~the partitions $\sigma$ and $\rho$ from \eref{par24quiv} get exchanged.

One can compute the dimension of the moduli space from the quiver diagram.  
The quaternionic dimension of the Higgs branch of the mirror theory is
\bea
\dim_{\BH} \text{Higgs}_{\text{mirror}} = 2(2 \times 1) +(2 \times 2) -  (2 \times 1^2) - 2^2= 2~. \label{dimHiggsmir24}
\eea
The quaternionic dimension of the Coulomb branch of the mirror theory is
\bea
\dim_{\BH} \text{Coulomb}_{\text{mirror}} =1+2+1 = 4~. \label{dimcoumir24}
\eea
The results are in agreement with the exchange of the Coulomb and Higgs branches of the $(2)-[4]$ theory predicted by mirror symmetry.

\subsubsection{The Coulomb branch of the $(2)-[4]$ theory}
\begin{figure}[htbp]
\begin{center}
\includegraphics[height=1 in]{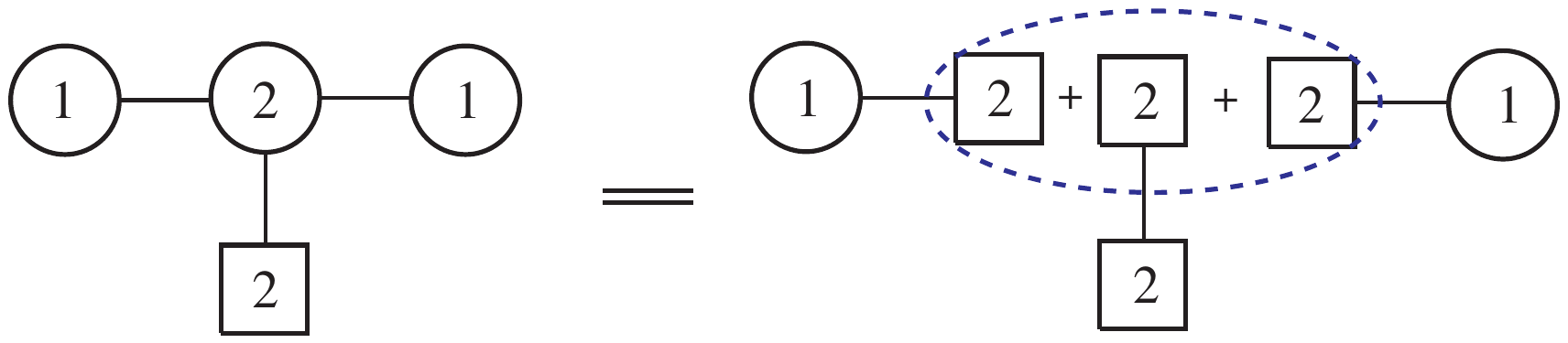}
\caption{The gluing process to obtain the mirror of the $(2)-[4]$ theory.}
\label{fig:gluemir24}
\end{center}
\end{figure}

By mirror symmetry, the Coulomb branch of the $(2)-[4]$ theory is identical to the Higgs branch of its mirror.  
The Hilbert series of the Higgs branch of the mirror theory can be obtained by gluing process \cite{Benvenuti:2010pq,Hanany:2010qu} schematically depicted in \fref{fig:gluemir24}.

Let $x$ be the $SU(2) \subset U(2)$ global fugacity.  The Hilbert series of the Higgs branch of the mirror of the $(2)-[4]$ theory (or, equivalently, the Coulomb branch of the $(2)-[4]$ theory) is given by
\bea
H^{C}_{(2)-[4]} (t, x) &=& (1-t^6)(1-t^{8}) \PE \left[ [2]_x t^2 + [2]_x t^4 \right] ~. \label{HSCou24}
\eea
%Note that the representation $[2]_x +[0]_x$ of $SU(2)$ is in fact the adjoint representation of the $U(2)$ global symmetry, and the representation  $\left( \frac{q_1}{q_2} + \frac{q_2}{q_1} \right) [1]_x$ is the bi-fundamental representation of $U(1) \times U(2)$ global symmetry.

The Hilbert series indicates that the Coulomb branch of the $(2)-[4]$ theory is indeed a complete intersection. There are $3$ generators at order $t^2$ and $3$ generators at order $t^4$.  These generators are subject to one relation at each order of $t^6$ and $t^{8}$.  These altogether give $3+3-2=4$ complex dimensional space or, equivalently, 2 quaternionic dimensional space -- in agreement with \eref{dimcou24} and \eref{dimHiggsmir24}.

\begin{figure}[htbp]
\begin{center}
\includegraphics[height=1.6 in]{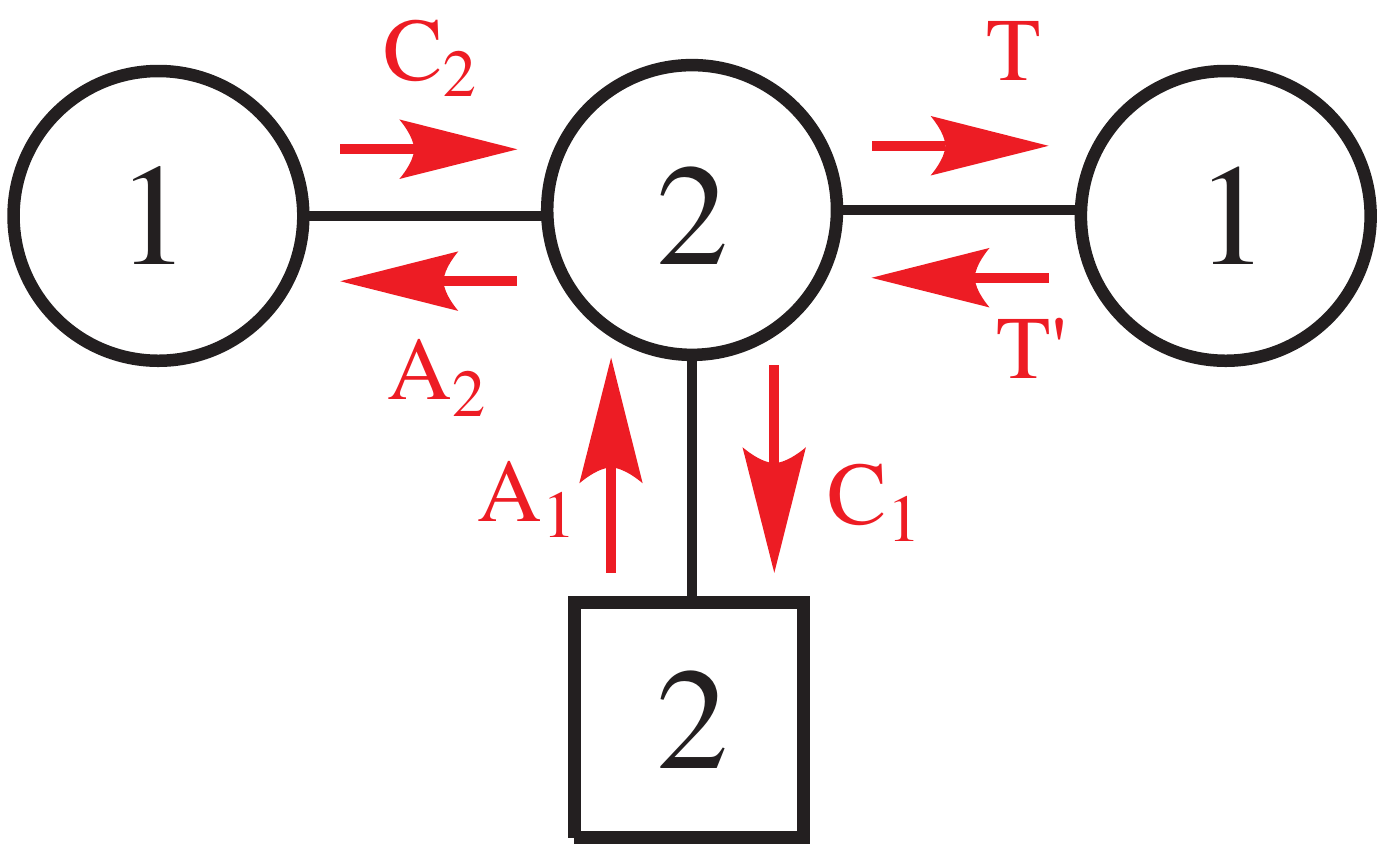}
\caption{The quiver diagram of the mirror of the $(2)-[4]$ theory, with the bi-fundamental chiral multiplets labelled. Note that $A$'s denote the chiral fields in the anti-clockwise direction, $C$'s denote the chiral fields in the clockwise direction, and $T,~T'$ denote the chiral fields in the tail with reducing ranks.}
\label{fig:mir24field}
\end{center}
\end{figure}

\paragraph{The $F$-terms.}  From \fref{fig:mir24field}, we see that the F-term constraints for the bi-fundamental chiral fields are
\bea
0 &=& (C_1)^a_{~i} (A_1)^i_{~b} + (A_2)^a (C_2)_{b} + T^a T'_{b}~, \label{Ftermsmir241} \\
0 &=& - (C_2)_{a} (A_2)^a = - T'_a T^a~, \label{Fterms2}
\eea
where the indices $a, b=1,2$ are the gauge indices corresponding to the $U(2)$ gauge group, and the indices $i, j =1,2$ are the global indices corresponding to the $U(2)$ global symmetry.

From Footnote \ref{fn:lemma}, the $F$-term constraint \eref{Fterms2} implies that the matrices $(A_2)^a (C_2)_{b}$ and $ T^a T'_{b}$ (with $a,b=1,2$) are nilpotent. Hence, using a $U(2)$ gauge transformation, one can put either of these matrices into an upper triangular matrix (with zero diagonal elements):
\bea
\begin{pmatrix} 0 & 1 \\ 0 & 0 \end{pmatrix}~. \label{JNF24}
\eea
This fact can be very useful for verifying the relations below.

\subsubsection*{Generators and relations of the Coulomb branch of the $(2)-[4]$ theory}    
\paragraph{Order $t^2$.} The generators at order $t^2$ can be written as
\bea
(G_2)^i_{~j} =  (A_1)^i_{~a} (C_1)^a_{~j}~.
\eea
Note that the trace of $G_2$ vanishes, \ie
\bea (G_2)^i_{~i} =0~.  \label{tracelessM24}
\eea
This follows immediately from the $F$-term constraints \eref{Ftermsmir241} and \eref{Fterms2}.  
Hence, $G_2$ indeed transforms under the adjoint representation of the global symmetry $SU(2)$, which is a subgroup of the global $U(2)$ symmetry.
This is in agreement with the information contained in Hilbert series \eref{HSCou24}.

Since $G_2$ is a traceless $2 \times 2$ matrix, the eigenvalues of $G_2$ are $\sqrt{-\det G_2}$ and $-\sqrt{-\det G_2}$.  It then follows that for any non-negative integer $n$,
\bea \tr  G_2^{2n+1}  = 0~. 
\eea

Another relation which follows immediately from the tracelessness of the $2 \times 2$ matrix $G_2$ is
\bea
\det G_2 = -\frac{1}{2} \tr (G_2^2)~.
\eea

\paragraph{Order $t^4$.} From \fref{fig:mir24field}, there are two possible candidates for the generators at order $t^4$ which carry two fundamental indices $i, j=1,2$ of the global symmetry $U(2)$:
\bea
(G_4)^i_{~j} = (A_1)^i_{~a} (A_2)^a (C_2)_{b} (C_1)^b_{~j}~, \qquad 
\widetilde{G_4}^i_{~j} = (A_1)^i_{~a} T^a T'_{b} (C_1)^b_{~j}~.
\eea
From \eref{Ftermsmir241}, it turns out that $G_4$ and $\widetilde{G_4}$ are related to each other by
\bea
G_4+ \widetilde{G_4} = - G_2^2~.
\eea
In other words, there is actually one independent generator -- we take $G_4$ to be the generator at order $t^4$.

Moreover, using \eref{Ftermsmir241} and \eref{Fterms2}, we find that
\bea
\tr G_4 &=&  (A_2)^a (C_2)_{b} (C_1)^b_{~i} (A_1)^i_{~a} \nn \\
&=& -(A_2)^a (C_2)_{b} \left[ (A_2)^b (C_2)_{a} + T^b T'_{a}\right] \nn \\
&=& - (A_2)^a (C_2)_{b}T^b T'_{a} \nn \\
&=&  \left[(C_1)^a_{~i} (A_1)^i_{~b} + T^a T'_{b}\right] T^b T'_{a}\nn \\
&=& (C_1)^a_{~i} (A_1)^i_{~b}T^b T'_{a} \nn \\
&=& \tr \widetilde{G_4}~. \label{GeqGtil24}
\eea
Thus, it follows that
\bea
0 = \tr G_2^2 + 2\tr G_4 ~; \label{trG424}
\eea
\ie, the trace of $G_4$ is fixed by the trace of $G_2^2$.
%(which is non-zero at a generic point on the moduli space).  
These facts about the generators at order $t^4$ are in agreement with the Hilbert series \eref{HSCou24}.

\paragraph{Order $t^6$.} The relation at order $t^6$ can be written as
\bea
\tr (G_2 \cdot G_4) =0~. \label{trMG024}
\eea

Note that $\tr G_2^3 =0$ is also a relation at order $t^6$.  However, as discussed in \eref{vanishingoddcasimir}, such a relation follows directly from the fact that $G_2$ is a traceless $2 \times 2$ matrix.  Therefore, we do not count it as an independent relation from \eref{tracelessM24}.

\paragraph{Order $t^8$.} Finally, the relation at order $t^8$ can be written as
\bea
\det G_4 = 0~, \label{detG024}
\eea

We can compute other relations which are not independent from the one above.  For example,
\bea
0 &=& \tr (G_4^2) - \left( \tr G_4 \right)^2~, \label{order824a} \\
0 &=& \tr (G_4^2)+  \tr (G_2^2 \cdot G_4)  ~, \label{order824} \\
0 &=& \tr (G_2^4) + 2 \tr(G_2^2 \cdot G_4)~. \label{trm4order824}
\eea
We emphasise that these relations are not independent from \eref{detG024} but can be derived from the aforementioned relations.

The relations \eref{trMG024}--\eref{trm4order824} are derived in Appendix \ref{app:rel24}.

\subsection{Example: The $(3)-[6]$ theory and its mirror}
\begin{figure}[htbp]
\begin{center}
\includegraphics[height=2.8 in]{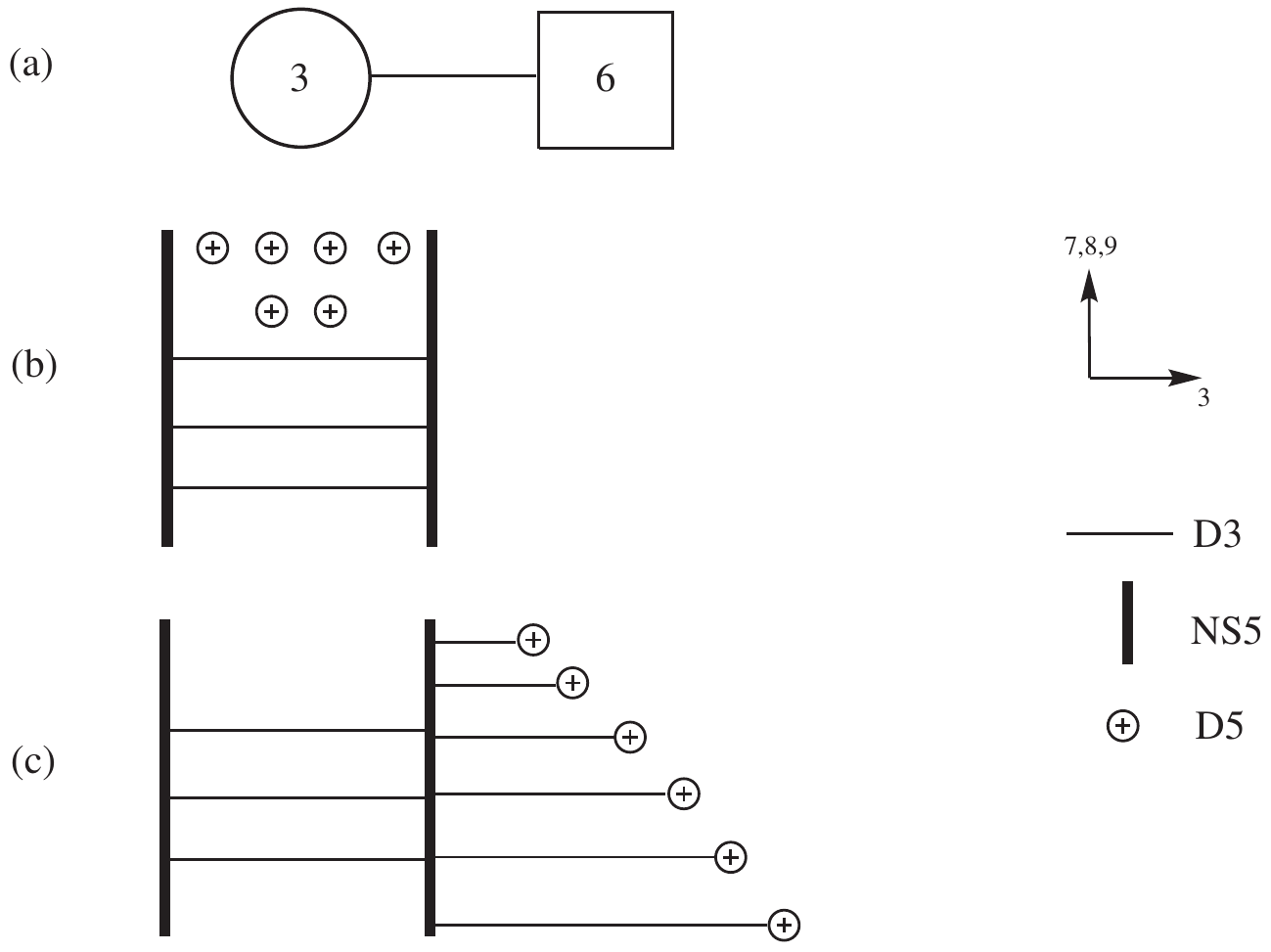}
\caption{(a) The quiver diagram of the $(3)-[6]$ theory. (b) The corresponding brane configuration. (c) The D5-branes are moved to the right of all NS5-branes.  The D3-branes are created according to \cite{Hanany:1996ie}.  The partitions $\sigma = (1,1,1,1,1,1)$ and $\rho = (3,3)$ are in one-to-one correspondence with this diagram.}
\label{fig:36quiv}
\end{center}
\end{figure}

The quiver diagram and the corresponding brane configuration are depicted in \fref{fig:36quiv}.  From diagram (c), this theory can be identified with $T^\sigma_{~\rho}(SU(6))$, where
\bea
\sigma = (1,1,1,1,1,1)~, \qquad \rho = (3,3)~, \label{par36quiv}
\eea
and the number 6 in $SU(6)$ indicates the total number of boxes in each partition and this is also the number of D5-branes present in \fref{fig:36quiv}.

One can compute the dimension of the moduli space from the quiver diagram.  
The quaternionic dimension of the Higgs branch of this theory is
\bea
\dim_{\BH} \text{Higgs}_{(3)-[6]} = (3 \times 6) - 3^2 = 9~. \label{dimhiggs36}
\eea
On the other hand, the quaternionic dimension of the Coulomb branch of this theory is
\bea
\dim_{\BH} \text{Coulomb}_{(3)-[6]} = 3 ~. \label{dimcou36}
\eea

\subsection*{The mirror theory}
\begin{figure}[htbp]
\begin{center}
\includegraphics[height=2.8 in]{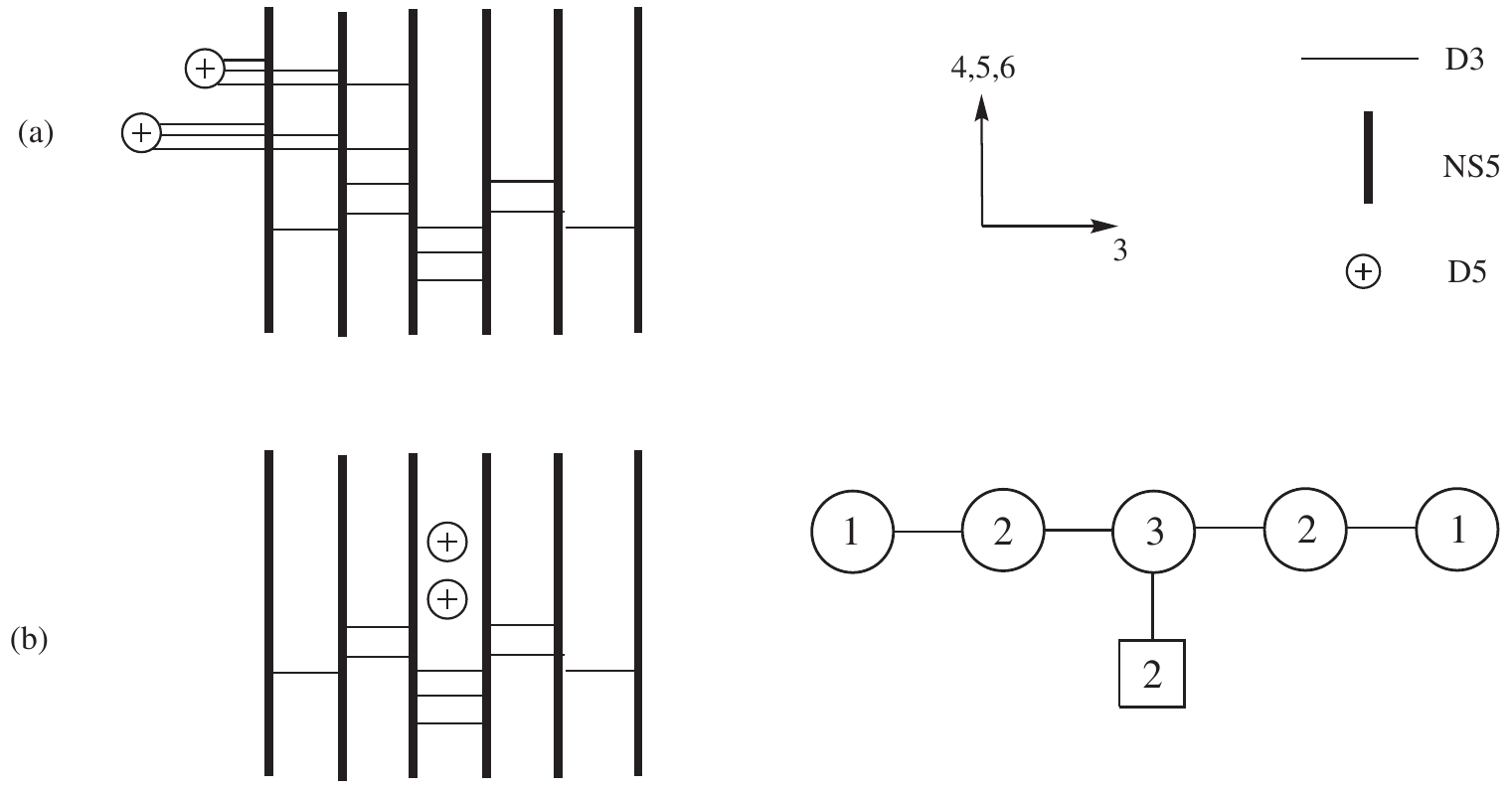}
\caption{{\it The mirror of the $(3)-[6]$ theory.} (a) From diagram (c) in \fref{fig:36quiv}, the NS5-branes and the D5-branes are exchanged and the directions $x^4, x^5, x^6$ are rotated into $x^7, x^8, x^9$ and {\it vice-versa}.  The partitions $\sigma = (3,3)$ and $\rho = (1,1,1,1,1,1)$ are in one-to-one correspondence with this diagram. (b) The D5-branes are moved across the NS5-branes. The D3-brane creation and annihilation are according to \cite{Hanany:1996ie}. The corresponding quiver diagram is also given next to the brane configuration.}
\label{fig:mirror36}
\end{center}
\end{figure}

Now let us consider the mirror of the $(3)-[6]$ theory.  The brane configuration of the mirror theory can be obtained as described in \cite{Hanany:1996ie} and is depicted in \fref{fig:mirror36}.  From diagram (c), it can be seen that 
\bea
\sigma = (3,3)~, \qquad \qquad \rho  = (1,1,1,1,1,1)~,
\eea
\ie~the partitions $\sigma$ and $\rho$ from \eref{par36quiv} get exchanged.

Let us compute the dimension of the moduli space. 
The quaternionic dimension of the Higgs branch of the mirror theory is
\bea
\dim_{\BH} \text{Higgs}_{\text{mirror}} = 2(1 \times 2) +3 (2 \times 3) -(2\times 1^2)- ( 2\times 2^2) -3^2= 3~. \label{dimHiggsmir36}
\eea
The quaternionic dimension of the Coulomb branch of the mirror theory is
\bea
\dim_{\BH} \text{Coulomb}_{\text{mirror}} =1+2+3+2+1= 9~. \label{dimcoumir36}
\eea
The results are in agreement with the exchange of the Coulomb and Higgs branches of the $(3)-[6]$ theory predicted by mirror symmetry.

\subsubsection{The Coulomb branch of the $(3)-[6]$ theory}
\begin{figure}[htbp]
\begin{center}
\includegraphics[height=2.1 in]{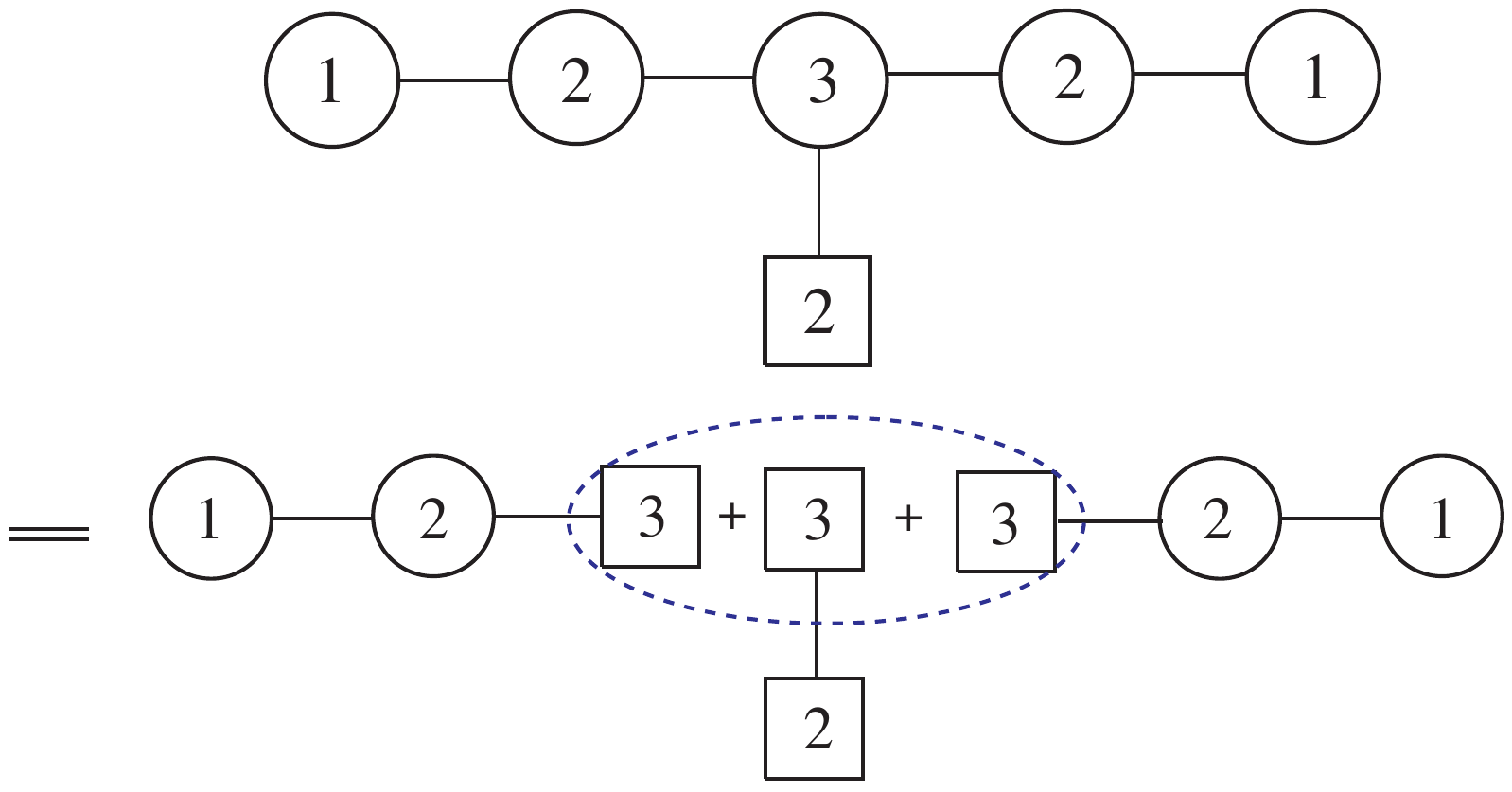}
\caption{The gluing process to obtain the mirror of the $(3)-[6]$ theory.}
\label{fig:gluemir36}
\end{center}
\end{figure}

By mirror symmetry, the Coulomb branch of the $(3)-[6]$ theory is identical to the Higgs branch of its mirror.   The Hilbert series of the Higgs branch of the mirror theory can be obtained by gluing process \cite{Benvenuti:2010pq,Hanany:2010qu} schematically depicted in \fref{fig:gluemir36}.

Let $x$ be the $SU(2) \subset U(2)$ global fugacity.  The Hilbert series of the Higgs branch of the mirror of the $(3)-[6]$ theory (or, equivalently, the Coulomb branch of the $(3)-[6]$ theory) is given by
\bea
H^{C}_{(3)-[6]} (t, x) &=& (1-t^8)(1-t^{10})(1-t^{12}) \PE \left[ [2]_x (t^2 + t^4 + t^6) \right] ~. \label{HSCou36}
\eea

The Hilbert series indicates that the Coulomb branch of the $(3)-[6]$ theory is indeed a complete intersection. There are $3$ generators at each order of $t^2$, $t^4$ and $t^6$.  These generators are subject to one relation at each order of $t^8$, $t^{10}$ and $t^{12}$.  These altogether give $9-3=6$ complex dimensional space or, equivalently, 3 quaternionic dimensional space -- in agreement with \eref{dimcou36} and \eref{dimHiggsmir36}.

\begin{figure}[htbp]
\begin{center}
\includegraphics[height=1.6 in]{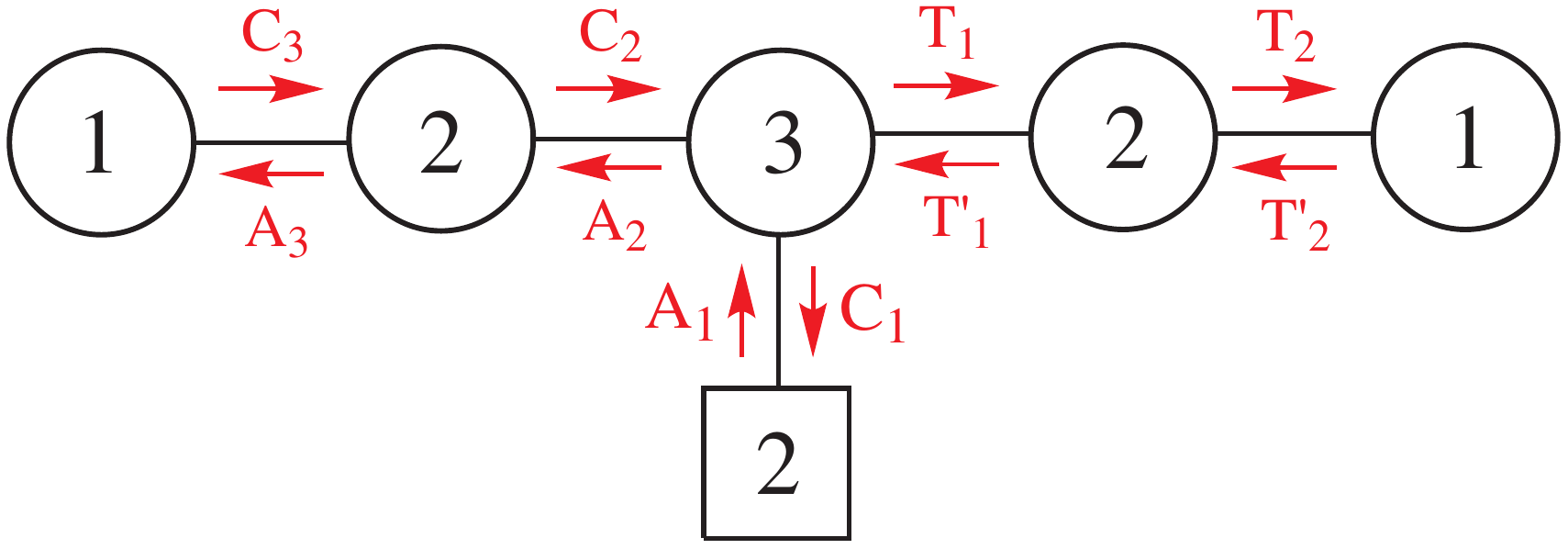}
\caption{The quiver diagram of the mirror of the $(3)-[6]$ theory, with the bi-fundamental chiral multiplets labelled. Note that $A$'s denote the chiral fields in the anti-clockwise direction, $C$'s denote the chiral fields in the clockwise direction, and $T$s,$T'$s denote the chiral fields in the tail with reducing ranks.}
\label{fig:mir36field}
\end{center}
\end{figure}

\paragraph{The $F$-terms.}  From \fref{fig:mir36field}, we see that the F-term constraints for the bi-fundamental chiral fields are
\bea
0 &=& (C_1)^a_{~i} (A_1)^i_{~b} + (A_2)^a_{~a_2} (C_2)^{a_2}_{~b} + (T_1)^a_{~a'_2} (T'_1)^{a'_2}_{~b}~, \label{Ftermsmir361} \nn \\
0 &=&  (A_3)^{a_2} (C_3)_{b_2} - (C_2)^{a_2}_{~a} (A_2)^a_{~b_2} \nn \\
0 &=&  (T_2)^{a'_2} (T'_2)_{b'_2} - (T'_1)^{a'_2}_{~a} (T_1)^a_{~b'_2} \nn \\
0 &=& - (C_3)_{a_2} (A_3)^{a_2} = - (T'_2)_{a'_2} (T_2)^{a'_2}~, \label{Fterms362}
\eea
where the indices $a, b =1,2,3$ are the gauge indices corresponding to the $U(3)$ gauge group, the indices $a_2, b_2 =1,2$ are the gauge indices corresponding to the $U(2)$ gauge group on the left,  the indices $a'_2, b'_2 =1,2$ are the gauge indices corresponding to the $U(2)$ gauge group on the right, and the indices $i, j =1,2$ are the global indices corresponding to the $U(2)$ global symmetry.

Using the lemma in Footnote \ref{fn:lemma}, the $F$-term constraint \eref{Fterms2} implies that the matrices $(A_2)^a_{~a_2} (C_2)^{a_2}_{~b}$ and $(T_1)^a_{~a'_1} (T'_1)^{a'_2}_{~b}$ (with $a,b=1,2,3$) are nilpotent. Hence, using a $U(2)$ gauge transformation, one can put either of these matrices into an upper triangular matrix (with zero diagonal elements):
\bea
\begin{pmatrix} 0 & 1 & 0 \\ 0 &0 & 1 \\ 0 & 0 & 0  \end{pmatrix}~.
\eea
This fact can be very useful for deriving the relations below.

\subsubsection*{Generators and relations of the Coulomb branch of the $(3)-[6]$ theory}    
\paragraph{Order $t^2$.} The generators at order $t^2$ can be written as
\bea
(G_2)^i_{~j} =  (A_1)^i_{~a} (C_1)^a_{~j}~.
\eea
Note that the trace of $G_2$ vanishes, \ie
\bea (G_2)^i_{~i} =0~.  \label{tracelessM36}
\eea
This follows immediately from the $F$-term constraints \eref{Ftermsmir361} and \eref{Fterms362}.  
Hence, $G_2$ indeed transforms under the adjoint representation of the global symmetry $SU(2)$, which is a subgroup of the global $U(2)$ symmetry.
This is in agreement with the information contained in Hilbert series \eref{HSCou36}.

Since $G_2$ is a traceless $2 \times 2$ matrix, the eigenvalues of $G_2$ are $\sqrt{-\det G_2}$ and $-\sqrt{-\det G_2}$.  It then follows that for any non-negative integer $n$,
\bea \tr  G_2^{2n+1}  = 0~. \label{vanishingoddcasimir}
\eea

Another relation which follows immediately from the tracelessness of the $2 \times 2$ matrix $G_2$ is
\bea
\det G_2 = -\frac{1}{2} \tr (G_2^2)~.
\eea

\paragraph{Order $t^4$.} The generators at order $4$ can be written as
\bea
(G_4)^i_{~j} = (A_1)^i_{~a} (A_2)^a_{~a_2} (C_2)^{a_2}_{b} (C_1)^b_{~j}~.
\eea
In Appendix \ref{app:rel36}, we show that
\bea
\tr G_2^2 + 2\tr G_4 = 0~; \label{trm236}
\eea
\ie, trace of $G_4$ is fixed by the trace of $G_2^2$.  These facts about the generators at order $t^4$ are in agreement with the Hilbert series \eref{HSCou36}.

\paragraph{Order $t^6$.} The generators at order $6$ can be written as
\bea
(G_6)^i_{~j} &=& (A_1)^i_{~a} (A_2)^a_{~a_2} (A_3)^{a_2} (C_3)_{b_2} (C_2)^{b_2}_{b} (C_1)^b_{~j} \nn \\
&=&  (A_1)^i_{~a} (A_2)^a_{~a_2}  (C_2)^{a_2}_{~a} (A_2)^a_{~b_2} (C_2)^{b_2}_{b} (C_1)^b_{~j}~.
\eea
Note that the trace of $G_6$ is fixed by the trace of $G_2 \cdot G_4$.  In particular, in Appendix \ref{app:rel36}, we show that
\bea
 \tr G_6 = \tr (G_2 \cdot G_4)~. \label{trG636}
\eea

\paragraph{Order $t^8$.} The relation at order $8$ can be written as
\bea
0&=& \tr (G_2^2 \cdot G_4)+2\tr(G_2 \cdot G_6) + \tr( G_4^2)~. \label{order836a}
\eea
Other relations at this order can be derived from the aforementioned relations, \eg,
\bea
0 &=& \tr G_2^4 + 2 \tr(G_2^2 \cdot G_4)~,  \label{order836b} \\
0 &=&  \tr (G_2^2 \cdot G_4) + \left(  \tr G_4 \right)^2~.  \label{order836c}
\eea

\paragraph{Order $t^{10}$.} The relation at order $10$ can be written as
\bea
0 &=&   \tr (G_2 \cdot G_4^2) +  \tr (G_2^2 \cdot G_6) + \tr(G_4 \cdot G_6) +  \tr (G_2^3 \cdot G_4)~.   \label{order1036a}
\eea
Other relations at this order follow from the aforementioned relations, \eg, 
\bea
0 &=&   \tr(G_2^3 \cdot G_4) + \tr (G_2 \cdot G_4^2) = \tr(G_2^3 \cdot G_4) + \tr(G_2^2 \cdot G_6)~. \label{order1036b} 
%0 &=& \tr(G_2^3 \cdot G_4) + \tr(G_2^2 \cdot G_6)~. \label{order1036c}
\eea

\paragraph{Order $t^{12}$.} The relation at order $12$ can be written as
\bea
\det G_6 = 0~. \label{detG636}
\eea
Other relations at this order can be derived from the aforementioned relations, \eg,
\bea
0 &=& 3 \tr(G_2^3 \cdot G_6)+ 3 \tr(G_2^4 \cdot G_4)  + \tr (G_2^6)   - \tr(G_4^3) ~, \label{order1236a } \\
0 &=& 3 \tr(G_2^2 \cdot G_4^2)+ 3 \tr(G_2^4 \cdot G_4)  + \tr (G_2^6)   + 2\tr(G_4^3) ~, \label{order1236b} \\
0 &=&  6 \tr (G_2\cdot G_4\cdot G_6) +6 \tr (G_4 \cdot G_2 \cdot G_6) +6 \tr (G_2^2\cdot G_4^2) +6 \tr (G_2^3\cdot G_6) + \nn \\
&& 6 \tr (G_2^4\cdot G_4)+3 \tr \left[ (G_2\cdot G_4)^2 \right] +3 \tr (G_6^2)+2 \tr (G_4^3)+ \tr (G_2^6)~. \label{order1236}
\eea

We derive relations \eref{order836a}--\eref{detG636} in Appendix \ref{app:rel36}.  Note that relations  \eref{order1236a }--\eref{order1236} can be derived in a similar way as \eref{order1036a} although the calculation is a little longer.

\subsection{Special case: The $(k)-[2k]$ theory and its mirror}
Let us now consider the $(k)-[2k]$ theory for general $k \geq 1$.  This theory can be identified with $T^\sigma_{~\rho}(SU(2k))$, where
\bea
\sigma =( \underbrace{1, \ldots,1}_{2k~\text{one's}})~, \qquad \qquad \rho = (k,k)~. \label{park2kquiv}
\eea
Note that the total number of boxes is $2k$.  Let us compute the dimension of the moduli space.  The quaternionic dimension of the Higgs branch of this theory is
\bea
\dim_{\BH} \text{Higgs}_{(k)-[2k]} &=& 2k^2 -k^2= k^2~.
\eea
The quaternionic dimension of the Coulomb branch of this theory is
\bea
\dim_{\BH} \text{Higgs}_{(k)-[2k]} &=& k~.
\eea

The Coulomb branch of the $(k)-[2k]$ theory can also be identified with the moduli space of magnetic monopoles.  In particular, this is the moduli space of $k$ $SU(2)$ monopoles with magnetic charge $(1,-1)$, in the presence of $2k$ fixed monopoles with magnetic charge $(0,1)$.

\subsection*{The mirror theory}
\begin{figure}[htbp]
\begin{center}
\includegraphics[height=1 in]{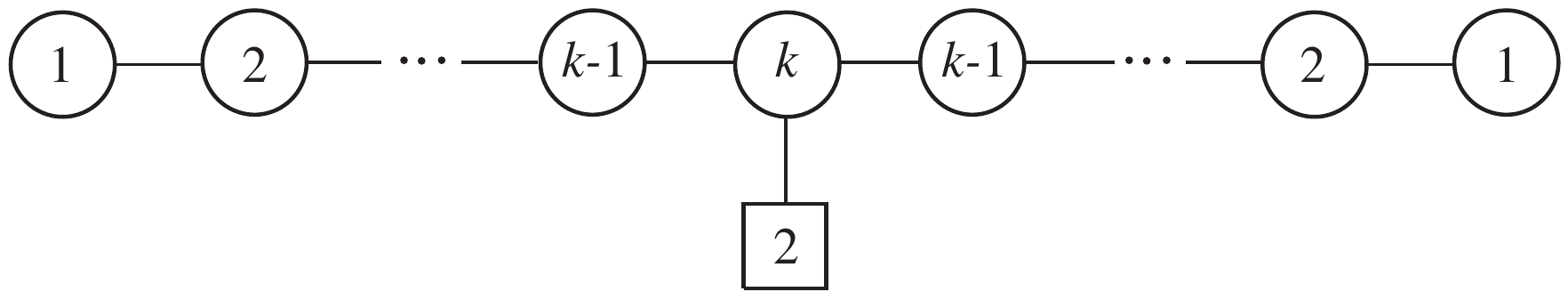}
\caption{The quiver diagram of the mirror of the $(k)-[2k]$ theory.}
\label{fig:mirk2k}
\end{center}
\end{figure}

The quiver diagram of the mirror theory is depicted in \fref{fig:mirk2k}.  This theory can be identified with $T^\rho_{~\sigma}(SU(2k))$, where the partitions $\rho$ and $\sigma$ are given by \eref{park2kquiv}.

One can compute the dimension of the moduli space from the quiver diagram.  The quaternionic dimension of the Higgs branch of this theory is
\bea
\dim_{\BH} \text{Higgs}_{\text{mirror}} &=& 2\sum_{i=1}^{k-1} i(i+1) +2k - 2\sum_{i=1}^{k-1} i^2 -k^2 = k~. \label{dimhiggsmirk2k}
\eea
 The quaternionic dimension of the Coulomb branch this theory is
 \bea
\dim_{\BH} \text{Coulomb}_{\text{mirror}}  &=& 2 \sum_{i=1}^{k-1} i +k  = k^2 ~. \label{dimcoumirk2k}
\eea
The results are in agreement with the exchange of the Coulomb and Higgs branches of the theory and its mirror predicted by mirror symmetry.

\subsubsection{The Coulomb branch of the $(k)-[2k]$ theory}
In this section, we compute the Hilbert series of the Coulomb branch of the $(k)-[2k]$ theory.  We make use of mirror symmetry and compute this from the Higgs branch of the mirror theory.

Let $x$ be a fugacity for the $SU(2) \subset U(2)$ global symmetry in the quiver diagram \fref{fig:mirk2k}.  The Hilbert series of the Higgs branch of the mirror of the $(k)-[2k]$ theory (or, equivalently, the Coulomb branch of the $(k)-[2k]$ theory) is given by
\bea
H^{C}_{(k)-[2k]} (t, x) = \PE \left[ [2]_x \sum_{p=1}^{k} t^{2p} \right] \prod_{q=k+1}^{2k} (1- t^{2q})~.
\eea
The Hilbert series indicates that the Coulomb branch of the $(k)-[2k]$ theory is indeed a {\it complete intersection}.  There are $3$ generators at each of the follwing order: $t^2, t^4, \ldots, t^{2k}$, and one generator at each of the following order: $t^{2(k+1)}, \ldots, t^{4k}$.  These altogether give $3k-k =  2k$ complex dimensional space or, equivalently, $k$ quaternionic dimensional space -- in agreement with \eref{dimhiggsmirk2k}.

\subsubsection*{Generators and relations of the Coulomb branch of the $(k)-[2k]$ theory}  

\begin{figure}[htbp]
\begin{center}
\includegraphics[height=1 in]{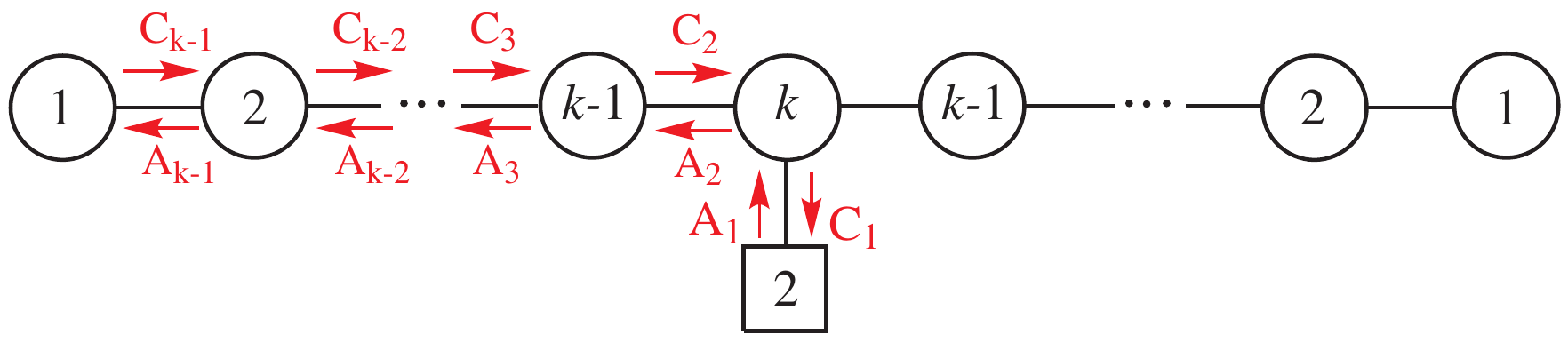}
\caption{The quiver diagram of the mirror of the $(k)-[2k]$ theory, with the bi-fundamental chiral multiplets labelled. Note that $A$'s denote the chiral fields in the anti-clockwise direction, $C$'s denote the chiral fields in the clockwise direction.}
\label{fig:mirk2kfield}
\end{center}
\end{figure}

\subsubsection*{The generators} 
Let us fix $k$. The generators at order $t^{2p}$ (with $1 \leq p \leq k$) can be written as
\bea
G_{2p} = \prod_{r=1}^p A_r  \prod_{s=1}^p C_s~.
\eea
For $p=1$, the operator $G_2$ is traceless:
\bea
\tr G_2 = 0~.
\eea
From the tracelessness condition and the fact that $G_2$ is a $2 \times 2$ matrix, we also have 
\bea \tr G_2^{2n+1} =0 \quad \text{for all $n=1,2,3,\ldots$} \label{trG2vanishodd} \eea

For $1 < p \leq k$, the trace of $G_{2p}$ is fixed by a relation consisting of operators of lower orders.  In particular, we compute the following relations directly from the $(k)-[2k]$ theories for $k=1, \ldots, 6$ (in the same way as in Appendix \ref{app:rel36}) and find that the traces $\tr G_{2p}$ satisfy
\bea
0 &=& \tr G_2~, \label{rel:trG2} \\
0&=&  2 \tr G_4 + \tr (G_2^2)~, \label{rel:trG4} \\
0 &=& 3 \tr G_6 + 3 \tr (G_2 \cdot G_4) +  \tr (G_2^3)~, \label{rel:trG6} \\
0 &=& 4 \tr G_8 + 4 \tr (G_2 \cdot G_6) + 4 \tr (G_2^2 \cdot G_4)  + 2 \tr(G_4^2) +  \tr (G_2^4)~, \label{rel:trG8} \\
0 &=& 5 \tr G_{10} + 5 \tr (G_2 \cdot G_8)+ 5 \tr (G_2 \cdot G_4^2) + 5 \tr (G_2^2 \cdot G_6) +5 \tr(G_4 \cdot G_6) + \nn \\
&& 5 \tr (G_2^3 \cdot G_4)  +  \tr (G_2^5)~,  \label{rel:trG10}  \\
0 &=& 6 \tr G_{12} +6 \tr (G_2\cdot G_{10}) +6 \tr (G_2\cdot G_4\cdot G_6) +6 \tr (G_4 \cdot G_2 \cdot G_6) +   \nn  \\
&& 6 \tr (G_2^2\cdot G_8) +6 \tr (G_2^2\cdot G_4^2) +6 \tr (G_2^3\cdot G_6) +6 \tr (G_2^4\cdot G_4)+ \nn \\
&& 6 \tr (G_4 \cdot G_8)  +3 \tr \left[ (G_2\cdot G_4)^2 \right] +3 \tr (G_6^2)+2 \tr (G_4^3)+ \tr (G_2^6)~, \label{rel:trG12} 
\eea
where it should be noted that for the $(k)-[2k]$ theory $\tr G_2^{2n+1} =0$.  Observe that the second equality \eref{rel:trG4} is identical to \eref{trG424} and \eref{trm236}, and the third equality \eref{rel:trG6} is identical to \eref{trG636}.

\paragraph{Conjecture.} From these examples, we conjecture that the trace $\tr G_{2p}$ satisfies
\bea
&& 0 = p \tr G_{2p} + p \sum_{1 \leq s_1 \leq \ldots \leq s_m \leq p-1} \delta \left( p -  \sum_{i=1}^{m} s_i\right)  \widehat{\sum}_{\substack{\sigma \in S_m}} \frac{\tr  \left( \prod_{i=1}^{m} G_{2 s_{\sigma(i)}} \right)}{\CM \left ( \tr  ( \prod_{i=1}^{m} G_{2 s_{\sigma(i)}} ) \right)}  \nn \\
&& +   \sum_{\substack{1\leq r \leq p/2 \\ r|p}} r \sum_{\substack{ 1 \leq s_1 \leq \cdots \leq s_l \leq r}} \delta \left( r - \sum_{j=1}^l s_j \right) \widehat{\sum}_{\substack{\rho \in S_l}} \frac{ \tr \left[ \left( \prod_{j=1}^l G_{2s_{\rho(j)}}  \right)^{p/r} \right] }{\CM \left ( \tr \left( \prod_{j=1}^l G_{2s_{\rho(j)}}  \right)^{p/r}  \right)} ~, \qquad
%&& 0 = p \tr G_{2p} + p \sum_{\substack{ 1 \leq s^{(j)}_{i} < s^{(j)}_{i+1} \leq p-1 \\ a_j, p^{(j)}_i \geq 0,~~s^{(j)}_{m_j} < s^{(j+1)}_{m_{j+1}}}} \delta \left( p - \sum_{j=1}^l \sum_{i=1}^{m_j} a_j p^{(j)}_i s^{(j)}_i \right)  \tr  \left[ \left( \prod_{i=1}^{m_1} G^{p^{(1)}_i}_{2 s^{(1)}_{i}} \right)^{a_1} \cdots \left(\prod_{i=1}^{m_l} G^{p^{(l)}_i}_{2 s^{(l)}_{i}} \right)^{a_l} \right] \nn \\
%&& +   \sum_{\substack{1\leq r \leq p/2 \\ r|p}} r \sum_{\substack{ 1 \leq s_1 < \cdots < s_n \leq r \\ p_1, \ldots, p_n \geq 0}} \delta \left( r - \sum_{j=1}^n p_j s_j \right) \sum_{\rho \in S_n} \frac{1}{|\CC(\rho)|} \tr \left[ \left( \prod_{j=1}^n G^{p_{\rho(j)}}_{2s_{\rho(j)}}  \right)^{p/r} \right]  ~,
\label{relk2keven}
\eea
where $\widehat{\sum}_{\sigma \in S_m}$ denotes the summation over $\sigma \in S_m$ such that $\prod_{i=1}^{m} G_{2 s_{\sigma(i)}}$ is not an operator with an integer power greater than 1, and $\CM \left ( \tr  ( \prod_{i=1}^{m} G_{2 s_{\sigma(i)}} ) \right)$ denotes the multiplicity of $\tr  \left( \prod_{i=1}^{m} G_{2 s_{\sigma(i)}} \right)$.

\subsubsection*{The relations}  
The relations \eref{relk2keven} also go through at higher orders.  Bearing in mind that the generators $G_{2p}$ with $ p \geq k+1$ do not exist and hence they can be set to zero in \eref{relk2keven}, we conjecture the relation at order $t^{2q}$ (with $k+1 \leq q \leq 2k$) to be as follows:
\bea
&& 0 = q \sum_{1 \leq s_1 \leq \ldots \leq s_m \leq k} \delta \left( q -  \sum_{i=1}^{m} s_i\right)  \widehat{\sum}_{\substack{\sigma \in S_m}} \frac{\tr  \left( \prod_{i=1}^{m} G_{2 s_{\sigma(i)}} \right)}{\CM \left ( \tr  ( \prod_{i=1}^{m} G_{2 s_{\sigma(i)}} ) \right)}  \nn \\
&& +   \sum_{\substack{1\leq r \leq q/2 \\ r|q}} r \sum_{\substack{ 1 \leq s_1 \leq \cdots \leq s_l \leq r}} \delta \left( r - \sum_{j=1}^l s_j \right) \widehat{\sum}_{\substack{\rho \in S_l}} \frac{ \tr \left[ \left( \prod_{j=1}^l G_{2s_{\rho(j)}}  \right)^{q/r} \right] }{\CM \left ( \tr \left( \prod_{j=1}^l G_{2s_{\rho(j)}}  \right)^{q/r}  \right)} ~. \qquad \label{releven}
\eea
%\item For even $q$,
%\bea
%0= q  \sum_{s=  q-k}^{ q-2 } \tr (G_2^s \cdot G_{2q-2s}) +  \frac{q-2}{2}  \sum_{r=1}^{q/2-1} (-1)^{r+1} \tr (G_{2 r}^{q/r})~. \label{releven}
%%0 = q  \sum_{s=  q-k}^{ q-2 } \tr (G_2^s \cdot G_{2q-2s}) + \tr (G_2^{q}) +2 \tr(G_q^2)~. \label{releven}
%\eea
%\item For odd $q$,
%\bea
%%0 &=&   \sum_{s=q-k}^{q-2} \tr (G_2^s \cdot G_{2q-2s}) + \frac{1}{q} \tr (G_2^{q}) \nn \\
%0&=& \sum_{s=q-k}^{q-2} \tr (G_2^s \cdot G_{2q-2s})~. \label{relodd}
%\eea
%%where the last equality follows from \eref{trG2vanishodd}.
Note that the relation at order $t^{4k}$ can alternatively be written as
\bea
\det G_{2k} = 0~.
\eea
Observe that these relations are in agreement with the following examples:  

\paragraph{The $(1)-[2]$ theory.} For $k=1$, we have $q=2$ and $\tr G_2 = 0$.  From \eref{releven}, the relation at order $t^4$ is simply $\tr (G_2^2)=0$, which is in agreement with \eref{relation123n}.

\paragraph{The $(2)-[4]$ theory.} For $k=2$, the relation at order $t^6$ (\ie, $q=3$) obtained above is the same as \eref{trMG024}. The relation at order $t^8$ (\ie, $q=4$) obtained above is the same as that derived from 2\eref{order824}+\eref{trm4order824},
\bea
0 = 4 \tr(G_2^2 \cdot G_4)  + 2 \tr(G_4^2) + \tr (G_2^4)~.
\eea

\paragraph{The $(3)-[6]$ theory.} For $k=3$, the relation at order $t^8$ (\ie, $q=4$) given by \eref{releven},
\bea
0 =  4 \tr (G_2 \cdot G_6) +4 \tr(G_2^2 \cdot G_4) + \tr (G_2^4) + 2 \tr(G_4^2)~,
\eea
can also be derived from 2\eref{order836a}+\eref{order836b}.  The relation at order $t^{10}$ (\ie, $q=5$) given by \eref{releven},
\bea
0 &=&  5 \tr (G_2 \cdot G_4^2) + 5 \tr (G_2^2 \cdot G_6) +5 \tr(G_4 \cdot G_6) + 5 \tr (G_2^3 \cdot G_4)  +  \tr (G_2^5)~, \nn \\
\text{or} \qquad 0 &=&   \tr (G_2 \cdot G_4^2) +  \tr (G_2^2 \cdot G_6) + \tr(G_4 \cdot G_6) +  \tr (G_2^3 \cdot G_4)
\eea
coincides with \eref{order1036a}. Finally, the relation at order $t^{12}$ (\ie, $q=6$) given by \eref{releven},
\bea
0 &=&  6 \tr (G_2\cdot G_4\cdot G_6) +6 \tr (G_4 \cdot G_2 \cdot G_6) +6 \tr (G_2^2\cdot G_4^2) +6 \tr (G_2^3\cdot G_6) + \nn \\
&& 6 \tr (G_2^4\cdot G_4)+3 \tr \left[ (G_2\cdot G_4)^2 \right] +3 \tr (G_6^2)+2 \tr (G_4^3)+ \tr (G_2^6)~,
\eea
coincides with \eref{order1236}. 

%\paragraph{The $(4)-[8]$ theory.} For $k=4$, the relation at order $t^{10}$ can be written as
%\bea
%\tr (G_2^3 \cdot G_4) + 2 \tr (G_2^2 \cdot G_6) + 3 \tr (G_4 \cdot G_6) + 3 \tr (G_2 \cdot G_8) + 2 \tr (G_2 \cdot G_4^2) = 0~. \qquad 
%\eea
%The relation at order $t^{12}$ can be written as
%\bea
%3 \tr (G_2^2 \cdot G_8) + 3 \tr(G_2^3 \cdot G_6) + 3 \tr(G_2^4 \cdot G_4)  +  \tr (G_2^6)  - \tr(G_4^3)=0~.
%\eea

\subsection{Example: The $(2)-(4)-[6]$ theory and its mirror}
\begin{figure}[htbp]
\begin{center}
\includegraphics[height=2.8 in]{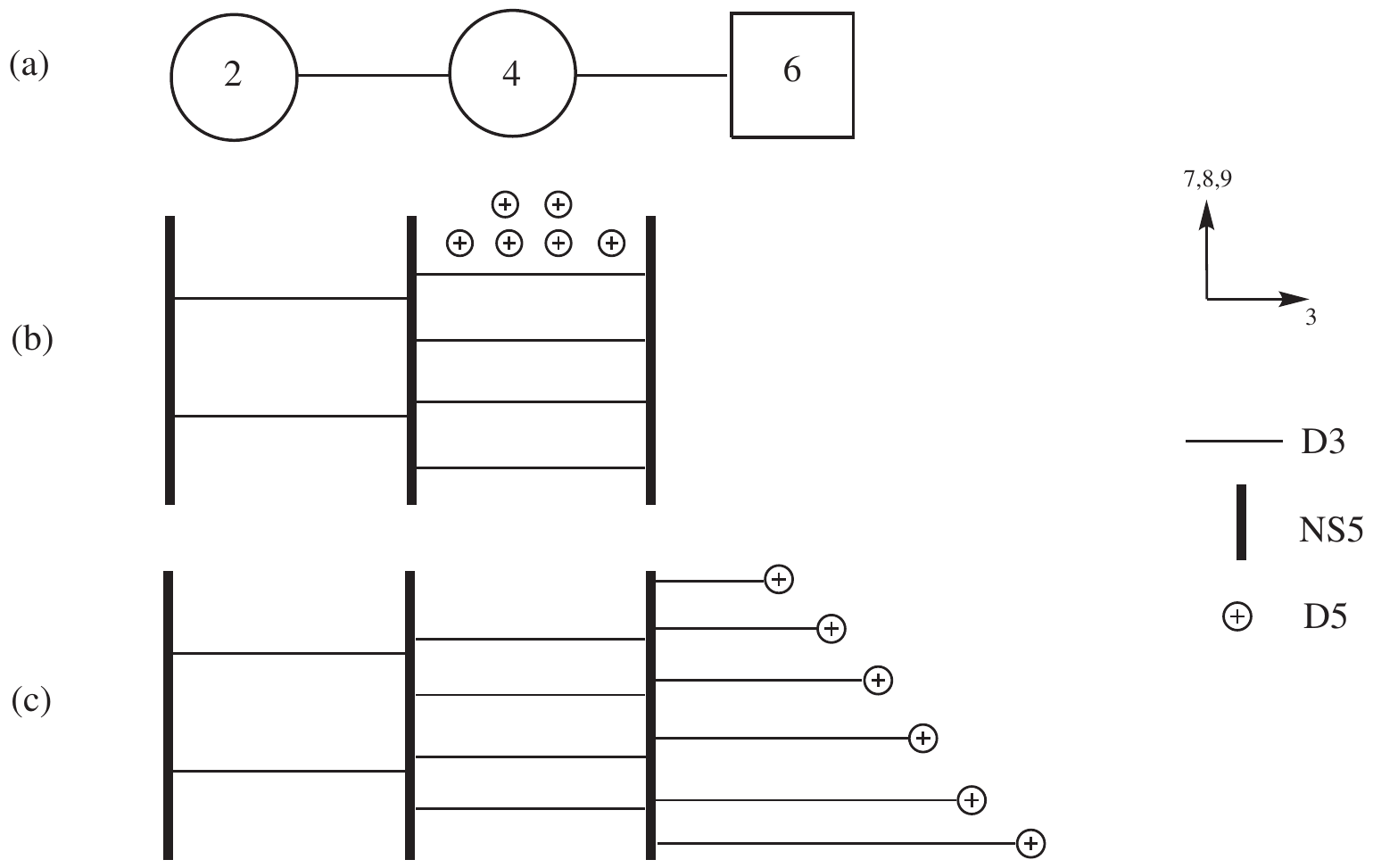}
\caption{(a) The quiver diagram of the $(2)-(4)-[6]$ theory. (b) The corresponding brane configuration. (c) The D5-branes are moved to the right of all NS5-branes.  The D3-branes are created according to \cite{Hanany:1996ie}.  The partitions $\sigma = (1,1,1,1,1,1)$ and $\rho = (2,2,2)$ are in one-to-one correspondence with this diagram.}
\label{fig:246quiv}
\end{center}
\end{figure}

The quiver diagram and the corresponding brane configuration are depicted in \fref{fig:246quiv}.  From diagram (c), this theory can be identified with $T^\sigma_{~\rho}(SU(6))$, where
\bea
\sigma = (1,1,1,1,1,1)~, \qquad \rho = (2,2,2)~, \label{par246quiv}
\eea
and the number 6 in $SU(6)$ indicates the total number of boxes in each partition.

One can compute the dimension of the moduli space from the quiver diagram.  
The quaternionic dimension of the Higgs branch of this theory is
\bea
\dim_{\BH} \text{Higgs}_{(2)-(4)-[6]} = (2 \times 4)+(4 \times 6) - 2^2-4^2 = 12~. \label{dimhiggs246}
\eea
On the other hand, the quaternionic dimension of the Coulomb branch of this theory is
\bea
\dim_{\BH} \text{Coulomb}_{(2)-(4)-[6]} = 2+4=6 ~. \label{dimcou246}
\eea

\subsection*{The mirror theory}
\begin{figure}[htbp]
\begin{center}
\includegraphics[height=2.5 in]{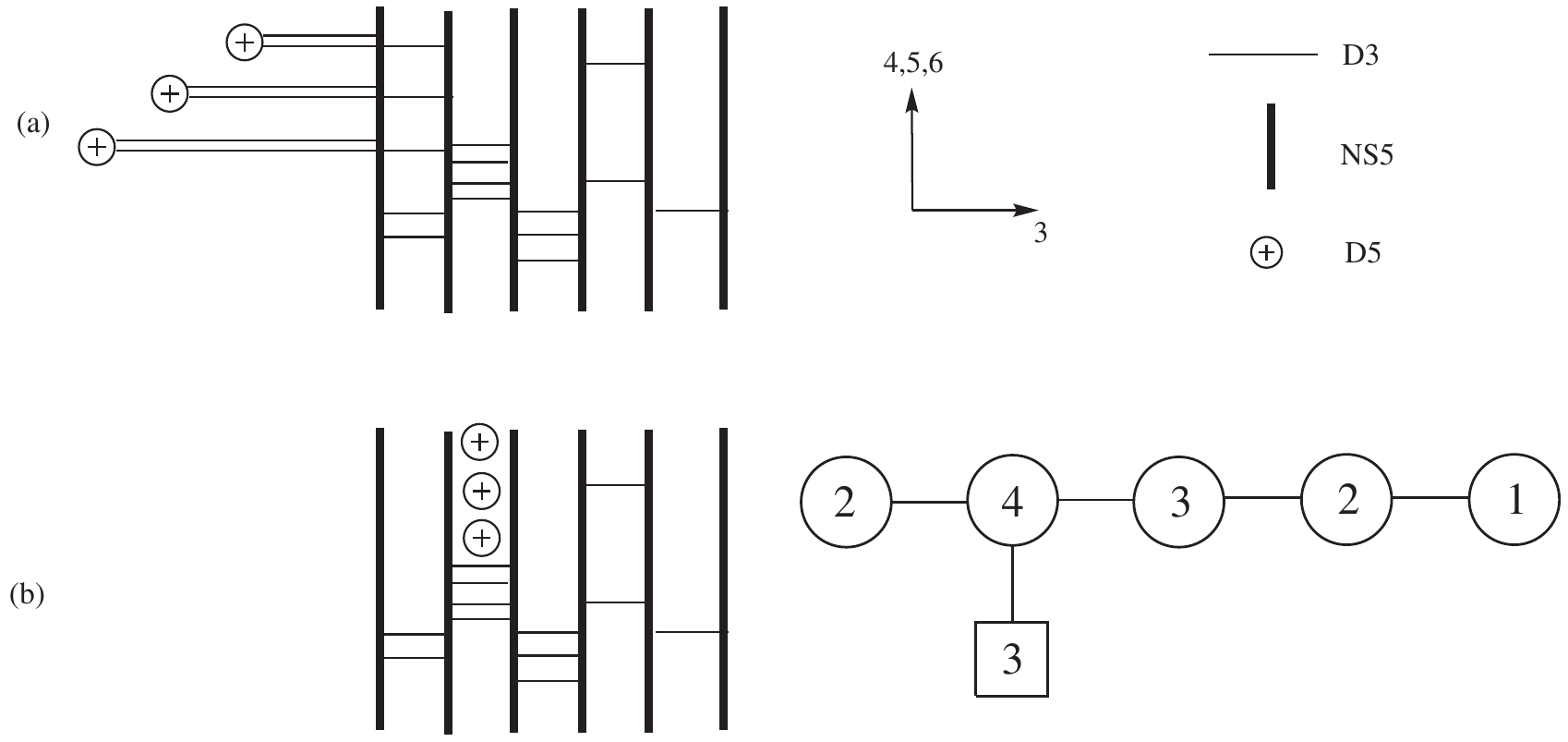}
\caption{{\it The mirror of the $(2)-(4)-[6]$ theory.} (a) From diagram (c) in \fref{fig:246quiv}, the NS5-branes and the D5-branes are exchanged and the directions $x^4, x^5, x^6$ are rotated into $x^7, x^8, x^9$ and {\it vice-versa}. The partitions $\sigma = (2,2,2)$ and $\rho = (1,1,1,1,1,1)$ are in one-to-one correspondence with this diagram. (b) The D5-branes are moved across the NS5-branes. The D3-brane creation and annihilation are according to \cite{Hanany:1996ie}. The corresponding quiver diagram is also given next to the brane configuration.}
\label{fig:mirror246}
\end{center}
\end{figure}

Now let us consider the mirror of the $(2)-(4)-[6]$ theory.  The brane configuration of the mirror theory can be obtained as described in \cite{Hanany:1996ie} and is depicted in \fref{fig:mirror246}.  From diagram (c), it can be seen that 
\bea
\sigma = (2,2,2)~, \qquad \qquad \rho  = (1,1,1,1,1,1)~,
\eea
\ie~the partitions $\sigma$ and $\rho$ from \eref{par246quiv} get exchanged.

One can compute the dimension of the moduli space from the quiver diagram.  
The quaternionic dimension of the Higgs branch of the mirror theory is
\bea
\dim_{\BH} \text{Higgs}_{\text{mirror}} = 8+(2 \times 12)+6+2 - 2^2-4^2-3^2-2^2 -1^2= 6~. \label{dimHiggsmir246}
\eea
The quaternionic dimension of the Coulomb branch of the mirror theory is
\bea
\dim_{\BH} \text{Coulomb}_{\text{mirror}} =1+2+3+4+2 = 12~. \label{dimcoumir246}
\eea
The results are in agreement with the exchange of the Coulomb and Higgs branches of the $(2)-(4)-[6]$ theory predicted by mirror symmetry.

\subsubsection{The Coulomb branch of the $(2)-(4)-[6]$ theory}
\begin{figure}[htbp]
\begin{center}
\includegraphics[height=1 in]{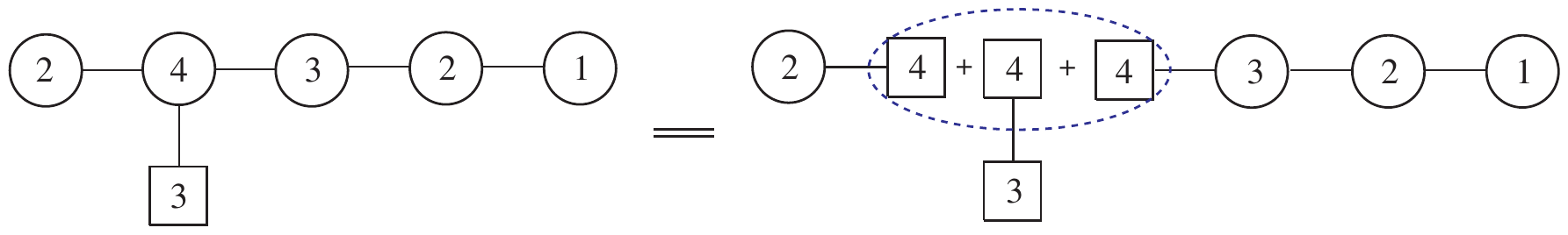}
\caption{The gluing process to obtain the mirror of the $(2)-(4)-[6]$ theory.}
\label{fig:gluemir246}
\end{center}
\end{figure}

By mirror symmetry, the Coulomb branch of the $(2)-(4)-[6]$ theory is identical to the Higgs branch of its mirror.  
The Hilbert series of the Higgs branch of the mirror theory can be obtained by gluing process \cite{Benvenuti:2010pq,Hanany:2010qu} schematically depicted in \fref{fig:gluemir246}.

Let $x_1, x_2$ be the $SU(3) \subset U(3)$ global fugacity.  The Hilbert series of the Higgs branch of the mirror of the $(2)-(4)-[6]$ theory (or, equivalently, the Coulomb branch of the $(2)-(4)-[6]$ theory) is given by
\bea
H^{C}_{(2)-(4)-[6]} (t, x_1, x_2) &=& (1-t^6)(1-t^{8})(1-t^{10})(1-t^{12}) \PE \left[ [1,1] t^2 + [1,1] t^4 \right] ~. \qquad \label{HSCou246}
\eea
%Note that the representation $[2]_x +[0]_x$ of $SU(2)$ is in fact the adjoint representation of the $U(2)$ global symmetry, and the representation  $\left( \frac{q_1}{q_2} + \frac{q_2}{q_1} \right) [1]_x$ is the bi-fundamental representation of $U(1) \times U(2)$ global symmetry.

The Hilbert series indicates that the Coulomb branch of the $(2)-(4)-[6]$ theory is indeed a {\it complete intersection}. There are $8$ generators at order $t^2$ and $8$ generators at order $t^4$.  These generators are subject to one relation at each order of $t^6$,$t^{8}$, $t^{10}$ and $t^{12}$.  These altogether give $8+8-4=12$ complex dimensional space or, equivalently, 6 quaternionic dimensional space -- in agreement with \eref{dimcou246} and \eref{dimHiggsmir246}.

\begin{figure}[htbp]
\begin{center}
\includegraphics[height=1.4 in]{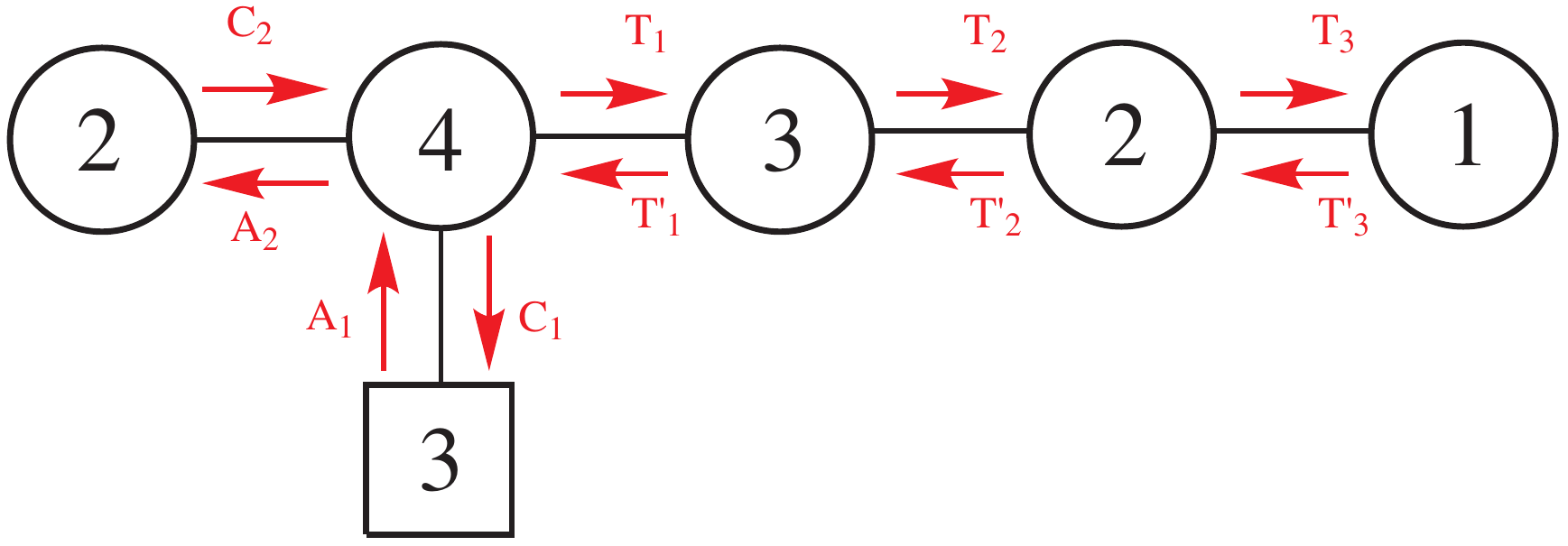}
\caption{The quiver diagram of the mirror of the $(2)-(4)-[6]$ theory, with the bi-fundamental chiral multiplets labelled. Note that $A$'s denote the chiral fields in the anti-clockwise direction, $C$'s denote the chiral fields in the clockwise direction, and $T,~T'$ denote the chiral fields in the tail with reducing ranks.}
\label{fig:mir246field}
\end{center}
\end{figure}

\paragraph{The $F$-terms.}  From \fref{fig:mir246field}, we see that the $F$-term constraints for the bi-fundamental chiral fields are
\bea
0 &=& - (C_2)^\alpha_{~a} (A_2)^a_{~\beta}~, \nn \\
0 &=& (C_1)^a_{~i} (A_1)^i_{~b} + (A_2)^a_{~\alpha} (C_2)^\alpha_{b} + (T_1)^a_{~a_3} (T'_1)^{a_3}_{~b}~, \nn \\
0 &=& -  (T'_1)^{a_3}_{~a} (T_1)^a_{~b_3} + (T_2)^{a_3}_{~a_2} (T'_2)^{a_2}_{~b_3}~, \nn \\
0 &=& - (T'_2)^{a_2}_{~a_3} (T_2)^{a_3}_{~b_2} + (T_3)^{a_2} (T'_3)_{b_2}~, \nn \\
0 &=& - (T'_3)_{a_2} (T_3)^{a_2}~. \label{FT246}
\eea
where the indices $a, b =1,2,3,4$ are the gauge indices corresponding to the gauge group $U(4)$, the indices $\alpha, \beta =1,2$ are the gauge indices corresponding to the leftmost $U(2)$ gauge group, the indices $i, j =1,2,3$ are the global indices corresponding the global symmetry $U(3)$, and the $a_2, a_3$ are the gauge indices corresponding to the gauge groups $U(2)$ and $U(3)$ in the tail.

Using the lemma discussed in Footnote \ref{fn:lemma}, we find that the matrices 
\bea
\begin{array}{ll}
(T'_2)^{a_2}_{~a_3} (T_2)^{a_3}_{~b_2} = (T_3)^{a_2} (T'_3)_{b_2}~, \qquad & (T'_1)^{a_3}_{~a} (T_1)^a_{~b_3} = (T_2)^{a_3}_{~a_2} (T'_2)^{a_2}_{~b_3}, \nn \\
 (T_1)^a_{~a_3} (T'_1)^{a_3}_{~b}~, \qquad & (A_2)^a_{~\alpha} (C_2)^\alpha_{b} \nn
\end{array}
\eea
are nilpotent.  By choosing appropriate bases in quiver gauge groups, one can transform these matrices into their Jordan normal form as follows.

Using a $U(2)$ gauge transformation, we can put the $2 \times 2$ matrix $(T'_2)^{a_2}_{~a_3} (T_2)^{a_3}_{~b_2} = (T_3)^{a_2} (T'_3)_{b_2}$ into a Jordan normal form:
\bea
\begin{pmatrix} 0 & 1 \\ 0 &0  \end{pmatrix}~.
\eea

Similarly, using a $U(3)$ gauge transformation, we can put the $3 \times 3$ matrix $(T'_1)^{a_3}_{~a} (T_1)^a_{~b_3} = (T_2)^{a_3}_{~a_2} (T'_2)^{a_2}_{~b_3}$ into a Jordan normal form:
\bea
\begin{pmatrix} 0 & 1 & 0 \\ 0 &0 & 1 \\ 0 & 0 & 0  \end{pmatrix}~.
\eea

Using a $U(4)$ gauge transformation, we can put either $(T_1)^a_{~a_3} (T'_1)^{a_3}_{~b}$ or $(A_2)^a_{~\alpha} (C_2)^\alpha_{b}$ into a Jordan normal form.  For the matrix $(T_1)^a_{~a_3} (T'_1)^{a_3}_{~b}$, we have
\bea
\begin{pmatrix} 0 & 1 & 0 & 0 \\ 0 &0 & 1 &0 \\ 0 & 0 & 0 & 1 \\ 0 & 0 & 0 & 0 \end{pmatrix}~. \label{JNFP3}
\eea
On the other hand, the Jordan normal form for the matrix $(A_2)^a_{~\alpha} (C_2)^\alpha_{b}$ contains two $2 \times 2$ blocks as follows:
\bea
\begin{pmatrix} 0 & 1 & 0 & 0 \\ 0 &0 & 0 &0 \\ 0 & 0 & 0 & 1 \\ 0 & 0 & 0 & 0 \end{pmatrix}~. \label{JNFP2}
\eea

It is convenient to use Jordan normal forms of these matrices to compute and verify the relations between generators below.

\subsubsection*{Generators and relations of the Coulomb branch of the $(2)-(4)-[6]$ theory}  
The generators and relations can be derived in a similar way to those of the $(2)-[4]$ theory discussed in the previous section.

\paragraph{Order $t^2$.} The generators at order $t^2$ can be written as
\bea
(G_2)^i_{~j} =  (A_1)^i_{~a} (C_1)^a_{~j}~.
\eea
It follows from the $F$-term constraints that 
\bea
\tr G_2 = (G_2)^i_{~i} =0~.
\eea
Thus, $G_2$ transforms in the adjoint representation of the global symmetry $SU(3)$.
Note that this is in agreement with \eref{rel:trG2}.

\paragraph{Order $t^4$.}  The generators at order $t^4$ can be written as
\bea
(G_4)^i_{~j} =   (A_1)^i_{~a} (A_2)^a_{~\alpha} (C_2)^\alpha_{~b} (C_1)^b_{~j}~.
\eea

Note that the operator
\bea
(\widetilde{G_4})^i_{~j} = (A_1)^i_{~a} (T_1)^a_{~a_3} (T'_1)^{a_3}_{~b} (C_1)^b_{~j}
\eea
is related to $(G_4)^i_{~j}$ by the formula
\bea
G_4+ \widetilde{G_4} = - G_2^2~.
\eea
Using the first relation in \eref{FT246} and the fact that $\tr (T_1 \cdot T_1')^2 = 0$ (as $T_1 \cdot T_1'$ is nilpotent), we can derive, in a similar way to \eref{GeqGtil24}, that
\bea
\tr G_4 &=&  (A_2)^a_{~\alpha} (C_2)^{\alpha}_{~b} (C_1)^b_{~i} (A_1)^i_{~a} \nn \\
&=& -(A_2)^a_{~\alpha} (C_2)^{\alpha}_{~b} \left[ (A_2)^b_{~\beta} (C_2)^\beta_{~a} + (T_1)^b_{~a_3} (T'_1)^{a_3}_{~a}\right] \nn \\
&=& - (A_2)^a_{~\alpha} (C_2)^{\alpha}_{~b}(T_1)^b_{~a_3} (T'_1)^{a_3}_{~a} \nn \\
&=&  \left[(C_1)^a_{~i} (A_1)^i_{~b} + (T_1)^a_{~b_3} (T'_1)^{b_3}_{~b}\right] (T_1)^b_{~a_3} (T'_1)^{a_3}_{~a} \nn \\
&=& (C_1)^a_{~i} (A_1)^i_{~b} (T_1)^b_{~a_3} (T'_1)^{a_3}_{~a} \nn \\
&=& \tr \widetilde{G_4}~.
\eea
Thus, we have
\bea
\tr G_2^2 + 2 \tr G_4 = 0~. \label{trm22trG246}
\eea
Therefore, the trace of the generator $G_4$ is fixed by that of $G_2^2$.
Note that this relation is in agreement with \eref{rel:trG4}.

\paragraph{Order $t^6$.}  The relation at order 6 can be written as
\bea
\tr G_2^3 + 3 \tr(G_2 \cdot G_4) = 0~. \label{trm3246}
\eea
Observe that this relation coincides with \eref{rel:trG6} after setting $G_6=0$ (since there is no generator $G_6$ in this theory).

\paragraph{Order $t^{8}$.} The relation at order 8 can be written as
\bea
4 \tr (G_2^2 \cdot G_4) + 2 \tr (G_4^2) + \tr(G_2^4) = 0~.  \label{trm4trg2}
\eea
Observe that this relation coincides with \eref{rel:trG8} after setting $G_6=G_8=0$ (since there is no generator $G_6$ and $G_8$ in this theory).

Other relations which are not independent from above can be computed.  For example,
\bea
\tr (G_2^4) - 2 \left( \tr G_4 \right)^2 = 0~. \label{trm4trg2a}
\eea

\paragraph{Order $t^{10}$.}  The relation at order 10 can be written as
\bea
\tr (G_2^5) +5 \tr (G_2^3 \cdot G_4) + 5 \tr (G_2 \cdot G_4^2) = 0~.  \label{trm5246}
\eea
Observe that this relation is in agreement with \eref{rel:trG10} when $G_6=G_8=G_{10}=0$ (this is because the generators $G_6$, $G_8$ and $G_{10}$ do not exist in this theory).

Other relations which are not independent from the one above can be computed.  For example,
\bea
\tr(G_2^3 \cdot G_4) + 2 \tr(G_2 \cdot G_4^2) = 0~. \label{trm5246a}
\eea

\paragraph{Order $t^{12}$.}  The relation at order 12 can be written as
\bea
\det G_4 &=& 0~. \label{detg0246}
\eea

Other relations which are not independent from the one above can be computed.  For example,
\bea
0 &=& 6 \tr (G_2^2\cdot G_4^2) +6 \tr (G_2^4\cdot G_4)+ 3 \tr \left[ (G_2\cdot G_4)^2 \right] +3 \tr (G_6^2) \nn \\
&& +2 \tr (G_4^3)+ \tr (G_2^6)~. \label{order12246} 
\eea
Observe that this relation is in agreement with \eref{rel:trG12} when we set $G_6=G_8=G_{10}=G_{12}=0$ (this is because the generators $G_6$, $G_8$, $G_{10}$ and $G_{12}$ do not exist in this theory).

We derive the relations \eref{trm3246}--\eref{detg0246} in Appendix \ref{app:rel246}.  Relation \eref{order12246} can be derived in a similar way to the others.

\subsection{General case: $(k)-(2k)- \cdots -(nk-k)-[nk]$}
Let us now consider the $(k)-(2k)- \cdots -(nk-k)-[nk]$ theory for general $k \geq 1$.
This theory can be identified with $T^\sigma_{~\rho}(SU(nk))$, where
\bea
\sigma =( \underbrace{1, \ldots,1}_{nk~\text{one's}})~, \qquad \qquad \rho = (\underbrace{k, \ldots, k}_{\text{$n$ $k$'s}})~. \label{park2knkquiv}
\eea
Note that the total number of boxes is $n k$.  Let us compute the dimension of the moduli space.  The quaternionic dimension of the Higgs branch of this theory is
\bea
\dim_{\BH} \text{Higgs}_{(k)-(2k)- \cdots -(nk-k)-[nk]} &=& k^2 \sum_{m=1}^{n-1} m(m+1) -k^2 \sum_{m=1}^{n-1} m^2 \nn \\
&=& \frac{1}{2} n (n-1) k^2~.
\eea
The quaternionic dimension of the Coulomb branch of this theory is
\bea
\dim_{\BH} \text{Higgs}_{(k)-(2k)- \cdots -(nk-k)-[nk]} &=& k\sum_{m=1}^{n-1} m = \frac{1}{2} n (n-1) k~.
\eea
The Coulomb branch of the $(k)-(2k)- \cdots -(nk-k)-[nk]$ theory can also be identified with the moduli space of $SU(n)$ magnetic monopoles in the presence of $nk$ fixed monopoles.  Among these $SU(n)$ monopoles, $m k$ of them (with $1 \leq m \leq n-1$) carry magnetic charge $\alpha_m$, where
\bea
\alpha_1 = (1,-1,0, \ldots,0), \quad \alpha_2 = (0,1,-1,0,\ldots,0), \quad \ldots, \quad \alpha_{n-1} = (0,\ldots,0,1,-1)~. \nn
\eea

%In particular, this is the moduli space of $k$ $SU(n+1)$ monopoles with magnetic charge $(1,-1,0, \ldots, 0)$, $2k, in the presence of $2k$ fixed monopoles with magnetic charge $(0,0,1)$.

\subsubsection*{The mirror theory}

\begin{figure}[htbp]
\begin{center}
\includegraphics[height=1.3 in]{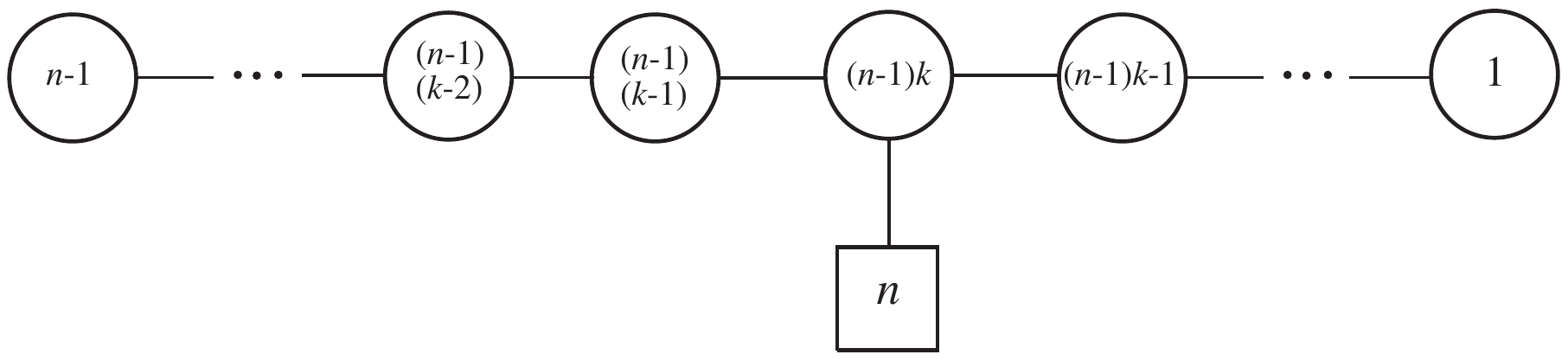}
\caption{The mirror of the $(k)-(2k)- \cdots -(nk-k)-[nk]$ theory}
\label{fig:mirk2nnk}
\end{center}
\end{figure}

The mirror of the $(k)-(2k)- \cdots -(nk-k)-[nk]$ theory is depicted in \fref{fig:mirk2nnk}.  This theory can be identified with $T^\rho_{~\sigma}(SU(n))$, where the partitions $\sigma$ and $\rho$ are defined as in \eref{park2knkquiv}.

One can compute the dimension of the moduli space from the quiver diagram.  The quaternionic dimension of the Higgs branch of this theory is
\bea
\dim_{\BH} \text{Higgs}_{\text{mirror}} &=& n(n-1)k + (n-1)^2 \sum_{i=1}^{k-1} (k-i+1)(k-i)  \nn \\
&& +\sum_{i=1}^{(n-1)k-1}  \left[ (n-1)k -i+1\right]  \left[ (n-1)k -i \right] \nn \\
&& - \sum_{i=1}^{(n-1)k} \left[(n-1)k-i+1\right]^2 - \sum_{i=1}^{(n-1)k} \left[(n-1)^2(k-i)\right]^2 \nn\\
&=& \frac{1}{2} n(n-1) k~. \label{dimhiggsmirk2knk}
\eea
 The quaternionic dimension of the Coulomb branch this theory is
 \bea
\dim_{\BH} \text{Coulomb}_{\text{mirror}}  &=&  \sum_{i=1}^{(n-1)k} \left[(n-1)k-i+1\right] +\sum_{i=1}^{(n-1)k} \left[(n-1)^2(k-i)\right]  \nn \\ 
&=&  \frac{1}{2} n(n-1) k^2~. \label{dimcoumirk2knk}
\eea
The results are in agreement with the exchange of the Coulomb and Higgs branches of the theory and its mirror predicted by mirror symmetry.

\subsubsection{The Coulomb branch of the $(k)-(2k)- \cdots -(nk-k)-[nk]$ theory}
In this section, we compute the Hilbert series of the Coulomb branch of the $(k)-(2k)- \cdots -(nk-k)-[nk]$ theory.  We make use of mirror symmetry and compute this from the Higgs branch of the mirror theory.

Let $x$ be a fugacity for the $SU(2) \subset U(2)$ global symmetry in the quiver diagram \fref{fig:mirk2k}.  The Hilbert series of the Higgs branch of the mirror of the $(k)-(2k)- \cdots -(nk-k)-[nk]$ theory (or, equivalently, the Coulomb branch of the $(k)-(2k)- \cdots -(nk-k)-[nk]$ theory) is given by
\bea
H^{C}_{(k)-(2k)- \cdots -(nk-k)-[nk]} (t, x_1, \ldots, x_{n-1}) = \PE \left[ [1,0, 
\ldots,0,1]_{SU(n)} \sum_{p=1}^{k} t^{2p} \right] \prod_{q=k+1}^{nk} (1- t^{2q})~. \nn \\
\eea
The Hilbert series indicates that the Coulomb branch of the $(k)-(2k)- \cdots -(nk-k)-[nk]$ theory is indeed a {\it complete intersection}.  There are $n^2-1$ generators at each of the follwing order: $t^2, t^4, \ldots, t^{2k}$, and one generator at each of the following order: $t^{2(k+1)}, \ldots, t^{2nk}$.  These altogether give $(n^2-1)k-(nk-k) =  n^2k-nk=n(n-1)k$ complex dimensional space -- in agreement with \eref{dimhiggsmirk2knk}.

\subsubsection*{Generators and relations of the Coulomb branch of the $(k)-(2k)- \cdots -(nk-k)-[nk]$ theory}  

\begin{figure}[htbp]
\begin{center}
\includegraphics[height=1.3 in]{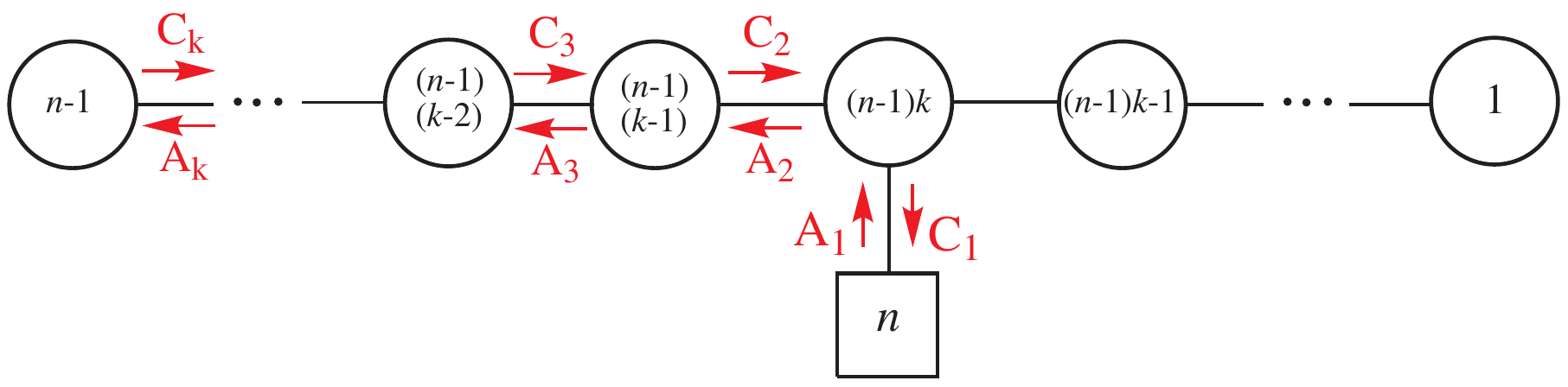}
\caption{The quiver diagram of the mirror of the $(k)-(2k)- \cdots -(nk-k)-[nk]$ theory, with the bi-fundamental chiral multiplets labelled. Note that $A$'s denote the chiral fields in the anti-clockwise direction, $C$'s denote the chiral fields in the clockwise direction.}
\label{fig:mirk2knkfield}
\end{center}
\end{figure}

\paragraph{The generators.} 
Let us fix $k$. The generators at order $t^{2p}$ (with $1 \leq p \leq k$) can be written as
\bea
G_{2p} = \prod_{r=1}^p A_r  \prod_{s=1}^p C_s~.
\eea
For $p=1$, the operator $G_2$ is traceless:
\bea
\tr G_2 = 0~.
\eea
It should be noted that the relation $\tr G_2^{2m+1} =0$ for $m\geq 1$ does not hold for the theories with $n \geq 3$, since $G_2$ is not a $2 \times 2$ matrix even though $\tr G_2 =0$.

For $1 < p \leq k$, the trace of $G_{2p}$ is fixed by a relation consisting of operators of lower orders.  For example, given a $k$ such that $1 < k \leq 6$, we find that the traces $\tr G_{2p}$ satisfy the relations \eref{rel:trG2}--\eref{rel:trG12}. In general, the formula for $\tr G_{2p}$ (with $1 < p \leq k$) is conjectured to be \eref{relk2keven}.

\paragraph{The relations.}  As can be seen from the Hilbert series, there is one relation at each of order $t^{2q}$ with $q = k+1, \ldots,nk$.  One can obtain these relations from \eref{relk2keven} by setting $G_{2p}$ with $ p \geq k+1$ to zero (since there is no generator $G_{2p}$ with $ p \geq k+1$).  Such a relation at order $t^{2q}$ (with $k+1 \leq q \leq nk$) are therefore given by \eref{releven}.
%%%%%%%%%%%%%%%%%%%%%%%%%%%%%%%%
\section*{Acknowledgements}
We would like to thank Yuji Tachikawa for useful discussions, James Gray for his advice in making use of {\tt STRINGVACUA} package \cite{StringvacuaBlock, Gray:2009fy} in checking various relations in this paper, and Sergio Benvenuti for a closely related collaboration and several discussions.

N.~M. is grateful to the following institutes and collaborators for their very kind hospitality during the completion of this paper:  Thomas Grimm, Oliver Schlotterer, Seok Kim, Aroonroj Mekareeya, Institute for the Physics and Mathematics of the Universe (IPMU), Yukawa Institute of Theoretical Physics (YITP), Seoul National University, and Korea Institute for Advanced Study (KIAS), and Imperial College London.  This research is supported by a research grant of the Max Planck Society, World Premier International Research Center Initiative (WPI Inititative), MEXT, Japan, and the DPST Project of the Royal Thai Government.

%%%%%%%%%%%%%%%%%%%%%%%%%%%%%%%%
\appendix
\section{Proof of Hilbert series \eref{hs123n}}  \label{app:HS123n}
This can be proven by induction.  For $n=2$, formula \eref{hs123n} becomes
\bea
H_{(1)-[2]} (t, x_1)= (1-t^4 )\PE \left[ [2]_{x_1} t^2 \right]~;
\eea
this is the Hilbert series of $\BC^2/\BZ_2$, or equivalently the moduli space of 1 $SU(2)$ instanton on $\BR^4$ 
(neglecting the translational degrees of freedom) -- this is in agreement with the result in \cite{Benvenuti:2010pq}.
Now suppose that formula \eref{hs123n} holds for $n=k$.  We shall derive the expression for $n=k+1$ using the gluing technique.
Note that the Hilbert series for the $[k]-[k+1]$ theory is
\bea
&& H_{[k]-[k+1]} (t, q_x, x_1, \ldots, x_{k-1}, q_y,y_{1}, \ldots, y_{k}) \nn \\
&=& \PE \left[ [1,0,\ldots,0]_{\{x_i\}} [0,\ldots,0,1]_{\{y_l\}} \frac{q_x}{q_y} t + [0,\ldots,0,1]_{\{x_i\}} [1,0,\ldots,0]_{\{y_l\}} \frac{q_y}{q_x} t \right]~,
\eea 
where $q_x$ is the fugacity of $U(1) \subset U(k)$, $x_1, \ldots, x_{k-1}$ are the fugacities of $SU(k) \subset U(k)$, $q_y$ is the fugacity of $U(1) \subset U(k+1)$, and $y_1, \ldots, y_{k}$ are the fugacities of $SU(k+1) \subset U(k+1)$.
The gluing factor is given by
\bea
H_{\text{glue}} (t, x_1, \ldots, x_{k-1}) &=& \frac{1}{\PE \left[ (1+[1,0,\ldots,0,1]_{\{x_i\}}) t^2\right]} \nn \\
&=& \frac{1-t^2}{\PE \left[ [1,0,\ldots,0,1]_{\{x_i\}} t^2\right]}~.
\eea

The Hilbert series for $(1)-(2)- \cdots -(k)-[k+1]$ is therefore
{\small
\bea
&& H_{(1)-(2)- \cdots -(k)-[k+1]} (t, y_1, \ldots,y_{k}) \nn \\
&& =  \int \ud \mu_{U(k)} (q_x,x_1, \ldots,x_{k-1})  \times \nn \\
&&  \quad H_{(1)-(2)- \cdots -[k]} (t, x_1, \ldots, x_{k-1}) H_{\text{glue}} (t, x_1, \ldots, x_{k-1}) H_{[k]-[k+1]} (t, q_x, x_1, \ldots, x_{k-1}, q_y, y_{1}, \ldots, y_{k}) \nn \\
&& = \prod_{q=1}^k (1-t^{2q})   \int \ud \mu_{U(k)} (q_x,x_1, \ldots,x_{k-1})  H_{[k]-[k+1]} (t, q_x,x_1, \ldots, x_{k-1}, q_y, y_{1}, \ldots, y_{k}) \nn \\
&& = \left(1-t^{2(k+1)} \right) \PE \left[ (1+[1,0, \ldots, 0,1]_{SU(k+1)} )t^2 \right] \times \prod_{q=1}^k (1-t^{2q}) \nn \\
&& =  \PE \left[ [1,0, \ldots, 0,1]_{SU(k+1)} t^2 \right] \prod_{q=2}^{k+1} (1-t^{2q}) ~,
\eea
}
as required.  Note that in the third equality we have used the fact that
\bea
&& \int \ud \mu_{U(k)} (q_x,x_1, \ldots,x_{k-1})  H_{[k]-[k+1]} (t, q_x,x_1, \ldots, x_{k-1}, q_y, y_{1}, \ldots, y_{k}) \nn \\
&& = \left(1-t^{2(k+1)} \right) \PE \left[ (1+[1,0, \ldots, 0,1]_{SU(k+1)} )t^2 \right]~.
\eea
This formula can be seen by consider the free theory $[k]-[k+1]$ and count all operators which are invariant under $U(k)$.
Let $q^i_{~a}$ and $\tilde{q}_i^{~j}$ be the bi-fundamental chiral multiplets in the free theory $[k]-[k+1]$, where $a= 1, \ldots, k$ and $i =1, \ldots,k+1$.
It is clear that the generators of the $U(k)$ invariant operators are $M^i_{~j} :=  q^i_{~a} \tilde{q}^a_{~j}$.  
Observe that $M$ transforms under the adjoint representation of $U(k)$, or equivalently $1+[1,0, \ldots, 0,1]$ of $SU(k+1)$.
The operator $M$ satisfies the relation
\bea
\det M &=& \frac{1}{(k+1)!} \epsilon^{i_1 i_2 \ldots i_{k+1}}  \epsilon_{i'_1 i'_2 \ldots i'_{k+1}} M^{i'_1}_{~i_1} \ldots M^{i'_{k+1}}_{~i_{k+1}}~ \nn \\
&=& \frac{1}{(k+1)!} \epsilon^{i_1 \ldots i_{k+1}}  \epsilon_{i'_1 \ldots i'_{k+1}} q^{i'_1}_{~a_1} \ldots q^{i'_{k+1}}_{~a_{k+1}} \tilde{q}^{a_1}_{~i_1} \ldots \tilde{q}^{a_{k+1}}_{~i_{k+1}} \nn \\
&=& 0~,
\eea
where we have used the fact that the epsilon symbol $ \epsilon_{i'_1 i'_2 \ldots i'_{k+1}}$ induces the anti-symmetrisation over the indices $a_1, \ldots, a_{k+1}$ and yields zero since $a_1, \ldots, a_{k+1}$ run over $1, \ldots, k$, \ie~ 
\bea \epsilon_{i'_1 \ldots i'_{k+1}} q^{i'_1}_{~a_1} \ldots q^{i'_{k+1}}_{~a_{k+1}} = 0~. \nn 
\eea

\section{Derivation of the relations} \label{app:derivations}
In this appendix, we derive the relations discussed in the main text.  

\subsection{The relations for the mirror of the $(1)-(2)-[4]$ theory} \label{app:rel124}
Let us focus on the mirror of the $(1)-(2)-[4]$ theory.  Before we proceed, let us define the following shorthand notation:
\bea
(\CP_1)^\alpha_{~\beta} := (A_2)^\alpha_{~a} (C_2)^a_{~\beta} ~, \quad (\CP_2)^\alpha_{~\beta} :=  (C_3)^\alpha (A_3)_\beta~, \quad (\CP_3)^\alpha_{~\beta} := T^\alpha T'_\beta~.
\eea
Then, from the second $F$-terms in \eref{Fterms124}, 
\bea
\CP_1= \CP_2 + \CP_3~.
\eea
Below we write various gauge invariant operators in terms of $\CP_2$ and $\CP_3$.  
Using matrix properties of $\CP_2$ and $\CP_3$, the relations can be derived in a straightforward way.

\paragraph{Traces.} It is easy to check that $\tr (\CP_2^m) = (\tr \CP_2)^m$ for all $m \geq 1$.  Note also that $\CP_3$ is nilpotent due to the third $F$-term in \eref{Fterms124} and the lemma in Footenote \ref{fn:lemma}. Therefore, we have $\CP_3^k=0$ for all $k \geq 2$ and $\tr (\CP_3^m) =0$ for all $m\geq 1$.  Using \eref{Fterms124}, we obtain
\bea \label{relset1}
\begin{array}{llll}
\tr M &= - \tr (\CP_1) &= -\tr (\CP_2+ \CP_3) &= -\tr(\CP_2)~,  \\
\tr (M^2) &=  \tr(\CP_1^2) &= \tr[ (\CP_2+\CP_3)^2] &= \tr(\CP_2^2) + 2 \tr(\CP_2 \CP_3)~, \\
\tr(M^3) &= - \tr (\CP_1^3) &=  - \tr[ (\CP_2+\CP_3)^3] &= -\tr(\CP_2^3) - 3 \tr(\CP_3 \CP_2^2)~, \\
\tr(M^4) &= \tr (\CP_1^4)  &=  \tr[ (\CP_2+\CP_3)^4]  &=\tr(\CP_2^4) + 4 \tr(\CP_3 \CP_2^3) + 2 \tr[ (\CP_2 \CP_3)^2]~.
\end{array}
\eea
Therefore, we see that
\bea \label{relset1a}
\begin{array}{lll}
(\tr M)(\tr M^2) &= -\tr(\CP_2) \left[ \tr(\CP_2^2) + 2 \tr(\CP_2 \CP_3) \right]  &= -\tr(\CP_2^3) - 2 \tr(\CP_3 \CP_2^2) \\
(\tr M)(\tr M^3) &= \tr(\CP_2) \left[\tr(\CP_2^3)+3 \tr(\CP_3 \CP_2^2)\right] &= \tr(\CP_2^4) +3 \tr(\CP_3 \CP_2^3)~.
\end{array}
\eea

\paragraph{Determinants.} Furthermore, applying the Cayley-Hamilton theorem to $M$, we see that
\bea \label{reldet}
\det M &=& \frac{1}{2} \left[ (\tr M)^2- (\tr M^2) \right]  \nn\\
&=& \frac{1}{2} \left[ (\tr \CP_2)^2- \tr(\CP_2^2) - 2 \tr(\CP_2 \CP_3)\right] 
= - \tr(\CP_2 \CP_3)~.
\eea
Thus, we obtain
{\small
\bea \label{relset2}
\begin{array}{lll}
(\tr M)(\det M) &= (-\tr \CP_2)[- \tr (\CP_2 \CP_3) ] &= \tr (\CP_3 \CP_2^2)~,  \\
(\det M)^2 &= [- \tr(\CP_2 \CP_3)]^2 &= \tr[ (\CP_2 \CP_3)^2]~, \\
(\tr M^2)(\det M) &=  \left[ \tr(\CP_2^2) + 2 \tr(\CP_2 \CP_3) \right][- \tr (\CP_2 \CP_3) ] &= -\tr (\CP_3 \CP_2^3) - 2 \tr[ (\CP_2 \CP_3)^2]~.
\end{array}
\eea}

\paragraph{Products of $L$, $M$ and $R$.} We also have
{\small
\bea
\begin{array}{llll}
L_i R^i &= -\tr \left( \CP_2 \CP_1^2 \right) &= -\tr \left[ \CP_2 (\CP_2+\CP_3) ^2 \right] &= -\tr(\CP_2^3) - 2 \tr(\CP_3 \CP_2^2)~,  \\
 L_i M^i_{~j} R^j  &= \tr ( \CP_2 \CP_1^3) &= \tr [ \CP_2 (\CP_2+\CP_3)^3] &= \tr(\CP_2^4) + 3 \tr(\CP_3 \CP_2^3) +  \tr[ (\CP_2 \CP_3)^2]~.
\end{array}
\eea}

\paragraph{Relations.} Using \eref{relset1} and \eref{relset2}, we arrive at relation \eref{rel6124} at order $t^6$, \ie
\bea
L_i R^i &=& (\tr M)(\tr M^2)  \\
&=& \tr (M^3) + (\tr M)(\det M)  \\
&=&  (\tr M)^3 -  2(\tr M)(\det M)~.
\eea
as required.  Similarly, we have relation \eref{rel8124} at order $t^8$, \ie
\bea
L_i M^i_{~j} R^j &=& (\tr M)(\tr M^3)+(\det M)^2  \\
&=& \tr (M^4) + \tr(M^2) (\det M)+(\det M)^2~. 
\eea

\subsection{The relations for the mirror of the $(1)-(2)-[5]$ theory} \label{app:rel125}
Let us define
\bea
(\CP_1)^\mu_{~\nu} := (A_3)^\mu_{~\alpha} (C_3)^\alpha_{~\nu} ~, \quad (\CP_2)^\mu_{~\nu} :=   (C_4)^\mu (A_4)_\nu~, \quad (\CP_3)^\mu_{~\nu} := T^\mu T'_\nu~.
\eea
Then, from the second $F$-terms in \eref{Fterms125}, 
\bea
\CP_1= \CP_2 + \CP_3~.
\eea

\paragraph{Traces and determinants.} It is easy to check that \eref{relset1}, \eref{relset1a}, \eref{reldet} and \eref{relset2} still hold. 
Therefore,
\bea
\tr M^4 &=& \tr(\CP_2^4)+4 \tr( \CP_3 \CP_2^3)+2\tr[ (\CP_2 \CP_3)^2]~, \nn \\
(\tr M)(\tr M^3) &=& \tr(\CP_2^4)+3 \tr( \CP_3 \CP_2^2)~, \nn \\
(\det M)^2 &=& \tr[ (\CP_2 \CP_3)^2]~.
\eea 
We also have
\bea
\tr M^4 &=& \tr(\CP_2^4)+4 \tr( \CP_3 \CP_2^3)+2\tr[ (\CP_2 \CP_3)^2]~, \nn \\
\tr M^5 &=& -\tr(\CP_2^5)-5 \tr( \CP_3 \CP_2^4)-5\tr[ \CP_2 (\CP_2 \CP_3)^2]~. \label{highercasimir125}
\eea

\paragraph{Products of $L$, $M$ and $R$.} In this theory, we have
\bea
\begin{array}{llll}
L_i R^i &= -\tr \left( \CP_2 \CP_1^3 \right)&= -\tr(\CP_2^4) - 3 \tr(\CP_3 \CP_2^3) - \tr[(\CP_2 \CP_3)^2]~,  \\
 L_i M^i_{~j} R^j  &= \tr ( \CP_2 \CP_1^4) &= \tr(\CP_2^5) + 4 \tr(\CP_3 \CP_2^4) +  3\tr[ \CP_2 (\CP_2 \CP_3)^2]~.
\end{array}
\eea

\paragraph{Relations.} Therefore, the relation at order $t^8$ can be written as
\bea
L_i R^i &=& - \left[ (\tr M)(\tr M^3) + (\det M)^2 \right]~.
%&=&  - \left[ \tr (M^4) + \tr (M^2) (\det M) +(\det M)^2 \right]~.
\eea
The relation at order $t^{10}$ can be written as
\bea
 L_i M^i_{~j} R^j  &=& -(\tr M) \left[ (\tr M^4)+ (\det M)^2 \right]~.
\eea

Using  \eref{relset1}, \eref{relset1a}, \eref{reldet}, \eref{relset2} and \eref{highercasimir125}, we find that these relations can also be rewritten as, \eg,
\bea
L_i R^i &=& - \left[ \tr (M^4) + \tr (M^2) (\det M) +(\det M)^2 \right]~, \nn \\
L_i M^i_{~j} R^j &=& - \left[ \tr (M^5) + \tr (M^3) (\det M) + (\tr M) (\det M)^2 \right]~.
\eea

\subsection{The relations for the mirror of the $(1)-(2)-(3)-[5]$ theory} \label{app:rel1235}
Let us define the following shorthand notation:
\bea
(\CP_1)^\alpha_{~\beta} := (A_2)^\alpha_{~a} (C_2)^a_{~\beta} ~, \quad (\CP_2)^\alpha_{~\beta} :=  (C_3)^\alpha (A_3)_\beta~, \quad (\CP_3)^\alpha_{~\beta} := (T_1)^\alpha_{~\mu} (T'_1)^\mu_{~\beta}~.
\eea
Then, from the second $F$-terms in \eref{fterms1235}, 
\bea
\CP_1= \CP_2 + \CP_3~.
\eea
Note that $\tr (\CP_2^m) = (\tr \CP_2)^m$ for all $m \geq 1$.  Furthermore, the $3 \times 3$ matrix $\CP_3$ is nilpotent due to the third $F$-term in \eref{fterms1235} and the lemma in Footenote \ref{fn:lemma}. Therefore, we have $\CP_3^k=0$ for all $k \geq 3$ and $\tr (\CP_3^m) =0$ for all $m\geq 1$.

Below we write various gauge invariant operators in terms of $\CP_2$ and $\CP_3$. For example,
\bea
\tr M &=& - \tr (\CP_2)~, \nn \\
\tr M^2 &=& \tr (\CP_1^2) = \tr(\CP_2^2) +2 \tr (\CP_2 \CP_3)~.
\eea
Using properties of $\CP_2$ and $\CP_3$, the relations can be derived in a straightforward way.  

\paragraph{The relation at order $t^6$.}  Observe that
\bea
\tr (M^3) &=& -\tr (\CP_1^3) = - \tr (\CP_2^3) - 3 \tr(\CP_2 \CP_3^2) - 3 \tr(\CP_3 \CP_2^2)~, \\
L_i R^i &=& -\tr (\CP_2 \CP_1)^2 = - \tr (\CP_2^3) -  \tr(\CP_2 \CP_3^2) - 2 \tr(\CP_3 \CP_2^2)~, \\
(\tr M)(\tr M^2) &=& -\tr (\CP_2^3) - 2\tr( \CP_3 \CP_2^2)~.
\eea
For the determinant of the $3 \times 3$ matrix $M$, we use the following identity:
\bea
\det M = \frac{1}{6} \left[ (\tr M)^3 + 2 \tr (M^3) - 3 (\tr M) \tr(M^2) \right]~.
\eea
Substituting the above traces into this identity, we obtain
\bea
\det M  = - \tr (\CP_2 \CP_3^2)~.
\eea
Thus, it is immediate that
\bea
L_i R^i = (\tr M)(\tr M^2)+ \det M~.
\eea

\paragraph{The relation at order $t^8$.}  Observe that
{\small
\bea
\tr (M^4) &=& \tr( \CP_1^4) = \tr( \CP_2^4) + 4 \tr(\CP_2^2 \CP_3^2) + 4 \tr( \CP_3 \CP_2^3)+2  \tr[ (\CP_2 \CP_3)^2]~, \\
L_i M^i_{~j} R^j &=& \tr( \CP_2 \CP_1^3) = \tr( \CP_2^4) + 2 \tr(\CP_2^2 \CP_3^2) + 3 \tr(\CP_3 \CP_2^3) + \tr[ (\CP_2 \CP_3)^2]~, \qquad \\
(\tr M)^2 (\tr M^2) &=& \tr(\CP_2^4) + 2 \tr(\CP_3 \CP_2^3)~.
\eea}
Therefore, we see that
\bea
L_i M^i_{~j} R^j = \frac{1}{2} \left[ \tr (M^4) + (\tr M)^2 (\tr M^2) \right]~.
\eea

\paragraph{The relation at order $t^{10}$.} Observe that
\bea
&& \tr M^5 = - \tr(\CP_1^5) \nn \\
&& =- \tr(\CP_2^5)- 5 \tr(\CP_3\CP_2^4)-5\tr (\CP_3^2 \CP_2^3) - 5 \tr[\CP_3 (\CP_2 \CP_3)^2] -5 \tr[\CP_2( \CP_3 \CP_2)^2], \qquad \\
&& L_i M^i_{~j} M^j_{~k} M^k = -\tr (\CP_2 \CP_1^4) \nn \\
&& = - \tr(\CP_2^5)- 4 \tr(\CP_3\CP_2^4)-3\tr (\CP_3^2 \CP_2^3) - 2\tr[\CP_3 (\CP_2 \CP_3)^2] -3 \tr[\CP_2( \CP_3 \CP_2)^2]. \qquad
\eea
Note also that
\bea
(\tr M)^2 \det M &=& -\tr (\CP_3^2 \CP_2^3), \\
(\tr M)^2 (\tr M^3) &=& - \tr (\CP_2^5) - 3 \tr(\CP_3 \CP_2^4) - 3 \tr(\CP_3^2 \CP_2^3) , \\
(\tr M^2) (\tr M^3) &=& -\tr(\CP_2)^5-5 \tr(\CP_3\CP_2^4) -9 \tr(\CP_3^2\CP_2^3) -6 \tr[\CP_3 (\CP_2 \CP_3)^2], \qquad \\
(\tr M)^3 (\tr M^2) &=& - \tr (\CP_2^5) - 2\tr( \CP_3 \CP_2^4)~. 
\eea
Therefore,
\bea
L_i M^i_{~j} M^{j}_{~k} R^k &=& \tr (M^5) - (\tr M)(\tr M^4) + \frac{1}{2} \left[5 (\tr M)^2 - (\tr M^2) \right] (\tr M^3) \nn \\
&& - (\tr M)^3 (\tr M^2) - (\tr M)^2 (\det M) ~.
\eea

\subsection{The relations for the mirror of the $(2)-[4]$ theory} \label{app:rel24}
Let us focus on the mirror of the $(2)-[4]$ theory.  The relations discussed in the main text can be derived in a similar way as the previous section, namely by writing various gauge invariant quantities in terms of matrices appearing the $F$-terms (which we previously called $\CP$'s) -- some of which are nilpotent.  However, in this section, we present a different method in obtaining those relation. In particular, we apply various matrix identities (\eg, the Cayley-Hamilton theorem) to the generators.  Although this method works well in this section, it becomes very cumbersome as the ranks of the groups in the theory increase.

\paragraph{The relation \eref{trMG024}.}  Consider
\bea
\tr (G_2 \cdot G_4) = (C_1)^a_{~j} (A_1)^j_{~c} (A_2)^c (C_2)_{d}  (C_1)^d_{~i} (A_1)^i_{~a}~. \label{trMG24a}
\eea
Using the $F$-term relation \eref{Ftermsmir241}, we find that
\bea
(C_1)^d_{~i} (A_1)^i_{~a} = -(A_2)^d (C_2)_{a} - T^d T'_{a}~.
\eea
Substituting this into \eref{trMG24a} and applying \eref{Fterms2}, we find that
\bea
\tr (G_2 \cdot G_4) &=& - (C_1)^a_{~j} (A_1)^j_{~c} (A_2)^c (C_2)_{d} T^d T'_{a} \nn \\
&=& \left[(A_2)^a (C_2)_{c} + T^a T'_{c}\right] (A_2)^c (C_2)_{d} T^d T'_{a} \nn \\
&=& 0~,
\eea
as required.

\paragraph{The relation \eref{detG024}.} Consider
\bea
\det G_4 &=&  \frac{1}{2} \epsilon^{i_1 i_2} \epsilon_{j_1 j_2} (G_4)^{j_1}_{~i_1} (G_4)^{j_2}_{~i_2} \nn \\
&=& \frac{1}{2} \epsilon^{i_1 i_2} \epsilon_{j_1 j_2} \left[(A_1)^{j_1}_{~a} (A_2)^a (C_2)_{b} (C_1)^b_{~i_1} \right] \left[ (A_1)^{j_2}_{~c} (A_2)^c (C_2)_{d} (C_1)^d_{~i_2}\right]~.
\eea
Now consider the factor $\epsilon_{j_1 j_2} (A_1)^{j_1}_{~a} (A_1)^{j_2}_{~c}$.  The epsilon symbol $\epsilon_{j_1 j_2}$ imposes the anti-symmetrisation on the indices $a, c$. The contraction of such a factor with $(A_2)^a(A_2)^c$ therefore yields zero.  Hence the determinant $\det G_4$ vanishes, as required.

\paragraph{The relations \eref{order824a}.}  
First of all, the characteristic equation for $G_4$ is $0=\lambda^2 - (\tr G_4) \lambda + (\det G_4) \BU = \lambda^2 - (\tr G_4) \lambda$.  Using the Cayley-Hamilton theorem, we find that $G_4^2 - (\tr G_4)G_4=0$. Taking the trace of both sides, we obtain
\bea
\tr (G_4^2) = (\tr G_4)^2~,
\eea
as stated in \eref{order824a}.

\paragraph{The relations \eref{order824}.}  Now we would like to show that
\bea
0 =  \tr (G_2^2 \cdot G_4)+ \left( \tr G_4 \right)^2~. \label{trG224}
\eea
Applying the Cayley-Hamilton theorem to $G_2$ and noting that $\tr G_2 =0$, we have $ G_2^2 +(\det G_2) \BU =0$.  Then, multiplying $G_4$ on the right to both sides, we have
\bea
G_2^2G_4 + (\det G_2) G_4 = 0~.
\eea
Taking trace of both sides and recalling that $\det G_2 = -\frac{1}{2} \tr (G_2^2) = \tr G_4$, we obtain \eref{trG224}, as required.

\paragraph{The relation \eref{trm4order824}.}
Now applying the Cayley-Hamilton theorem to the matrix $G_2^2$, we find that
$G_2^4 - (\tr G_2^2) G_2^2 + (\det G_2)^2 \BU =0$.  Taking the trace of both sides and using the relations which have been established, we obtain
\bea
\tr G_2^4 &=& (\tr G_2^2)^2 - 2 (\det G_2)^2 \nn \\
&=& 4 (\tr G_4)^2 - 2 \left[ - \frac{1}{2} \tr (G_2^2) \right]^2 \nn \\
&=& 4 (\tr G_4)^2 - 2 (\tr G_4)^2  \nn \\
&=& 2 (\tr G_4)^2~.  \label{trM4trGsq24}
\eea
Using \eref{order824}, we arrive at \eref{trm4order824}, as required.

%%%%%%%%%%%%%%%%%%%%%%%%%%%%%%%%
\subsection{The relations for the mirror of the $(3)-[6]$ theory} \label{app:rel36}
Let us focus on the mirror of the $(3)-[6]$ theory.  Before we proceed, let us defined the following shorthand notation:
\bea
(\CP_1)^a_{~b} := (C_1)^a_{~i} (A_1)^i_{~b}~, \quad (\CP_2)^a_{~b} := (A_2)^a_{~\alpha} (C_2)^\alpha_{b}~, \quad (\CP_3)^a_{~b} := (T_1)^a_{~a_3} (T'_1)^{a_3}_{~b}~.
\eea
The first $F$-term constraints in \eref{Fterms362} imply that
\bea
\CP_1 = - \CP_2 - \CP_3 ~. \label{P1P2P3036}
\eea
Recall that the $3 \times 3$ matrices $\CP_2$ and $\CP_3$ are nilpotent, \ie~ $\CP_2^k =0$ and $\CP_3^k=0$ for all $k \geq 3$.

\paragraph{The relation \eref{trm236}.} Observe that
\bea
\tr G_2^2 &=& \tr(\CP_1^2) = \tr \left[ (\CP_2 +\CP_3)^2 \right]~, \nn \\
\tr G_4 &=&  \tr( \CP_2 \CP_1) = -\tr [ \CP_2 (\CP_2 +\CP_3) ]~.
\eea
Since $\CP_2^k =0$ and $\CP_3^k=0$ for all $k \geq 3$, we also have $\tr(\CP_2^m) = \tr(\CP_3^m) = 0$ for all $m \geq 1$. Thus, we obtain
\bea
\tr G_2^2 &=& 2 \tr(\CP_2 \CP_3)~, \nn \\
\tr G_4 &=& -\tr(\CP_2 \CP_3)~.
\eea
Therefore, we see that 
\bea
\tr G_2^2 + 2\tr G_4 = 0~,
\eea
as required.

\paragraph{The relation \eref{trG636}.}  Observe that
\bea
\tr G_2^3 &=& \tr(\CP_1^3) = \tr \left[ (\CP_2 +\CP_3)^3 \right]~, \nn \\
\tr (G_2 \cdot G_4) &=& \tr( \CP_1^2 \CP_2) = \tr \left[ (\CP_2 +\CP_3)^2 \CP_2 \right]~, \nn \\
\tr G_6 &=&  \tr( \CP_2^2 \CP_1) = -\tr [ \CP_2^2 (\CP_2 +\CP_3) ]~.
\eea
Since $\CP_2$ and $\CP_3$ are nilpotent, we have $\tr(\CP_2^k) = \tr(\CP_3^k) = 0$ for all $k \geq 1$. Thus, we obtain
\bea
\tr G_2^3 &=&  -3\tr(\CP_2 \CP_3^2) -3 \tr(\CP_3 \CP_2^2)~, \nn \\
\tr (G_2 \cdot G_4) &=& \tr(\CP_2 \CP_3^2) + 2 \tr(\CP_3 \CP_2^2)~, \nn \\
\tr G_6 &=& - \tr(\CP_2 \CP_3^2)~.
\eea
Therefore, we have
\bea
2 \tr G_2^3 + 3 \tr (G_2 \cdot G_4) - 3 \tr G_6 =0~.
\eea
From \eref{vanishingoddcasimir}, we have $\tr G_2^3 =0$ and so
\bea
\tr G_6 = \tr (G_2 \cdot G_4)~.
\eea

\paragraph{The relations \eref{order836a},\eref{order836b} and \eref{order836c}.} We proceed in the same way as before.  Using \eref{P1P2P3036} and the fact that $\CP_2^k = \CP_3^k=0$ for all $k \geq 3$ and $\tr(\CP_2^m) = \tr(\CP_3^m) = 0$ for all $m \geq 1$, we obtain
\bea
\begin{array}{lll}
\tr G_2^4 &= \tr(\CP_1^4) &= 4 \tr( \CP_2^2  \CP_3^2 )+2  \tr [ (\CP_2  \CP_3)^2]~, \nn \\
\tr (G_2^2 \cdot G_4) &= \tr(\CP_1^3 \CP_2) &= -2\tr( \CP_2^2  \CP_3^2 )-  \tr [ (\CP_2  \CP_3)^2]~, \nn \\
\tr(G_2 \cdot G_6) &=  \tr (\CP_1^2 \CP_2^2) &=\tr( \CP_2^2  \CP_3^2 )~, \nn \\
\tr (G_4^2) &= \tr [ (\CP_1  \CP_2)^2] &=\tr [ (\CP_2  \CP_3)^2]~. \nn \\
\end{array}
\eea
Thus, it is easy to see that
\bea
0 &=& \tr G_2^4 + 2 \tr(G_2^2 \cdot G_4)~, \\
0&=& \tr (G_2^2 \cdot G_4)+2\tr(G_2 \cdot G_6) + \tr( G_4^2)~.
\eea
Thus, we obtain \eref{order836a} and \eref{order836b} as required.  
To derive \eref{order836c}, we simply use \eref{order836b} with \eref{trM4trGsq24}.

\paragraph{The relations \eref{order1036a} and \eref{order1036b}.}
Using \eref{P1P2P3036} and the fact that $\CP_2^k = \CP_3^k=0$ for all $k \geq 3$ and $\tr(\CP_2^m) = \tr(\CP_3^m) = 0$ for all $m \geq 1$, we obtain
\bea
\begin{array}{lll}
\tr G_2^5 &= \tr(\CP_1^5) &= -5 \tr( \CP_2 \CP_3 \CP_2 \CP_3^2 )-5 \tr( \CP_3 \CP_2 \CP_3 \CP_2^2 )~, \nn \\
\tr (G_2^3 \cdot G_4) &= \tr(\CP_1^3 \CP_2 \CP_1) &= 2 \tr( \CP_2 \CP_3 \CP_2 \CP_3^2 )+3 \tr( \CP_3 \CP_2 \CP_3 \CP_2^2 )~, \nn \\
\tr (G_2^2 \cdot G_6) &= \tr(\CP_1^2 \CP_2^2 \CP_1) &=  - \tr( \CP_3 \CP_2 \CP_3 \CP_2^2 )~, \nn \\
\tr (G_2 \cdot G_4^2) &=\tr[ (\CP_1 \CP_2)^2 \CP_1] &= - \tr( \CP_2 \CP_3 \CP_2 \CP_3^2 ) -2 \tr( \CP_3 \CP_2 \CP_3 \CP_2^2 )~, \nn \\
\tr(G_4 \cdot G_6) &= \tr( \CP_2 \CP_1 \CP_2^2 \CP_1) &= \tr( \CP_3 \CP_2 \CP_3 \CP_2^2 )~.
\end{array}
\eea
Then, it follows that
\bea
0 &=&  \tr G_2^5 +  5\tr(G_2^3 \cdot G_4) + 5\tr (G_2 \cdot G_4^2)~, \nn \\
0 &=& 2\tr G_2^5 + 5\tr(G_2^3 \cdot G_4) + 5\tr(G_2^2 \cdot G_6)~, \nn \\
0 &=& 5 \tr (G_2 \cdot G_4^2) + 5 \tr (G_2^2 \cdot G_6) +5 \tr(G_4 \cdot G_6) + 5 \tr (G_2^3 \cdot G_4)  +  \tr (G_2^5)~. \qquad
\eea
Using the fact that $\tr(G_2^5)=0$, we arrive that \eref{order1036a} and \eref{order1036b}.

\paragraph{The relation \eref{detG636}.}  This can be derived in a similar way as \eref{detG024}.
Observe that $\det G_6$ contains the following factor:
\bea
\epsilon_{i i'} (A_1)^i_{~a} (A_1)^{i'}_{~a'} (A_2)^a_{~a_2} (A_2)^{a'}_{~a'_2}(A_3)^{a_2}(A_3)^{a'_2}~.
\eea
The epsilon tensor imposes an anti-symmetrisation on the indices $i, i'$ and hence $a, a'$ and hence $a_2, a'_2$.  Since the anti-symmetrisation $(A_3)^{[a_2}(A_3)^{a'_2]}$ yields zero, it follows that the determinant $\det G_6$ is zero.

%%%%%%%%%%%%%%%%%%%%%%%%%%
\subsection{The relations for the mirror of the $(2)-(4)-[6]$ theory} \label{app:rel246}
Let us focus on the mirror of the $(2)-(4)-[6]$ theory.  Before we proceed, let us defined the following shorthand notation:
\bea
(\CP_1)^a_{~b} := (C_1)^a_{~i} (A_1)^i_{~b}~, \quad (\CP_2)^a_{~b} := (A_2)^a_{~\alpha} (C_2)^\alpha_{b}~, \quad (\CP_3)^a_{~b} := (T_1)^a_{~a_3} (T'_1)^{a_3}_{~b}~.
\eea
The second $F$-term constraints in \eref{FT246} imply that
\bea
\CP_1 = - \CP_2 - \CP_3 ~. \label{P1P2P30}
\eea
Recall that the $4 \times 4$ matrices $\CP_2$ and $\CP_3$ are nilpotent -- their Jordan normal forms are respectively given by \eref{JNFP2} and \eref{JNFP3}.

\paragraph{The relation \eref{trm3246}.}  The trace $\tr(G_2 \cdot G_4)$ can be written as
\bea
\tr(G_2 \cdot G_4) = \tr( \CP_1 \CP_2 \CP_1)~.
\eea
Substituting \eref{P1P2P30} into it, we find that
\bea
\tr(G_2 \cdot G_4) = \tr \left[  (\CP_2+\CP_3)^2 \CP_2 \right]~. \label{trmga}
\eea
Observe that this expression involves only nilpotent matrices.
Similarly, we can rewrite $\tr(G_2^3)$ as
\bea
\tr(G_2^3) = \tr( \CP_1^3) = -\tr \left[ (\CP_2+\CP_3)^3\right]~. \label{trm3a}
\eea

%For convenience, let us work on a basis with respect to which $\CP_2$ takes its Jordan normal form \eref{JNFP2}.  (Note that we do \emph{not} assume that $\CP_3$ takes its Jordan normal form with respect to this basis.)  

From \eref{JNFP2}, it is clear that $\CP_2^k =0$ for all $k \geq 2$ (note that this this true with respect to any basis, not just for the basis in which $\CP_2$ takes its Jordan normal form).  Since $\CP_3$ is nilpotent, we have $\tr(\CP_3^k) = 0$ for all $k \geq 1$. Then, from \eref{trmga}, we have
\bea
\tr(G_2 \cdot G_4) = \tr( \CP_2 \CP_3^2)~,
\eea
and, from \eref{trm3a}, we also find that
\bea
\tr(G_2^3) = - 3 \tr( \CP_2 \CP_3^2)~.
\eea
Therefore, we find that
\bea
\tr G_2^3 + 3 \tr(G_2 \cdot G_4) = 0~.
\eea

\paragraph{The relation \eref{trm4trg2}.} We write the trace of $G_2^2 \cdot G_4$ as
\bea
\tr(G_2^2G_4) = \tr( \CP_1^2 \CP_2) = \tr \left[ (\CP_2+\CP_3)^2 \CP_2 \right]~,
\eea
and similarly we have
\bea
\tr (G^2) &=& \tr( \CP_2 \CP_1 \CP_2 \CP_1) = \tr \left[ \left \{ \CP_2 (\CP_2+\CP_3)\right \}^2 \right]~, \nn \\
\tr (G_2^4) &=& \tr(\CP_1^4) = \tr \left[(\CP_2+\CP_3)^4 \right]~.
\eea

%Now we pick a basis with respect to which $\CP_2$ takes its Jordan normal form \eref{JNFP2}.  (Note that we do \emph{not} assume that $\CP_3$ takes its Jordan normal form with respect to this basis.)

From \eref{JNFP2}, it is clear that $\CP_2^k =0$ for all $k \geq 2$ (note that this this true with respect to any basis, not just for the basis in which $\CP_2$ takes its Jordan normal form).  Since $\CP_3$ is nilpotent, we have $\tr(\CP_3^k) = 0$ for all $k \geq 1$.  Thus, we have
\bea
\tr(G_2^4) &=& 4\tr( \CP_2 \CP_3^3) + 2\tr\left[ (\CP_2 \CP_3)^2 \right]~, \nn \\
\tr(G_2^2G_4) &=& -\tr( \CP_2 \CP_3^3) - \tr\left[ (\CP_2 \CP_3)^2 \right]~, \nn \\
\tr(G_4^2) &=& \tr\left[ (\CP_2 \CP_3)^2 \right]~.
\eea
Therefore, it is easy to see that
\bea
\tr(G_2^4) +4 \tr (G_2^2 \cdot G_4) + 2 \tr (G_4^2) = 0~.
\eea

\paragraph{The relation \eref{trm4trg2a}.}  Let $\lambda_1, \lambda_2, \lambda_3$ be the eigenvalues of $G_2$.  Then, by direct expansion, one can easily verify that
\bea
 \left( \sum_{i=1}^3 \lambda_i  \right)^2 = \sum_{i=1}^3 \lambda_i^2 + 2 e_2~,
\eea
where $e_2$ is the 2nd order elementary symmetric polynomial:
\bea
e_2 : = \sum_{1 \leq i < j \leq 3} \lambda_i \lambda_j~.
\eea
Recalling that $\tr G_2 =0$, we can rewrite this as
\bea
0 = \left[ \tr(G_2) \right]^2 = \tr(G_2^2) + 2 e_2~. \label{e2temp}
\eea

The Newton identity tells us that $e_2$ appears in the characteristic equation of $G_2$:
\bea
0 = \det(t I - G_2) &=&  t^3 - (\tr G_2) t^2 +e_2 t - (\det G_2) \nn \\
&=& t^3  +e_2 t - (\det G_2)~.
\eea
Applying the Cayley-Hamilton theorem to $G_2$, we find that
\bea
0 = G_2^3 + e_2 G_2 - (\det G_2)\BU~.
\eea
Multiplying by $G_2$ and taking the trace of both sides (with $\tr G_2 =0$), we obtain
\bea
e_2 \tr(G_2^2) = -\tr (G_2^4)~. 
\eea

Substituting $e_2$ into \eref{e2temp}, we find that
\bea
0 = \left[ \tr(G_2^2) \right]^2 - 2 \tr (G_2^4)~.
\eea
Using the relation \eref{trm22trG246}, we have
\bea
\tr G_2^4 - 2 \left[ \tr (G_4) \right]^2 = 0~.
\eea

\paragraph{The relations \eref{trm5246} and \eref{trm5246a}.}  The trace $\tr(G_2 \cdot G_4^2)$ can be written as
\bea
\tr(G_2 \cdot G_4^2) &=& \tr( \CP_1 \CP_2\CP_1 \CP_2 \CP_1 ) \nn \\
&=& - \tr \left[ (\CP_2+\CP_3) \CP_2 (\CP_2+\CP_3) \CP_2 (\CP_2+\CP_3) \right]~. \label{trmg2a}
\eea
Similarly, we can write
\bea
\tr(G_2^5) = \tr( \CP_1^5) = -\tr \left[ (\CP_2+\CP_3)^5\right]~, \label{trm5a}
\eea
and 
\bea
\tr (G_2^3 \cdot G_4) = \tr( \CP_1^3 \CP_2 \CP_1 )~. \label{trm3ga}
\eea

%For convenience, let us work on a basis with respect to which $\CP_2$ takes its Jordan normal form \eref{JNFP2}.  Note that we do \emph{not} assume that $\CP_3$ takes its Jordan normal form with respect to this basis.  

From \eref{JNFP2}, it is clear that $\CP_2^k =0$ for all $k \geq 2$ (note that this this true with respect to any basis, not just for the basis in which $\CP_2$ takes its Jordan normal form). Since $\CP_3$ is nilpotent, $\CP_3^4 = 0$.  Then, from \eref{trmg2a}, we have
\bea
\tr(G_2 \cdot G_4^2) = - \tr( \CP_2 \CP_3\CP_2 \CP_3^2)~,
\eea
from \eref{trm5a}, we also find that
\bea
\tr(G_2^5) = - 5 \tr(  \CP_2 \CP_3\CP_2 \CP_3^2)~,
\eea
and, from \eref{trm3ga}, we obtain
\bea
\tr (G_2^3 \cdot G_4) = 2 \tr(  \CP_2 \CP_3\CP_2 \CP_3^2)~.
\eea

Therefore, we find that
\bea
0 &=&  \tr (G_2^5) +5 \tr (G_2^3 \cdot G_4) + 5 \tr (G_2 \cdot G_4^2) ~, \nn \\
0 &=&  2 \tr(G_2 \cdot G_4^2) + \tr(G_2^3 \cdot G_4)~.
\eea

\paragraph{The relation \eref{detg0246}.}   The derivation is similar to that of the $(2)-[4]$ theory.  
\bea
\det G_4 &=&  \frac{1}{3!} \epsilon^{i_1 i_2 i_3} \epsilon_{j_1 j_2 j_3} (G_4)^{j_1}_{~i_1} (G_4)^{j_2}_{~i_2} (G_4)^{j_3}_{~i_3} \nn \\
&=& \frac{1}{3!} \epsilon^{i_1 i_2 i_3} \epsilon_{j_1 j_2 j_3} \left[(A_1)^{j_1}_{~a} (A_2)^a_{~\alpha} (C_2)^\alpha_{~b} (C_1)^b_{~i_1} \right] \left[ (A_1)^{j_2}_{~c} (A_2)^c_{~\beta} (C_2)^\beta_{~d} (C_1)^d_{~i_2}\right]  \nn \\
&& \qquad \times \left[ (A_1)^{j_3}_{~e} (A_2)^e_{~\gamma} (C_2)^\gamma_{~f} (C_1)^f_{~i_3}\right]~.
\eea
Consider the factor $\epsilon_{j_1 j_2 j_3}(A_1)^{j_1}_{~a}(A_1)^{j_2}_{~c}(A_1)^{j_3}_{~e}$.  We see that the epsilon tensor $\epsilon_{j_1 j_2 j_3}$ imposes the totally anti-symmetrisation on the indices $a, c, e$.  The contraction of such a factor with $(A_2)^a_{~\alpha} (A_2)^c_{~\beta} (A_2)^e_{~\gamma}$ yields zero, because the indices $\alpha, \beta, \gamma$ run from 1 to 2.  Thus, we have $\det G_4 =0$.

\section{The Hilbert series of Higgs branches of certain theories} 
In this section, we focus on Higgs branches of certain theories where the Hilbert series can be written in terms of character expansions.

\subsection{The Higgs branch of the $(1)-(2)-[3]$ theory} \label{App:Higgs123}
As discussed in Section \ref{sec:123n}, the Higgs branch and Coulomb branch of the $(1)-(2)-[3]$ theory are identical.
From \eref{hs123n}, the Hilbert series of the Higgs/Coulomb branch of the $(1)-(2)-[3]$ theory can be written in terms of $SU(3)$ representations as
\bea
&& H_{(1)-(2)- [3]} (t,  x_1, x_2)= (1-t^4) (1-t^6) \PE[[1,1]t^2]  \\
&& = \sum_{n_1, n_2 =0}^\infty [n_1+n_2,n_1+n_2]t^{2n_1+4n_2} + \nn \\
&& \quad \sum_{n_1, n_2,n_3 =0}^\infty  \Big( [n_1+n_2+3n_3+3,n_1+n_2] + [n_1+n_2, n_1+n_2+3n_3+3] \Big) t^{2n_1+4n_2+6n_3+6} . \nn
\eea
Setting $x_1=x_2=1$, we obtain the unrefined Hilbert series
\bea
H_{(1)-(2)- [3]} (t)= \frac{\left(1-t^4\right)\left(1-t^6\right)}{\left(1-t^2\right)^8} = \frac{\left(1+t^2\right) \left(1+t^2+t^4\right)}{\left(1-t^2\right)^6}~.
\eea
This indicates that the Higgs/Coulomb branch is a 6 complex dimensional complete intersection; this is in agreement with \eref{dimhiggs123n}.

\subsection{The Higgs branch of the $(1)-(2)-[4]$ theory} \label{App:Higgs124}
The Hilbert series of the Higgs branch of the $(1)-(2)-[4]$ theory can be computed using the gluing technique:
\bea
H^{\text{Higgs}}_{(1)-(2)- [4]} (t,  x_1, x_2, x_3) &=&   \int \ud \mu_{U(2)} (z_1,z_2) H_{(1)-[2]} (t, z_1,z_2)  H_{\text{glue}} (t,z_1,z_2)  \times \nn \\
&& \quad H_{[2]-[4]} (t, z_1,z_2, q, x_1, x_2, x_3)~,
\eea
where $(z_1, z_2)$ are $U(2)$ fugacities, $(x_1,x_2,x_3)$ are $SU(4) \subset U(4)$ fugacities, and $q$ is a $U(1) \subset U(4)$ fugacity.  The Haar measure is given by \eref{HaarU2}, the Hilbert series $H_{(1)-[2]}$ is given by \eref{hs12w1w2}, and the Hilbert series $H_{[2]-[4]}$ is given by
\bea
H^{\text{Higgs}}(t, z_1,z_2, q, x_1, x_2, x_3) &=& \PE \left[ q \left(\frac{1}{z_1}+\frac{1}{z_2} \right) \left( x_1 +\frac{x_2}{x_1}+ \frac{1}{x_2} \right) t +  \nn \right. \nn \\
&& \qquad \left.  \frac{1}{q} \left(z_1+z_2 \right) \left( \frac{1}{x_1} +\frac{x_1}{x_2}+ x_2 \right) t\right]~.
\eea
This integral can be written in terms of $SU(4)$ representations as
\bea
&& H^{\text{Higgs}}_{(1)-(2)- [4]} (t,  x_1, x_2, x_3)=  \nn \\
&& \quad \sum_{n_1, n_2, n_3 =0}^\infty [n_1 + n_3, 2 n_2, n_1 + n_3] t^{2 n_1 + 4 n_2 + 4 n_3} +  \nn \\
&& \sum_{n_1, n_2, n_3, n_4 =0}^\infty \Big( [n_1 + n_3, 2 n_2 + n_4 + 1, n_1 + n_3 + 2 n_4 + 2] \nn \\
&& \qquad \qquad \quad + [n_1 + n_3 + 2 n_4 + 2, 2 n_2 + n_4 + 1, n_1 + n_3] \Big) t^{2 n_1 + 4 n_2 + 4 n_3 + 6 n_4 + 6}~.
\eea
Setting $x_1=x_2=x_3=1$, we obtain the unrefined Hilbert series
\bea
H^{\text{Higgs}}_{(1)-(2)- [4]} (t)= \frac{1+5 t^2+14 t^4+14 t^6+5 t^8+t^{10}}{\left(1-t^2\right)^{10}}~.
\eea
This indicates that the Higgs branch is 10 complex dimensional; this is in agreement with \eref{dimhiggs124}.

\subsection{The Higgs branch of the $(1)-(2)-[5]$ theory} \label{App:Higgs125}
The Hilbert series of the Higgs branch of the $(1)-(2)-[5]$ theory can be written in terms of $SU(5)$ representations as
\bea
&& H^{\text{Higgs}}_{(1)-(2)- [5]} (t,  x_1, x_2, x_3,x_4)=  \nn \\
&& \quad \sum_{n_1, n_2, n_3 =0}^\infty [n_1 + n_3, n_2, n_2, n_1 + n_3] t^{2 n_1 + 4 n_2 + 4 n_3}+ \nn \\
&& \sum_{n_1, n_2, n_3, n_4 =0}^\infty  \Big( [n_1 + n_3, n_2 + n_4 + 1, n_2, n_1 + n_3 + 2 n_4 + 2] + \nn \\
&& \qquad \qquad \quad   [n_1 + n_3 + 2 n_4 + 2, n_2, n_2 + n_4 + 1, n_1 + n_3] \Big) t^{2 n_1 + 4 n_2 + 4 n_3 + 6 n_4 + 6}~.
\eea
Setting $x_1=x_2=x_3=x_4=1$, we obtain the unrefined Hilbert series
\bea
H^{\text{Higgs}}_{(1)-(2)- [5]} (t)= \frac{1+10 t^2+54 t^4+110 t^6+110 t^8+54 t^{10}+10 t^{12}+t^{14}}{\left(1-t^2\right)^{14}}~.
\eea
This indicates that the Higgs branch is 14 complex dimensional; this is in agreement with \eref{dimhiggs125}.

\end{document}